\documentclass[manuscript, nonacm]{acmart}

\AtBeginDocument{%
  \providecommand\BibTeX{{%
    \normalfont B\kern-0.5em{\scshape i\kern-0.25em b}\kern-0.8em\TeX}}}

\usepackage{hyperref}
\usepackage[framemethod=tikz]{mdframed} % 
\usepackage{enumitem}
\usepackage{stfloats}
\usepackage{url}
\usepackage{soul}
\usepackage{colortbl}
\usepackage{adjustbox}
\usepackage{multirow}
\usepackage{tikz}
\usepackage{xcolor}
\usepackage{refcount}
\usepackage{tabularx}
\usepackage{threeparttable}
\usepackage{multicol}
\usepackage{float}
\usepackage{siunitx}
\usepackage{booktabs}
\usepackage{graphicx}
\usepackage[most]{tcolorbox} 

\usepackage[normalem]{ulem}
\definecolor{vlightgray}{gray}{0.9}
\definecolor{myBlue}{RGB}{200,213,231}  %NOTE: "RGB" and colorname are CASE-SENSITIVE
\definecolor{myYellow}{RGB}{255,242,204}  %NOTE: "RGB" and colorname are CASE-SENSITIVE
\definecolor{purple}{RGB}{128,0,128}

\definecolor{TekheletPurple}{HTML}{572F96}

\definecolor{TakeawayTeal}{HTML}{3D8B8B}
\newtcolorbox{takeawayBox}{
  breakable,
  enhanced,
  sharp corners,
  colback=TakeawayTeal!15!white,      % Light teal background
  colframe=TakeawayTeal,              
  boxrule=0.8pt,
  left=3pt,
  right=3pt,
  top=3pt,
  bottom=3pt,
  before skip=5pt,
  after skip=5pt,
  fonttitle=\bfseries\small,
  coltitle=TakeawayTeal,    % Darker teal for title
  attach title to upper={},
  separator sign={\ },
}

\hypersetup{
 colorlinks=true,   
 citecolor=black,
 urlcolor=blue,
 linkcolor=black,
 bookmarksnumbered=true,     
 bookmarksopen=true,         
 bookmarksopenlevel=1,          
 pdfstartview=Fit,           
 pdfpagemode=UseOutlines,    
 pdfpagelayout=TwoPageRight 
}

\newcolumntype{P}[1]{>{\raggedright\arraybackslash}p{#1}}

\newcommand{\boldification}[1]{\ifdraft\indent **\textbf{#1}**\\\indent\else\relax\fi}

\usepackage{xcolor}
\usepackage{framed}
\newenvironment{bluebox}{%
    \MakeFramed{\advance\hsize-\width\FrameRestore}%
    \itshape % Italic text
}{%
    \endMakeFramed
}

\newif\ifdraft
\draftfalse %hides boldifications and comments
\newif\ifrevising
   \revisingfalse %makes revisions normal appearance
 \newcommand{\deleted}[1]{{\ifrevising{\relax}\else\relax\fi}}

\begin{document}

\title{Why Johnny Can't Think: GenAI Induces a Cognitive Debt Cycle}
\title{Why Johnny Can't Think: GenAI's Cognitive Debt Cycle}
\title{Why Johnny Can't Think: GenAI's Vicious Cycle of  Cognitive Debt}
\title{Why Johnny Can't Think: The Why's, How's, and Who's}
\title{Why Johnny Can't Think: GenAI's Impacts on Cognitive Engagement}
\title{Thinking Less, Trusting More: GenAI's Impacts on Students' Cognitive Habits}
%\title{Why Johnny Can't Think: \\GenAI's Why/How/Who of Spiraling Cognitive Debt}
%\title{Why Johnny Can't Think: How GenAI Affects Students’ Reflection, Understanding, and Critical Thinking}
% \\ commented
% **or: AI-induced atrophy of human cognition\\
% **or: how to accelerate cognitive atrophy\\
% **or: from Artificial Intelligence to Declining Intelligence: An Endangered Ecosystem \\
% **or: more title ideas here**}

% \title{Atrophy by Automation: The Cognitive Consequences of GenAI for Reflective and Critical Thinking in CS Education}

% \author{
% \IEEEauthorblockN{Anonymous Authors% % \vspace{-3mm}}
% }

\author{Rudrajit Choudhuri$^*$}
\affiliation{%
  \institution{Oregon State University}
  \city{Corvallis}
  \state{OR}
  \country{USA}
}
\email{choudhru@oregonstate.edu}
\thanks{$^*$Corresponding author. Email: choudhru@oregonstate.edu}

\author{Christopher A. Sanchez}
\affiliation{%
  \institution{Oregon State University}
  \city{Corvallis}
  \state{OR}
  \country{USA}
}
\email{christopher.sanchez@oregonstate.edu}

\author{Margaret Burnett}
\affiliation{%
  \institution{Oregon State University}
  \city{Corvallis}
  \state{OR}
  \country{USA}
}
\email{burnett@eecs.oregonstate.edu}

\author{Anita Sarma}
\affiliation{%
  \institution{Oregon State University}
  \city{Corvallis}
  \state{OR}
  \country{USA}
}
\email{anita.sarma@oregonstate.edu}

%
% By default, the full list of authors will be used in the page
% headers. Often, this list is too long, and will overlap
% other information printed in the page headers. This command allows
% the author to define a more concise list
% of authors' names for this purpose.
% \renewcommand{\shortauthors}{Choudhuri et al.}

% \newcommand{\explaintwo}[1]{
% % % \vspace{2mm}%
% \par%
% \noindent\fbox{%
%     \parbox{\dimexpr\linewidth \fbo \fboxrule}{#1}%
% }%
% % % \vspace{2mm}%
% }

\newcommand{\grayline} {\arrayrulecolor{gray}\hline\arrayrulecolor{black}}

\newcommand{\explaintwo}[1]{%
\par%
\noindent\fbox{%
    \parbox{\dimexpr\linewidth-2\fboxsep-2\fboxrule}{#1}%
}%
}

\renewcommand{\shortauthors}{Choudhuri et al.}

\settopmatter{printacmref=false}
\setcopyright{none}
\renewcommand\footnotetextcopyrightpermission[1]{}
% \pagestyle{plain}

% \newcounter{guidelineno}
%\newcommand{\guideline}[1]{\noindent\textcolor{blue}{\textbf{DG \refstepcounter{guidelineno}\theguidelineno:} #1}}

%\newcounter{fullguidelineno}
% \newcommand{\fullguideline}[2]{\noindent\textcolor{blue}{\textbf{DG \refstepcounter{guidelineno}\theguidelineno: #1} #2}}
\newcommand{\blue}[1]{\textcolor{blue}{#1}}

%\newcommand{\guideline}[1]{}

% \maketitle
% \IEEEpeerreviewmaketitle

\begin{abstract}

\noindent\textbf{Objectives}: 
When students use generative AI in coursework, what are its persistent effects on their intellectual development? We investigate (RQ1-How) how students’ trust in and routine use of genAI affect their cognitive engagement habits in STEM coursework, and (RQ2-Who) which students are particularly vulnerable to cognitive disengagement.

\noindent\textbf{Method}: 
Drawing on dual-process, cognitive offloading, and automation bias theories, we developed a statistical model explaining how and to what extent students’ trust-driven routine genAI use affected their cognitive engagement---specifically, reflection, the need for understanding, and critical thinking in coursework, and how these effects differed across students’ cognitive styles. We empirically evaluated this model using Partial Least Squares Structural Equation Modeling on survey data from \textbf{299} STEM students across five North American universities.

\noindent\textbf{Results}: 
Students who trusted and routinely used genAI reported significantly lower cognitive engagement. Unexpectedly, students with higher technophilic motivations, risk tolerance, and computer self-efficacy---traits often celebrated in STEM---were more prone to these effects. Interestingly, students' prior experience with genAI or academia did not protect them from cognitively disengaging.
%Neither academic seniority nor prior experience mitigated these effects.

\noindent\textbf{Implications:} 
Our findings suggest a potential \textit{cognitive debt cycle} where routine genAI use weakens students' intellectual habits, potentially driving and escalating over-reliance. This poses challenges for curricula and genAI system design, requiring interventions that actively support cognitive engagement.
\end{abstract}

%%
%% The code below is generated by the tool at http://dl.acm.org/ccs.cfm.
%% Please copy and paste the code instead of the example below.
%%
% \begin{CCSXML}
% <ccs2012>
% <concept>
% <concept_id>10003120.10003121.10011748</concept_id>
% <concept_desc>Human-centered computing~Empirical studies in HCI</concept_desc>
% <concept_significance>500</concept_significance>
% </concept>
% </ccs2012>
% \end{CCSXML}

% \ccsdesc[500]{Human-centered computing~Empirical studies in HCI}
%%
%% Keywords. The author(s) should pick words that accurately describe
%% the work being presented. Separate the keywords with commas.
% \begin{IEEEkeywords}
% \end{IEEEkeywords}

\maketitle

\section{Introduction}
\label{sec:intro}

\begin{bluebox}
``People are neither fully rational nor completely selfish — and they are often lazy thinkers.'' --- Daniel Kahneman
\end{bluebox}

Human cognition evolved to be lazy and for good reason. Our ancestors who conserved mental energy for genuine threats survived; those who exhaustively analyzed every decision did not. This ``law of least effort'' pervades human behavior~\cite{risko2016cognitive, sparrow2011google, kahneman2011thinking}. We outsource memory to shopping lists, arithmetic to calculators, and navigation to GPS, trading internal cognitive work for external support. Generative AI (genAI) tools (e.g., ChatGPT, Claude, Copilot) tend to amplify this tendency by delivering fluent, confident support on demand. As a result, interacting with genAI involves less ``thinking by doing'' and more ``choosing from outputs''~\cite{lee2025impact}. In many domains, this reallocation saves time, improves productivity, and creates a compelling sense of progress~\cite{nfw2025}.

But what works for productivity can be problematic for learning. In STEM education, for example, genAI may remove the desirable difficulties students need~\cite{barcaui2025chatgpt}: slow, effortful, and often uncomfortable cognitive work that builds intuition and transferable skills~\cite{chi2014icap}. Students may still produce equal (or even better) quality work yet learn less~\cite{walker2025learningOutcomes}. Recent studies indeed show that genAI assistance reduced depth of inquiry, evidence use, and self-regulatory behaviors in students~\cite{stadler2024cognitive, fan2025beware, kosmyna2025your, barcaui2025chatgpt, yan2025beyond, kreijkes2025effects}. 

Moreover, automation bias research shows that when systems appear competent, users trust outputs with limited scrutiny---even when those outputs conflict with their own knowledge or judgment~\cite{goddard2012automation, lyell2017automation}. In human-AI interaction (HAI) contexts, this tendency can manifest as over-reliance~\cite{choudhuri2024far, hamid2025inclusive}, with users incorporating AI suggestions without much critical evaluation~\cite{buccinca2021trust, lee2025impact}.
% ~\citet{lee2025impact} found that knowledge workers with higher confidence in genAI reported less critical thinking and cognitive effort in tasks.
These effects are amplified in settings that reward immediate efficiency, including professional work~\cite{nfw2025, lee2025impact}, and increasingly academic contexts~\cite{walker2025learningOutcomes}, where genAI use is becoming routine.

Consider these two scenarios:

\begin{itemize} [topsep=1pt]
    \item \textbf{Scenario A (episodic)}: A student uses genAI to bypass a single difficult problem, showing momentary \textit{overreliance} and \textit{reduced cognitive effort}.

    \item \textbf{Scenario B (routine)}: Through trust-driven routine genAI use, that student develops a general unwillingness to \textit{engage} in reflection, understanding, and critical thinking \textit{across coursework}---fundamentally recalibrating what cognitive effort investment feels ``worth it''.
\end{itemize}

Prior work has largely focused on Scenario A~\cite{stadler2024cognitive, fan2025beware, kosmyna2025your, barcaui2025chatgpt, yan2025beyond, chen2025more, kreijkes2025effects}. 
In contrast, this paper focuses on Scenario B---how and to what extent \textit{routine} use of genAI affects students' cognitive engagement habits.
If such effects exist, does it mean that STEM education is cultivating a generation adept at querying intelligence but unwilling to exercise their own?
Will they enter professional work having never developed the cognitive habits that prior generations built through practice? 
As a first step towards answering these long-term questions, we ask: 
\textbf{RQ1-How: \textit{To what extent does students' trust in and routine use of genAI affect their cognitive engagement habits in STEM coursework?}}

Drawing on decades of research in educational psychology and learning sciences~\cite{dewey1933we, biggs2001revised, pintrich1990motivational, kember2000development, bryce2003engaging, li2023study, gong2025impact}, we conceptualize cognitive engagement as three intellectual habits: 
(1) \textit{Reflection} (the metacognitive monitoring and evaluation of one’s understanding)~\cite{dunlosky2005self, kember2000development, silver2023using}, 
(2) \textit{Need for understanding} (the intrinsic motivation to seek conceptual coherence rather than surface-level correctness)~\cite{chi2009active, cacioppo1982need, biggs2001revised}, and 
(3) \textit{Critical thinking} (the analytic evaluation of evidence, assumptions, and alternatives)~\cite{facione1990critical, sosu2013development, facione2011critical} (detailed in Sec.~\ref{sec:cognitive-engagement}). 
Foundational education research shows that these habits foster durable learning and skill development~\cite{boekaerts1999handbook, evans2003approaches}; 
RQ1-How investigates how/how much all three are simultaneously affected when students trust and routinely offload cognitive work to genAI.

However, these effects are unlikely to be the same for all individuals. Prior work shows that differences in individuals' cognitive styles~\cite{sternberg1997cognitive, burnett2016gendermag} substantively shape the ways they use technology~\cite{martinez2017assessing, anderson2024-GMstylesInAI-TiiS, burnett2016-GMfieldStudy-CHI, vorvoreanu2019-gmInMicrosoftApp-CHI}. 
(We expand upon the specific cognitive styles in Sec.~\ref{model-dev}).
Further, recent HAI studies suggest that these differences, alongside trust, influence how individuals adopt genAI~\cite{choudhuri2025needs, choudhuri2025ai}. 
%Since these differences naturally exist among students too, we investigate whether/to what extent disengagement effects are unevenly distributed across diverse individuals. 
Thus, we ask: \textbf{RQ2-Who: Who (i.e., students with which cognitive styles) are particularly prone to cognitive disengagement?
}

\boldification{Our contributions are threefold. And speaks about a broader concern beyond STEM education}

Our study makes three contributions: 
(1) a theoretically grounded statistical model explaining (for RQ1-How) how students' trust-driven routine use of genAI affects their cognitive engagement habits in coursework, and (for RQ2-Who) who are particularly affected. 
We then empirically evaluate this model using Partial Least Squares–Structural Equation Modeling (PLS-SEM) on survey data from 299 STEM students across five North American universities.
This model and its empirical evaluation also provide (2)~a psychometrically validated survey instrument for measuring these constructs in student–genAI interaction contexts.
Finally, we derive (3)~implications and open challenges for educators and researchers seeking to support diverse individuals' cognitive engagement in genAI-augmented STEM education.

% \textbf{* The remainder of this paper is organized as follows:}

% The remainder proceeds as follows. Sec.\ref{sec: related-work} reviews prior work on genAI's effects on cognition and learning, and situates our contribution. Sec.\ref{sec:theory} introduces the theoretical foundations of cognitive engagement and develops the proposed model. Sec.\ref{sec:method} describes study methodology, including survey design, data collection, and analysis procedures. Sec.\ref{results} reports findings. Sec.\ref{sec:discussion} discusses implications for genAI-augmented learning and identifies open challenges. Sec.\ref{sec:conclusion} concludes with calls for future research.
% % \vspace{-3mm}
\section{Related Work}
\label{sec: related-work}

The past few years have seen a rapid adoption of genAI in educational settings, prompting growing interest in how these systems shape students’ learning processes and outcomes. In response, a substantial body of work examined the effects of genAI use on cognition, learning, and task performance across educational levels and domains~\cite{gerlich2025ai, kosmyna2025your, walker2025learningOutcomes, barcaui2025chatgpt, qu2025generative, prather2025beyond, chen2025more}.

Much of this work reports a consistent pattern: genAI supported short-term efficiency gains (e.g., faster ideation, improved writing fluency, note-taking, and problem-solving) while coinciding with weaker higher-order learning outcomes (e.g., creativity, metacognition, self-regulation)~\cite{walker2025learningOutcomes, yan2025beyond, kreijkes2025effects, chen2025more}. For instance, in secondary education, \citet{kreijkes2025effects} found that students who relied on AI-generated notes perceived comprehension tasks as easier and expended less effort, yet demonstrated weaker retention than peers who generated their own notes.

Related findings were reported in higher education contexts. Several studies observed that while genAI use reduced perceived task difficulty and mental effort, it was also associated with diminished depth of inquiry, evidence use, reasoning quality, and self-regulatory behaviors~\cite{stadler2024cognitive, fan2025beware, barcaui2025chatgpt}. For example, \citet{barcaui2025chatgpt} found that undergraduates given unrestricted ChatGPT access for learning performed substantially worse on assessments administered weeks later, attributing this decline to the bypassing of “desirable difficulties” (e.g., self-quizzing, note-taking) essential for consolidating knowledge. Similarly, \citet{fan2025beware} reported that students revising essays with ChatGPT showed reduced motivation to rely on source material, exhibiting “metacognitive laziness” when progress felt effortless.

Other work examined how these effects varied with how genAI was incorporated into learning activities~\cite{choudhuri2025insights, qin2025role}. In one such study, \citet{choudhuri2025insights} found that software engineering students benefited when genAI was used to clarify concepts or resolve impasses following independent effort, but experienced breakdowns when it was used prematurely---during initial learning or for complex implementations without sufficient foundational understanding.

Beyond immediate task performance, several studies suggested that genAI use could create an ``illusion of competence’’ and a sense of progress that outpaced actual understanding~\cite{zhang2025paradox}. For example, Prather and colleagues~\cite{prather2024widening, prather2023s} found that students using genAI tools reported higher confidence and completed tasks more quickly (sometimes producing higher-quality artifacts), yet faced metacognitive difficulties (e.g., challenges in debugging, explaining their own solutions, or adapting when AI-generated suggestions failed)~\cite{prather2024widening, prather2025beyond}.

Complementing these behavioral findings, \citet{kosmyna2025your} examined neurophysiological correlates of cognitive offloading. In a controlled lab study, college students writing with genAI assistance showed reduced activation in brain networks associated with creativity and cognitive control, alongside weaker subsequent memory recall and a lower sense of ownership over their written contributions. 

Notably, these dynamics extend beyond classrooms. Studies of genAI use among knowledge workers reported similar patterns of cognitive offloading~\cite{lee2025impact, nfw2025}. Trust in AI appeared to further amplify these effects: when systems were perceived as competent, users were more likely to accept outputs with limited scrutiny~\cite{buccinca2021trust, lepp2025does}. For example, \citet{lee2025impact} found that workers with higher confidence in genAI engaged less in creativity and critical thinking during tasks, shifting from deliberate reasoning toward passive evaluation of AI-generated content. Broader population studies likewise linked frequent reliance on genAI with increased offloading, work-slop, and overconfidence~\cite{gerlich2025ai, miklian2025new}.

Overall, prior work has largely focused on episodic cognitive offloading and its immediate effects on learning and task performance. What remains underexplored is how routine genAI use during formative education reshapes students’ enduring habits of cognitive engagement. Our study addresses this gap by examining how students’ trust in and routine use of genAI affect students’ reflection, need for understanding, and critical thinking in STEM coursework (\textit{RQ1-How}). Further, while prior work has shown that variations in individuals’ cognitive styles affect technology adoption~\cite{anderson2024-GMstylesInAI-TiiS, anderson2025llm, choudhuri2025needs, choudhuri2025ai}, our work is the first, to our knowledge, to investigate how these individual differences condition students’ vulnerability to genAI-related cognitive disengagement (\textit{RQ2-Who}).

\section{Foundations and Model Development}
\label{sec:theory}
We draw on foundational work in HCI and cognitive psychology to develop a theoretical model explaining two pathways of influence (see Fig.~\ref{fig:theory-model}): (1) how students' trust in and routine genAI usage affects their cognitive engagement habits (RQ1-How), while (2) accounting for variations in their cognitive styles (RQ2-Who).

% Fig.~\ref{fig:theory-model} depicts the proposed model.

%, where Each path in corresponds to a hypothesis (e.g., Trust$\rightarrow$Usage is H1). (Fig.~\ref{fig:theory-model})

\subsection{Cognitive engagement}
\label{sec:cognitive-engagement}

Consider what happens when you encounter a confusing passage in a textbook. One response is to highlight it and move on, trusting that re-reading will somehow clarify things later. Another response is to stop, notice the confusion, and deliberately work to resolve it: \textit{``Wait—how does this claim follow from the previous one? What am I missing?''} The second response reflects \textit{cognitive engagement}. It involves making sense of ideas, recognizing gaps in your own understanding, and investing effort to close them, rather than simply collecting information or reproducing facts~\cite{fredricks2004school, lamborn1992significance, boekaerts1999handbook}.

Educational psychology has long characterized cognitive engagement habits across three core intellectual dimensions: (a) \textit{Reflection}, (b) \textit{Need for understanding}, and (c) \textit{Critical thinking}~\cite{dewey1933we, biggs2001revised, pintrich1990motivational, kember2000development, bryce2003engaging, li2023study, gong2025impact}. Learners who cultivate these habits are more likely to develop durable understanding and transferable skills than those who rely on more superficial strategies~\cite{biggs2001revised, pintrich1990motivational, kember2000development, gong2025impact}. In this study, we therefore conceptualize cognitive engagement along these dimensions. Below, we define each dimension and describe how it is operationalized in our model.

\textbf{Reflection} is the metacognitive habit of monitoring one's own understanding, evaluating assumptions and strategies, and recognizing gaps in thinking~\cite{dunlosky2005self}. For example, when you read that ocean temperatures are rising by 0.13°C per decade, do you catch yourself wondering whether that's fast or slow, and what it implies? Education reformer John Dewey described reflection as ``overcoming the inertia that inclines one to accept suggestions at their face value''~\cite{dewey1933we}. Research shows that reflective learners are better positioned to catch their own errors, recognize when they're stuck, and adjust learning strategies adaptively~\cite{dunlosky2005self, kember2000development, silver2023using}. 

\begin{figure*}[!h]
\centering
% % \vspace{-10px}
\includegraphics[width=0.85\textwidth]{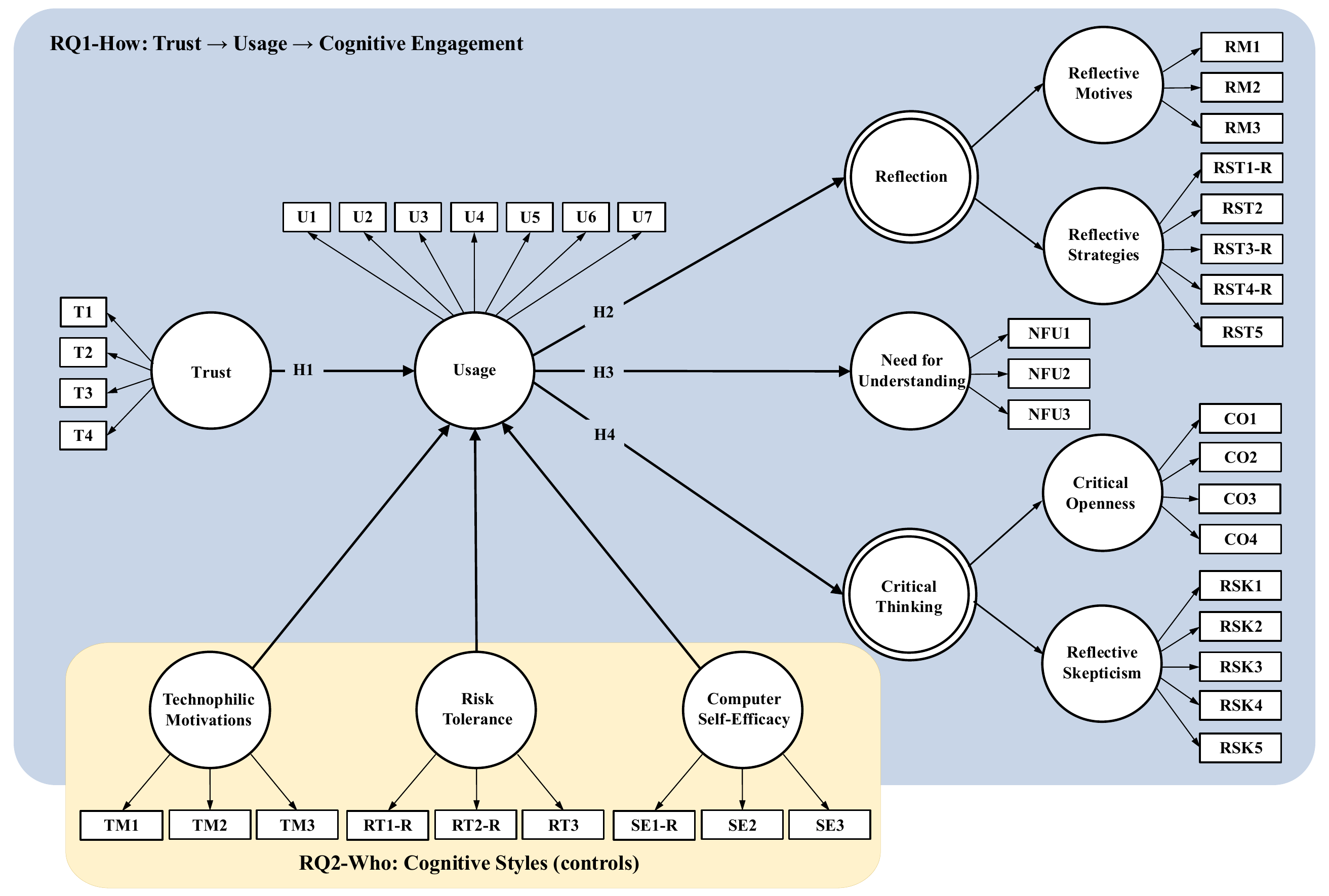}
% \vspace{-5px}
\caption{Proposed theoretical model and measurement specification. The blue region (RQ1-How) models how \textit{Trust} and routine \textit{Usage} affect cognitive engagement: (a) \textit{Reflection}, (b) \textit{Need For Understanding}, and (c) \textit{Critical Thinking}.
The yellow region (RQ2-Who) controls for cognitive styles (\textit{Technophilic Motivations}, \textit{Risk Tolerance}, \textit{Computer Self-Efficacy}). Latent constructs (circles) are reflectively measured by survey items (squares)~\cite{russo2021pls}; higher-order constructs (double circles) are reflectively measured via their lower-order constructs (e.g., \textit{Reflection} is measured via \textit{Reflective Motives} \& \textit{Strategies}). Reverse-coded items carry the “-R” suffix (e.g., RT1-R). Directed arrows represent hypothesized relationships (H1–H4) and cognitive-style controls. Item wordings (questionnaire) are in~\cite{supplemental}.}
% \vspace{-5px}
\label{fig:theory-model}
% \vspace{-7px}
\end{figure*}

Reflection is a latent psychological construct (subjective, intangible, and not directly observable) and is therefore captured through validated self-report questionnaires~\cite{raykov2011introduction, biggs2001revised, kember2000development}. In our study, we adopt Biggs’ validated framework~\cite{biggs2001revised}, which conceptualizes \textit{Reflection} as a combination of \textit{Reflective Motives} (why learners reflect) and \textit{Reflective Strategies} (how they do so). 
Consistent with this framework, we model Reflection as a higher-order construct (HOC) composed of these two lower-order constructs (LOCs)~\cite{hair2014primer}, represented as circles in Fig.~\ref{fig:theory-model}. Each LOC is itself latent and measured through multiple self-report items (questions); in our model, Reflective Motives and Reflective Strategies are measured using three and five items respectively (denoted as squares in Fig.~\ref{fig:theory-model}).

%\footnote{Latent constructs such as reflective motives are measured via multiple self-report items (questions) each capturing different facets of the construct \hl{[CITE]}.}
\textbf{Need for Understanding} captures a learner’s intrinsic drive to make sense of things~\cite{chi2009active, cacioppo1982need}. Some learners are satisfied once they can execute a procedure correctly. Others experience cognitive discomfort until they grasp \textit{why} it works. This is not stubbornness or perfectionism---it is a stable difference in what feels intellectually satisfying~\cite{cacioppo1982need}. Learners high in this trait habitually press for explanations and expend cognitive effort to construct meaning, not because they're told to, but because surface-level answers feel incomplete. Unsurprisingly, research shows they learn more effectively: their constructive sensemaking processes foster effective self-regulation that produces durable learning~\cite{chi2014icap,evans2003approaches}.
Following prior work~\cite{biggs2001revised}, we model Need for Understanding as a unidimensional construct, measured using items adapted from Kember’s validated instrument~\cite{kember2000development} (Fig.~\ref{fig:theory-model}).

\textbf{Critical Thinking} is the analytic work learners do in inspecting assumptions, weighing evidence, and considering alternatives~\cite{facione1990critical, sosu2013development, facione2011critical}. For example, when a solution fails, some may fixate on the most obvious cause. A critical thinker, on the other hand, considers multiple hypotheses: \textit{Is the error due to my inputs, or an unstated assumption in the problem? What evidence would distinguish between these possibilities?} 

Critical thinking has been assessed in multiple ways spanning self-report scales, performance tasks, observations, and expert evaluations~\cite{facione1990critical, paul2014assessment, ennis1993critical, kobylarek2022critical, kember2000development, zuriguel2022nursing}. The APA Delphi consensus~\cite{facione1990critical} characterizes critical thinking through two complementary dimensions: (1) \textit{Critical Openness}, the willingness to entertain alternative perspectives and/or competing ideas, and interpret information from multiple angles; and (2) \textit{Reflective Skepticism}, the discipline of questioning evidence and scrutinizing claims. Effective critical thinking requires both~\cite{ennis2018critical}: openness without skepticism produces gullibility; skepticism without openness produces rigidity. Accordingly, we model Critical Thinking as an HOC composed of these two LOCs (Fig.~\ref{fig:theory-model})~\cite{facione1990critical, facione2011critical, facione1995disposition}, using a validated questionnaire adapted from~\cite{sosu2013development}.

We adopt this definition over alternatives such as Bloom's taxonomy~\cite{bloom1956taxonomy} or the Paul-Elder framework~\cite{paul1997california} for three reasons: First, the Delphi-report formulation is grounded in empirical consensus among multiple experts in psychology and education. Second, its parsimonious two-factor structure (unlike the Paul-Elder framework~\cite{paul1997california}, which spans eight ``elements of thought'', ten ``intellectual standards'', and eight ``intellectual virtues'') aligns with our survey design and structural modeling goals. Third, it lowers measurement complexity and multicollinearity (highly fragmented constructs become unwieldy~\cite{hair2014primer}), yielding a model that is more straightforward to estimate and interpret.

% These three habits—reflection, need for understanding, and critical thinking—are not innate; they develop through practice and reinforcement during formative learning experiences. Students who cultivate them build understanding that endures and transfers. Those who don't build fragile knowledge that collapses when challenged~\cite{biggs2001revised, pintrich1990motivational, kember2000development, gong2025impact}. The critical question, then, is whether genAI use, undertaken with trust, systematically disrupts this development, recalibrating what intellectual effort feels necessary or worthwhile.

\subsection{Model development: How trust, usage, and cognitive styles affect cognitive engagement}
\label{model-dev}

To understand how genAI tools affect cognitive engagement, we must first understand something fundamental about how human cognition works—or more precisely, how it avoids working. These theories form the basis of our hypotheses:

\subsubsection{\textbf{The laziness of thought} (dual process and cognitive offloading theory)} Our minds did not evolve to think carefully about everything. That would be metabolically ruinous and strategically foolish~\cite{kahneman2011thinking, risko2016cognitive}. According to Kahneman's dual-process theory~\cite{kahneman2011thinking}, human cognition operates through two systems: System 1 (\textit{``fast thinking''}) is intuitive, automatic, and low-effort, drawing rapidly on associations and patterns. For example, it is what delivers the answer \textit{``2 + 2 = ?''} without deliberation, or completes the phrase \textit{``bread and...''}. System 2 (\textit{``slow thinking''}), on the other hand, is deliberate, controlled, and effortful. It is invoked when, for example, you calculate 17 × 24, evaluate a complex argument, or consciously monitor your behavior in a social situation.

System 2 is lazy. It avoids effort whenever System 1 can provide a plausible solution. This is not a bug; it is an evolutionary feature. Our ancestors who conserved cognitive resources for genuine threats survived; those who exhaustively analyzed every decision were outcompeted. The result is what psychologists call the \textit{``law of least effort''}~\cite{kahneman2011thinking}---a pervasive tendency to accept the path of least cognitive resistance.

This explains why cognitive offloading is so natural~\cite{risko2016cognitive}. When external aids can shoulder cognitive demand, we instinctively let them---\textit{why waste cognitive resources solving problems that have already been solved?} In learning contexts, however, it can cause students to skip over mental processes that are integral to the formation of knowledge. In the past, we have seen the \textit{``Google effect''}~\cite{sparrow2011google}: people became less likely to remember information they believed would be available online. Recent HAI work shows similar offloading patterns with genAI, where users lean on AI-generated suggestions at the expense of their own thinking~\cite{lee2025impact, kosmyna2025your}.  

\subsubsection{\textbf{Automation bias}} The tendency of cognitive offloading is compounded by trust~\cite{goddard2012automation, lyell2017automation}. When automated systems appear competent, users may accept outputs with limited scrutiny---even when those outputs conflict with their own knowledge or judgment~\cite{goddard2012automation, lyell2017automation}. In HAI contexts, this manifests as over-reliance: users defer judgment to AI, rapidly integrating its suggestions while failing to consider the likelihood or occurrence of errors/problems~\cite{buccinca2021trust, hamid2025inclusive, lee2004trust}. 

\subsubsection{\textbf{From laziness of thought and automation bias to RQ1-How's hypotheses}}

% \subsubsection{\textbf{RQ1 asks:} \textit{To what extent does students’ trust in and routine use of genAI affect their cognitive engagement habits in coursework?}}

We posit, drawing on dual-process, cognitive offloading, and automation bias theories, a set of cascading relationships through which students' trust in and routine use of genAI substantively affect their cognitive engagement habits in coursework.

First, \textit{trust drives routine usage}. We adopt a widely used definition of trust in AI: \textit{``the attitude that an agent will help achieve an individual's goals in a situation characterized by uncertainty and vulnerability''}~\cite{lee2004trust, liao2022designing, vereschak2021evaluate, wang2023investigating}. 
Across automation and HAI research, trust has been shown to predict adoption~\cite{lee2004trust, wischnewski2023measuring, choudhuri2025guides}. Students who view genAI as reliable are likely to consult (and defer to) it more frequently for their work. 
Accordingly, 
\begin{itemize}
    \item \textbf{H1:} \textit{Trust positively affects routine genAI usage}.
\end{itemize}

Second, we hypothesize that \textit{routine usage affects cognitive engagement habits}. Repeated offloading to genAI reinforces a new System-1 habit: \textit{``problem $\rightarrow$ offload to genAI''}, a fast response that bypasses deliberate cognitive engagement:

\begin{itemize}
 \item \textbf{H2a.} \textit{Routine genAI usage negatively affects Reflection}.
 \item \textbf{H3a.} \textit{Routine genAI usage negatively affects the Need for Understanding}.
 \item \textbf{H4a.} \textit{Routine genAI usage negatively affects Critical Thinking}.
\end{itemize}

Third, we posit that \textit{trust acts as an accelerant}, increasing the readiness with which students defer to genAI, which in turn trains the System~1 response described above and erodes cognitive engagement habits. Accordingly:

\begin{itemize}
    \item \textbf{H2b.} \textit{Trust transitively reduces Reflection through increased routine usage}.
    \item \textbf{H3b.} \textit{Trust transitively reduces the Need for Understanding through increased routine usage}.
    \item \textbf{H4b.} \textit{Trust transitively reduces Critical Thinking through increased routine usage}.
\end{itemize}

Following prior work, we adapt validated instruments to measure students' trust (4-item TXAI instrument~\cite{perrig2023trust, choudhuri2025guides}) and routine use of genAI (academic genAI use instrument~\cite{abbas2024harmful}).

\subsubsection{\textbf{Beyond how's to who's (RQ2-Who)}} Individuals are not identical in their propensity to use tools. Instead, they differ in their \emph{cognitive styles}, i.e., \textit{the way they perceive and process information, problem solve, and work with technology}~\cite{sternberg1997cognitive, burnett2016gendermag, martinez2017assessing}, which affects their technology adoption~\cite{anderson2024-GMstylesInAI-TiiS, burnett2016-GMfieldStudy-CHI, vorvoreanu2019-gmInMicrosoftApp-CHI, anderson2025llm}. Recent HAI work shows that certain cognitive styles: \textit{technophilic motivations} (intrinsic interest in engaging with technology), \textit{risk tolerance} (comfort with uncertainty), and \textit{computer self-efficacy} (confidence in using technology) significantly predict genAI adoption~\cite{choudhuri2025needs, choudhuri2025ai}.

Thus, to avoid conflating these stable traits with the effects of trust modeled in RQ1-How, we include technophilic motivations, computer self-efficacy, and risk tolerance as \textbf{control variables} on genAI usage: they help explain baseline differences in \textit{who} is most likely to heavily adopt genAI~\cite{choudhuri2025guides}, and in turn, may be prone to cognitive disengagement. We capture these cognitive styles using the validated GenderMag facet survey~\cite{hamid2024-GMsurveyValidated}, following prior work~\cite{choudhuri2025guides}.

Finally, given evidence that academic level and prior genAI experience can influence usage~\cite{strzelecki2025chatgpt, singer2025generative, choudhuri2025insights}, we performed \textbf{group analyses} to test whether our model relationships varied across these groups.

% % \vspace{-2mm}

\section{Method} 
\label{sec:method}

\subsection{Survey design}

To address our RQs, we conducted a large-scale survey with STEM students across five North American institutions. The study was approved by the University IRB. We followed Kitchenham’s guidelines for designing empirical surveys~\cite{kitchenham2008personal}, and drew on established frameworks from educational psychology and Human–AI Interaction studies to measure our constructs (Tab.~\ref{tab:mm-instruments}). Each construct in our model (circles in Fig.~\ref{fig:theory-model}) represents a latent variable, i.e., a variable not directly observable but inferred through validated indicators (e.g., survey questions, squares in Fig.~\ref{fig:theory-model}). 
We adapted existing questions to the genAI context, where applicable. For example, a risk-tolerance item reading \textit{``I avoid advanced [technology] features or options''} was revised to \textit{``I avoid advanced \textit{genAI} features or options''} (see supplemental~\cite{supplemental}). 

Our survey comprised three sections:

(1) \textbf{Cognitive engagement.} After providing informed consent, participants reported their cognitive engagement habits in coursework, including their reflection (reflective motives and strategies), need for understanding, and critical thinking (critical openness and reflective skepticism), captured using validated instruments listed in Tab.~\ref{tab:mm-instruments}.

(2) \textbf{Trust, usage, and cognitive styles.} Participants reported their trust in and usage of genAI tools in coursework, alongside their cognitive styles, including risk tolerance, technophilic motivations, and computer self-efficacy (Tab.~\ref{tab:mm-instruments}).

(3) \textbf{Background and genAI experience.} Finally, participants rated their prior experience with genAI, academic level, and (optionally) their major, age, and gender (see supplemental material~\cite{supplemental}). During the pilot phase, participants were also invited to share open-ended feedback on the survey. 

% % \vspace{-3mm}
\begin{table}[!h]
% \footnotesize
\caption{Constructs in the theoretical model and validated instruments used to measure them}
\label{tab:mm-instruments}
% \vspace{-8px}
\centering
\begin{tabular}{>{\raggedright\arraybackslash}m{4cm} >{\raggedright\arraybackslash}m{6cm}}
\hline
\textbf{Construct} & \textbf{Instrument} \\
\hline \hline
Trust & TXAI instrument \cite{perrig2023trust, choudhuri2025guides} \\ 

Usage &  Academic genAI usage instrument~\cite{abbas2024harmful}\\

Reflection & Bigg's instrument~\cite{biggs2001revised} \\

Need for Understanding & Kember's instrument~\cite{kember2000development}\\

Critical Thinking & CTDS instrument~\cite{sosu2013development} \\

Cognitive Styles & GenderMag facet survey \cite{hamid2024-GMsurveyValidated, choudhuri2025guides} \\
\bottomrule
\end{tabular}
% \vspace{-3mm}
\end{table}

We administered the survey in Qualtrics~\cite{qualtrics2025}. All closed-ended questions used a five-point Likert scale (1 = \textit{strongly disagree}, 5 = \textit{strongly agree}) with a neutral option and a sixth ``\textit{I’m not sure}'' option to distinguish ignorance from indifference \cite{grichting1994meaning}. The survey took 7-10 minutes to complete. We included attention-check questions and randomized item order within blocks to reduce response bias. The survey questionnaire is available in the supplemental material~\cite{supplemental}.

\textbf{Sandbox and pilot.} We sandboxed the survey one-on-one with psychology/HCI/CS-education researchers ($n = 9$) and STEM instructors ($n = 4$) to assess interpretability, clarity, and realism; revising ambiguous question phrasing based on their feedback. We then piloted it with 67 STEM undergraduates, confirming item clarity and construct validity. We made minor wording edits and excluded pilot responses from the final analysis.

\textbf{Validation.} We psychometrically validated our survey instrument following~\cite{hair2014primer, russo2021pls} (see details in Sec.~\ref{sec:analysis-validation}).

\subsection{Data collection}

\textbf{Distribution.}
The survey was available for three months (April-June, 2025). We recruited STEM instructors at our institution and via education conferences and networks, who then distributed the survey to students in their classes. We asked instructors to encourage participation and, where appropriate, offer it as an optional, non-graded exercise. Contacting instructors directly, rather than students, broadened our reach and aligned with our IRB guidelines.

\textbf{Sample size determination.} We determined the minimum sample size for validating the model via power analysis using the G*Power tool~\cite{faul2009statistical}. 
We specified an F-test with multiple linear regression to detect even small effect sizes ($f^2 = 0.10$) at a significance level of ($\alpha = 0.05$) with power = 0.95, and assuming at most four predictors per construct.\footnote{In Fig.~\ref{fig:theory-model}, the construct with the most incoming arrows is Usage, predicted by (1) Trust, (2) Technophilic Motivations, (3) Risk Tolerance, and (4) Computer Self-Efficacy.} The analysis suggested a minimum sample size of 191.

% We performed an F test with multiple linear regression with the following parameters (acceptable effect size (f2)=.15, acceptable significance level =.05, power (1-$\beta$)=.95, max. predictors to any constructs=4).%
% \footnote{
% The maximum number of predictors (incoming arrows in Fig.~\ref{fig:model}) to any construct in our model is the four that point to Usage, namely Trust, Technophilic Motivations, Risk Tolerance, and Computer Self-Efficacy.
% }

\textbf{Responses.} We received a total of 392 responses. We removed patterned responses ($n = 20$), partial submissions ($n = 39$) and those that failed attention checks ($n = 34$). We treated ``\textit{I'm not sure}'' selections as missing data, and did not impute missing values given the unproven efficacy of imputation in SEM group contexts~\cite{sarstedt2017treating}. We retained \textbf{\textit{n}=299 valid responses} after filtration, which exceeded the \textit{apriori} sample size requirement of 191. Respondents reported a wide distribution of genAI experience, usage frequencies, academic level, age, and gender, consistent with prior studies~\cite{lee2025impact, russo2024navigating}. A summary of respondent demographics is available in the supplemental material~\cite{supplemental}.

\subsection{Data analysis \& survey validation}
\label{sec:analysis-validation}

We analyzed the survey data using \emph{Partial Least Squares–Structural Equation Modeling} (PLS-SEM) to empirically evaluate our theoretical model. PLS-SEM suits exploratory studies like ours because it estimates multiple relationships among constructs while accounting for measurement errors in one comprehensive analysis~\cite{russo2024navigating,trinkenreich2023belong,russo2021pls,choudhuri2025guides}. Importantly, it relaxes multivariate distributional assumptions. Instead, it calculates the statistical significance of path coefficients (i.e., relationships between constructs) through \textit{bootstrapping}: resampling thousands of subsamples (5,000 in our case) to derive inferences~\cite{hair2019use}.

We conducted the analyses using \emph{SmartPLS 4}~\cite{smartpls_website} in two phases. First, we evaluated the \emph{measurement model to validate our survey}: i.e., verify that the items (questions) reliably and validly capture the intended constructs. Second, we evaluated the \emph{structural model} to test the hypothesized relationships among these constructs. Below, we detail the \textit{measurement model} evaluation (\textbf{Steps 1-5}). The \textit{structural model} evaluation and results are detailed in Sec.~\ref{results}.

% MM eval tells us did we measure the right thing?
\textbf{Survey validation} \textit{(measurement model evaluation)}: We used a two-stage procedure to validate our survey instrument, i.e., assess both lower-order and higher-order reflective constructs (circles and double circles in Fig.~\ref{fig:theory-model}) following recommended PLS-SEM practices~\cite{becker2012hierarchical,sarstedt2019specify,hair2019use}. The data met standard assumptions for measurement-model analysis: Bartlett’s test of sphericity was significant for all constructs ($\chi^2(946)=7167.278$, $p<.001$), and the KMO measure of sampling adequacy was $0.901$, exceeding the $0.60$ threshold~\cite{howard2016review}.

\textsc{\textbf{\ul{Stage 1:}} Lower-order reflective construct evaluation:} We first assessed the lower-order reflective constructs (Trust, Usage, Reflective Motives, Reflective Strategies, Need for Understanding, Critical Openness, and Reflective Skepticism, circles in Fig.~\ref{fig:theory-model}) using tests for \textbf{(1)} convergent validity, \textbf{(2)} internal consistency reliability, \textbf{(3)} discriminant validity, and \textbf{(4)} indicator collinearity~\cite{russo2021pls,hair2019use}.

%-------- CONVERGENT VALIDITY ----------
\boldification{Do all indicators measure the same construct?}
\textbf{(1) Convergent validity} examines whether indicators (questions) intended to measure the same construct share sufficient variance (for reflective constructs, changes in the construct should be reflected in its indicators)~\cite{kock2014advanced}. We assessed convergent validity using (a) factor loadings and (b) Average Variance Extracted (AVE)~\cite{hair2019use}.

\begin{itemize}
    \item Factor loadings capture the association between indicators and their constructs (e.g., T1–T4 for \emph{Trust}); values $\ge .60$ suffice for exploratory studies~\cite{hair2019use}. We removed two weak indicators---U6 of \emph{Usage} and CO1 of \emph{Critical Openness}---that did not adequately reflect their constructs.\footnote{Removing CO1 and U6 also increased AVE from .48 to .52 for \emph{Critical Openness} and from .64 to .67 for \emph{Usage}.} All retained indicators exceeded the threshold, with loadings ranging from $.60$ to $.925$.
    \item AVE captures the shared variance across a construct’s indicators and should exceed $.50$, indicating the construct explains at least 50\% of shared variance among its indicators~\cite{hair2019use}. All constructs met this criterion (Tab.~\ref{table:internalreliability}).
\end{itemize}

% \vspace{-2mm}
\begin{table}[h]
% \scriptsize
% \footnotesize
\centering
\caption{Internal consistency reliability, and convergent validity for reflective lower-order and higher-order constructs in the theoretical model. These metrics reflect whether each construct is measured reliably and captures sufficient shared variance among its indicators. Cronbach’s $\alpha$, Composite Reliability (CR($\rho_a$), CR($\rho_c$)), and Average Variance Extracted (AVE) values indicate internal consistency and convergent validity respectively, with acceptable thresholds of $\ge$.7 (for reliability) and $\ge$.5 (for AVE). Higher values reflect stronger reliability and validity of the questionnaire.}

% \vspace{-8px}
\begin{tabular}{>{\raggedright\arraybackslash}m{3.65cm} >{\centering\arraybackslash}m{1.8cm} >{\centering\arraybackslash}m{0.8cm} >{\centering\arraybackslash}m{0.8cm} >{\centering\arraybackslash}m{0.8cm}}

\hline
\textbf{} & \textit{Cronbach's $\alpha$} & \textit{CR($\rho_a$)} & \textit{CR($\rho_c$)} & \textit{AVE} \\ \hline \hline
\textbf{Lower-order constructs} &&&&\\
Critical Openness & .746 & .802 & .735 & .524 \\
Need for Understanding & .902 & .905 & .938 & .835 \\
Reflection Motives & .896 & .899 & .937 & .833 \\
Reflective Skepticism & .828 & .903 & .845 & .511 \\
Reflective Strategies & .789 & .797 & .855 & .542 \\
Trust & .818 & .831 & .883 & .655 \\ 
Usage & .896 & .897 & .923 & .665 \\
Technophilic Motivations & .731 & .772 & .805 & .685\\
Risk Tolerance & .721& .752& .773 & .672\\
Computer Self-Efficacy & .711 & .796 & .821 & .645\\\hline
\textbf{Higher-order constructs} &&&&\\
Critical Thinking & .700 & .710 & .803 & .671 \\
Reflection & .801 & .809 & .909 & .834 \\
\hline
\end{tabular}
\label{table:internalreliability}
\begin{tablenotes}
\small
\item Cronbach's $\alpha$ tends to underestimate reliability, whereas composite reliability (CR: $\rho_c$) tends to overestimate it. The true reliability typically lies between these two estimates and is effectively captured by CR($\rho_a$) \cite{russo2021pls}.
\end{tablenotes}
% \vspace{-6px}
\end{table}

%-------- INTERNAL CONSISTENCY ----------
\boldification{Do the indicators consistently capture the construct?}
\textbf{(2) Internal consistency reliability} evaluates whether indicators are internally consistent and reliably measure the same construct. We evaluated internal consistency reliability using Cronbach’s $\alpha$ and Composite Reliability ($\rho_a$, $\rho_c$)~\cite{russo2021pls}. All constructs met established reliability thresholds ($.70$–$.95$)~\cite{hair2019use} (see Tab.~\ref{table:internalreliability}).

%-------- DISCRIMINANT VALIDITY ----------
\boldification{Are the constructs distinct?}
\textbf{(3) Discriminant validity} assesses the distinctness of constructs. We used the Heterotrait–Monotrait (HTMT) ratio~\cite{henseler2015new}, with adequacy indicated by values $<.90$. HTMT ratios among all constructs met this criterion, supporting discriminant validity. We report HTMT ratios for higher-order constructs and other model constructs in Tab.~\ref{tab:htmt-hoc}; ratios for the LOCs of Reflection and Critical Thinking, as well as the Fornell–Larcker criterion and cross-loadings, alongside confirmatory factor analysis, are provided in the supplemental for completeness~\cite{supplemental}.

\begin{table}[h]
\centering
\small
\setlength{\tabcolsep}{3pt}
\renewcommand{\arraystretch}{1}
\caption{Heterotrait–Monotrait (HTMT) ratios among model constructs (ranges between 0 and 1). Values $< .90$ indicate that constructs are sufficiently distinct from each other. HTMT ratios for the LOCs of Reflection (Reflective Motives, Reflective Strategies) and Critical Thinking (Critical Openness, Reflective Skepticism) are reported in~\cite{supplemental}.}
\begin{tabular}{lccccccc}
\toprule
 & \textbf{Critical Thinking} & \textbf{Tech. Motivations} & \textbf{NFU} & \textbf{Reflection} & \textbf{Risk Tolerance} & \textbf{Computer S.E.} & \textbf{Trust} \\
\midrule

        Tech. Motivations & .597 & ~ & ~ & ~ & ~ & ~ & ~ \\
        NFU & .07 & .179 & ~ & ~ & ~ & ~ & ~ \\ 
        Reflection & .66 & .823 & .322 & ~ & ~ & ~ & ~ \\ 
        Risk Tolerance & .552 & .735 & .217 & .669 & ~ & ~ & ~ \\ 
        Computer S.E. & .587 & .603 & .063 & .414 & .635 & ~ & ~ \\ 
        Trust & .529 & .711 & .266 & .843 & .608 & .489 & ~ \\ 
        Usage & .587 & .661 & .228 & .806 & .786 & .526 & .725 \\ 
\bottomrule
\end{tabular}
\label{tab:htmt-hoc}
\begin{tablenotes}
\item Tech. Motivations: Technophilic Motivations; NFU: Need for Understanding; Computer S.E.: Computer Self-Efficacy
\end{tablenotes}
\end{table}

%-------- COLLINEARITY ----------
\boldification{Are the indicators non-redundant (no problematic multicollinearity)?}
\textbf{(4) Indicator collinearity} assessment checks for multicollinearity among reflective indicators. We used the Variance Inflation Factor (VIF) to check for multicollinearity. All indicator VIFs were $< 3$, well below the cutoff of $5$~\cite{hair2019use}.

% ----------COMMON METHOD BIAS ----------
\textit{Common method bias.} Since we collected data via a single survey, we tested for Common Method Bias (CMB)~\cite{russo2021pls}. To do so, we first applied Harman's single-factor test on the latent constructs~\cite{podsakoff2003common}. No single factor explained more than 21.9\% variance. An unrotated exploratory factor analysis with a forced single-factor solution explained 26.6\% of variance, well below the 50\% threshold.
Additionally, we used Kock's collinearity check~\cite{kock2015common}: VIFs for all latent constructs were $< 2$, well under the cut-off. These results indicate CMB was not a concern in our study.

%--------HOC VALIDITY----------
\boldification{Do the high-order constructs work?} 
\textsc{\textbf{\ul{Stage 2:}} Higher-order reflective construct evaluation:} Our model includes two reflective higher-order constructs (HOCs): (1) Reflection and (2) Critical Thinking. Recall, reflection is a motive-strategy pair that acts together (see Sec.~\ref{sec:theory}). Following ~\citet{biggs2001revised}'s work, we modeled \emph{Reflection} as an HOC with \emph{Reflective Motives} and \emph{Reflective Strategies} as its lower-order constructs (LOCs). Similarly, following the APA Delphi consensus~\cite{facione1990critical}, we modeled \emph{Critical Thinking} as an HOC comprising \emph{Critical Openness} and \emph{Reflective Skepticism}, as in prior work~\cite{sosu2013development}. Collapsing the LOCs into HOCs reduces measurement complexity, 
aligns the model with theory, and lowers collinearity, yielding a more parsimonious, easier-to-test structure.

\textbf{(5) HOC assessment.} We validated HOCs using the disjoint two-stage approach recommended by \citet{sarstedt2019partial}. First, we estimated latent scores for the lower-order constructs in SmartPLS as linear combinations of indicator scores and outer weights. We then used these scores as indicators for the HOCs: (i) Reflective Motives and Reflective Strategies for \emph{Reflection}, and (ii) Critical Openness and Reflective Skepticism for \emph{Critical Thinking}. Finally, we evaluated the HOCs using the same assessment criteria applied to the LOCs:

\begin{itemize}

\item \textit{Convergent validity.} All factor loadings exceeded .70 (Fig.~\ref{fig:RQ1model}), and AVE values were above .50 for both HOCs (Tab.~\ref{table:internalreliability}), indicating sufficient explained variance of its indicators (the LOCs).

\item \textit{Internal consistency reliability.} Cronbach’s $\alpha$ and Composite Reliability values exceeded .70 for both HOCs, indicating sufficient reliability and internal consistency of the constructs (Tab.~\ref{table:internalreliability}).

\item \textit{Discriminant validity}. HTMT ratios were below .90 for all construct pairs (Tab.~\ref{tab:htmt-hoc}), indicating empirical distinctness.

\item \textit{Collinearity.} All VIF values were below 2 (see supplemental material~\cite{supplemental}), indicating no multicollinearity concerns.

\end{itemize}

Together, these stages (1\&2) \textit{validated our survey instrument}: questions reliably capture their intended constructs, and these constructs satisfy convergent and discriminant validity, enabling us to test structural relationships next (Sec.~\ref{results}).

% % \vspace{-2mm}
\subsection{Limitations}
\label{sec:limitations}

As with any empirical study, our work has limitations~\cite{ko2015practical}.
% 
% \textbf{Construct validity:}
All constructs were measured using validated self-report instruments grounded in established theory. Even then, surveys can introduce response biases or misinterpretations. We guarded against these threats by involving instructors in survey design, sandboxing and piloting, randomizing question blocks to reduce order effects, adding attention checks, and screening patterned responses. All constructs met standard thresholds for reliability and validity. 

% \textbf{Internal validity:}
As a cross-sectional study~\cite{stol2018abc}, we report associations rather than causation. Self-selection bias remains possible, as students with stronger views may have been more likely to participate. As with all survey-based research, reported experiences may not align perfectly with in-situ behavior. Nonetheless, given the internal consistency of the results and their triangulation with established theory, we consider this approach sufficient to support the reliability of our conclusions. Further, a theoretical model cannot capture an exhaustive set of factors, i.e., unmeasured confounds may remain. Accordingly, our findings should be interpreted as a theoretically grounded starting point; future work should consider longitudinal designs and replication in varied contexts to strengthen causal claims and generalize broadly.

% \textbf{External validity:}
Finally, while the study was designed to be region-agnostic and included students with varied demographics and levels of genAI experience, the sample reflects a primary US-based educational context. Its size and composition are comparable to prior empirical studies in STEM education~\cite{amoozadeh2024trust, margulieux2024self}, providing a reasonable foundation for theory development. Consistent with this goal, we emphasize theoretical rather than statistical generalizability~\cite{shull2007guide}. Future work should examine transferability across broader populations and contexts.

%--------------------------------

% % % \vspace{-1mm}
% \vspace{-2mm}
\section{Results}
\label{results}

After validating our survey (measurement model), we test our hypotheses by evaluating the structural model (recall Fig.~\ref{fig:theory-model}). From here on, when we say `significant', we mean `statistically significant'.

%----------------------------------
\subsection{RQ1-How/How much: Trust $\rightarrow$ Usage $\rightarrow$ Cognitive Engagement?
%How does students' trust-driven routine genAI use affect their cognitive engagement habits?
}

RQ1-How examines whether and to what extent students’ \textit{trust in genAI predicts their routine usage of these tools}, and, in turn, \textit{how that trust and usage affect their cognitive engagement habits} in STEM coursework---specifically reflection, need for understanding, and critical thinking (recall Sec.~\ref{sec:theory} for how we model these constructs).
We report these findings in a deliberate sequence: we first describe how routine genAI usage affects cognitive engagement, and then turn to how trust affects these outcomes.

%-----
\subsubsection{\textbf{Usage and cognitive engagement}} 
\begin{figure*}[!hbt]
\centering
% % % \vspace{-10px}
\includegraphics[width=\textwidth]{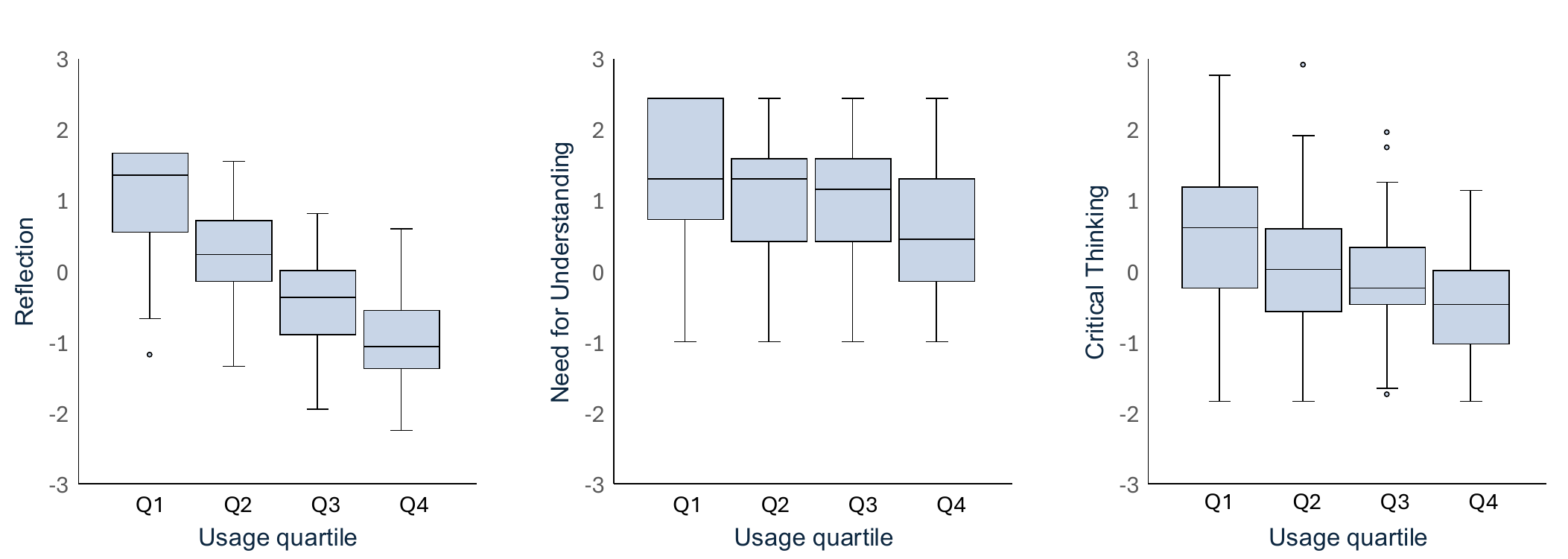}
% % \vspace{-5px}
\caption{Cognitive engagement scores by genAI usage. Boxplots show standardized latent scores for Reflection, Need for Understanding, and Critical Thinking across quartiles of reported genAI usage (Q1 = lowest 25\%; Q4 = highest 25\%). This figure provides a close-up view of data distributions; subsequent figures examine patterns using median-split groups. 
% \textcolor{red}{MMB to MMB: in here say that this is a close-up of the distributions, but from now on we'll switch to medians. }
}
\label{fig:usage-boxplot}
\end{figure*}

\begin{figure*}[h] 
\centering
% % \vspace{-10px}
\includegraphics[width=\textwidth]{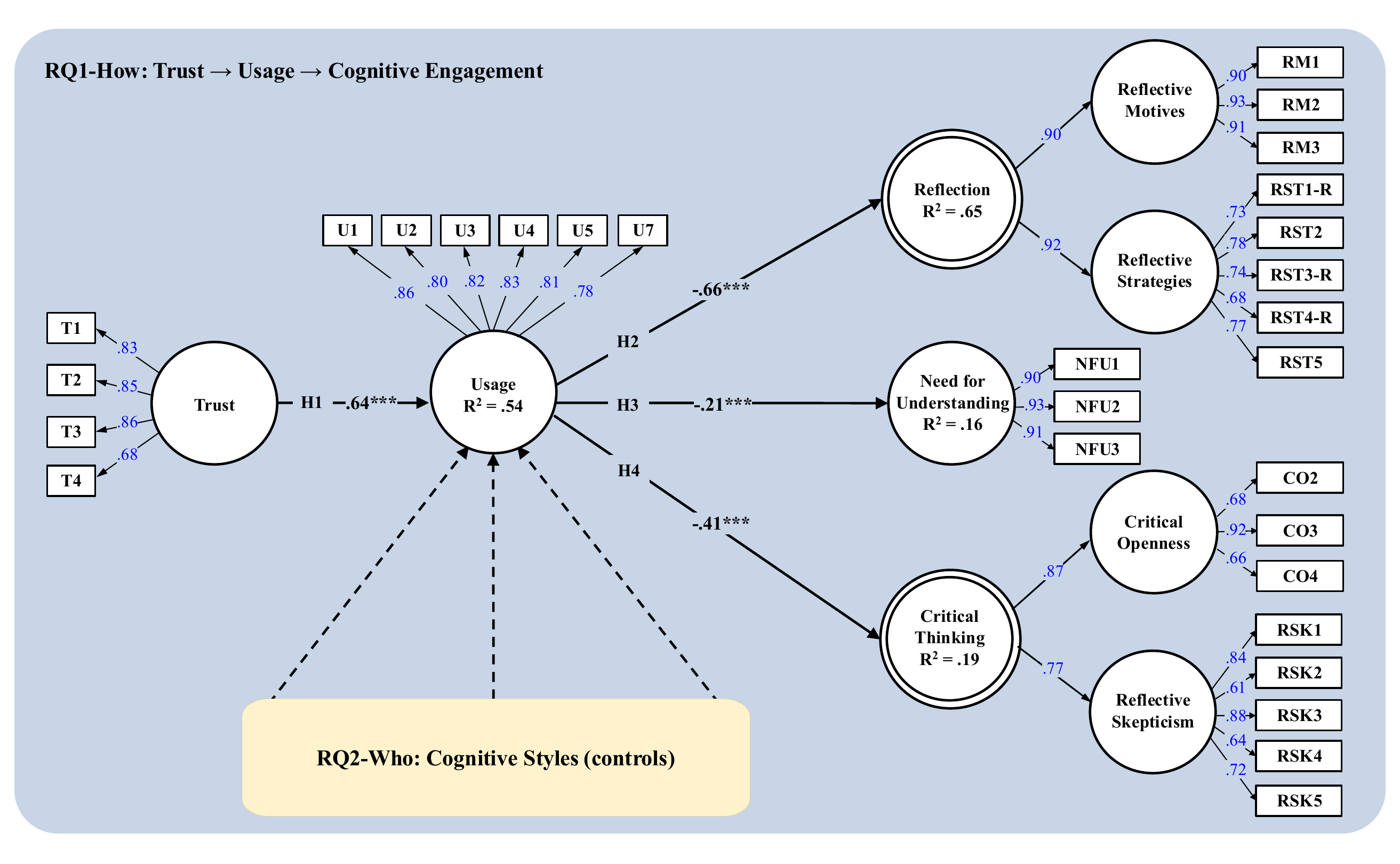}
% \vspace{-5px}
\caption{\small Structural model results (RQ1-How): Standardized path coefficients (H1–H4) on arrows between constructs (circles) show how Trust and routine genAI Usage affect Reflection, Need for Understanding, and Critical Thinking, with significance ***p<.001, **p<.01, and *p<.05. Corresponding transitive effects are summarized in Tab.~\ref{tab:new_path_analysis}. Arrows from constructs to survey items (squares), and from HOCs (double circles) to their lower-order constructs, indicate factor loadings (in \textcolor{blue}{\textit{blue}}; all $\ge$ threshold of 0.6~\cite{hair2019use}). Dotted arrows and the collapsed yellow region correspond to RQ2-Who results (omitted here for clarity), which are reported in Sec.~\ref{sec-RQ2}.}
\label{fig:RQ1model}
\end{figure*}

Fig.~\ref{fig:usage-boxplot} shows distributions of cognitive engagement by usage quartile (Q1 = lowest 25\%; Q4 = highest 25\%).\footnote{Each score is a model-estimated latent factor score: the measurement model computes how strongly each question signals the underlying construct, then uses these item weights to aggregate the factor score~\cite{hair2014primer}.} 
Across all three dimensions, median cognitive engagement factor scores declined monotonically from Q1 to Q4. Participants reporting the highest levels of genAI use (Q4) consistently reported the lowest levels of reflection, need for understanding, and critical thinking.
Tab.~\ref{tab:usage-engagement} reports the statistical significance of these relationships (visualized as directed paths among constructs in Fig.~\ref{fig:RQ1model}. E.g., Usage$\rightarrow$Reflection is H2). 

%  ---- Figures always go right AFTER they are first mentioned.

%  ---- Tables always go right AFTER they are first mentioned.
\begin{table}[!h] %MMB: use [!h] until just before ship
% \footnotesize
% \scriptsize
\centering
\caption{Effects of students' routine genAI usage on their cognitive engagement habits (H2a-H4a): Standardized path coefficients ($\beta$), standard deviations (SD), 95\% confidence intervals (CI), \textit{p}-values, and effect sizes ($f^2$). We consider $f^2< 0.02$ no effect, $f^2 \in [0.02, 0.15)$ small, $f^2 \in [0.15, 0.35)$ medium, and $f^2> 0.35$ large~\cite{cohen2013statistical}.}

% \vspace{-8px}
\robustify{\bfseries}
\sisetup{
    mode=text,
    group-digits = false ,
    input-signs ={-},
    input-symbols = ( ) [ ] - + *,
    detect-weight=true, 
    detect-family=true,
    table-format=0.2,
    add-decimal-zero=false, %% doesn't seem to work :-(
    add-integer-zero=false,
    round-mode=places, 
    round-precision=2, %% change this for precision.
    parse-numbers = true
}
% \begin{tabular}{P{4.3cm}
%                 S
%                 S[table-format=0.2]
%                 >{\centering\arraybackslash}p{1.3cm}
%                 S[table-format=0.3,round-precision=3]}
% \toprule
% & {\textit{B}} & {SD} & {95\% CI} & {\textit{p}}\\
% \midrule
\begin{tabular}{P{5.9cm}  % Adjusted width for the first column
                S
                S[table-format=0.2]
                >{\centering\arraybackslash}p{1.7cm}
                S[table-format=0.3,round-precision=3]
                S[table-format=0.3,round-precision=2]} % New column for f^2
\toprule
& {\textit{B}} & {SD} & {95\% CI} & {\textit{p}} & {$f^2$} \\
\midrule \midrule
%\textsc{Hypotheses}&&&&\\
%\midrule
% \rowcolor{myBlue} \hangindent1em \textbf{H2a} Usage$\rightarrow$Reflection & -.659 & .048 & (-.76, -.55) & \bfseries .000 & .51\\
\rowcolor{myBlue}
\hangindent1em \textbf{H2a} Usage$\rightarrow$Reflection & -0.66 & 0.05 & (-0.76, -0.55) & \bfseries 0.000 & 0.51\\
\rowcolor{myBlue} \hangindent1em \textbf{H3a} Usage$\rightarrow$Need for Understanding & -0.21 & 0.06 & (0.09, 0.34) & \bfseries 0.000 & 0.12\\
\rowcolor{myBlue} \hangindent1em \textbf{H4a} Usage $\rightarrow$Critical Thinking & -0.41 & 0.02 & (-0.46, -0.38) & \bfseries 0.001 & 0.20\\
\bottomrule

\end{tabular}
\begin{tablenotes}
  % \footnotesize
\item \textit{Note: PLS-SEM estimates all structural paths jointly within a single model. Significance tests (and their Type-I error rates) are defined within this joint estimation; separate family-wise error corrections are not required}~\cite{hair2014primer, hair2019use}.
\end{tablenotes}
\label{tab:usage-engagement}
\end{table}

%---Figures always go right AFTER the first mention
%--- this is the RQ1 model

\textbf{(H2a) Usage and reflection:} GenAI usage had a significant negative association with reflection ($\beta=-0.66$, \textit{p}$<.001$), with a large effect size ($f^2=0.51$), supporting \textbf{H2a}. Specifically, a unit (variance) increase in routine genAI usage predicted a 0.66-unit reduction in reflection, as indicated by the path coefficient (Tab.~\ref{tab:usage-engagement}). 
Reflection requires monitoring oneself---noticing when something does not make sense or when understanding feels superficial~\cite{dunlosky2005self, silver2023using}. The magnitude of this association suggests that routine genAI usage substantively affected such process---the pausing, puzzling, and noticing that something requires deeper thought. 

\textbf{(H3a) Usage and need for understanding.} GenAI usage had a significant negative association with need for understanding ($\beta=-0.21$, \textit{p} $<.001$), supporting \textbf{H3a}. The effect size ($f^2$ = 0.12) suggests that while routine genAI use meaningfully predicted need for understanding, stable individual differences (e.g., epistemic curiosity that students bring with them) remain important contributors to this construct. Still, the pattern is notable: participants who routinely used genAI reported significantly less intrinsic drive to grasp why's and how's.

\textbf{(H4a) Usage and critical thinking.} GenAI usage had a significant negative association with critical thinking ($\beta=-0.41$, \textit{p}=.001; $f^2=0.20$), supporting \textbf{H4a}. From a dual-process perspective~\cite{kahneman2011thinking}, critical thinking involves System~2 work---the deliberate questioning of claims, weighing of alternatives, and scrutiny of assumptions~\cite{sosu2013development, facione1990critical}. This finding shows that routine genAI use predicted reductions in such deliberate processes, consistent with the tendency for fluent, seemingly ``complete’’ genAI outputs to invite System~1 acceptance (as discussed in Sec.~\ref{sec:theory}).

%--------
\subsubsection{\textbf{Trust and cognitive engagement}} 
H1 hypothesized that students’ trust in genAI would predict how frequently they used it. For our participants, this turned out to be the case: trust showed a large, significant positive association with routine genAI usage ($\beta = 0.64$, \textit{p}<.001), with a moderate effect size ($f^2 = 0.25$), supporting \textbf{H1}. This finding is consistent with prior automation-bias accounts~\cite{parasuraman1997humans, buccinca2021trust}, showing that trust lowers the reluctance towards (AI) delegation, making it feel reasonable rather than risky or provisional.

Since routine usage was negatively associated with all three dimensions of cognitive engagement \textbf{(H2a–H4a)}, we examined the extent to which trust affected the same three dimensions transitively through increased usage \textbf{(H2b–H4b)}.
%Because trust predicted higher usage, and usage predicted lower cognitive engagement, we examined whether and to what extent trust transitively affected engagement.

\begin{figure*}[t] %trust
\centering
% % % \vspace{-10px}
\includegraphics[width=\textwidth]{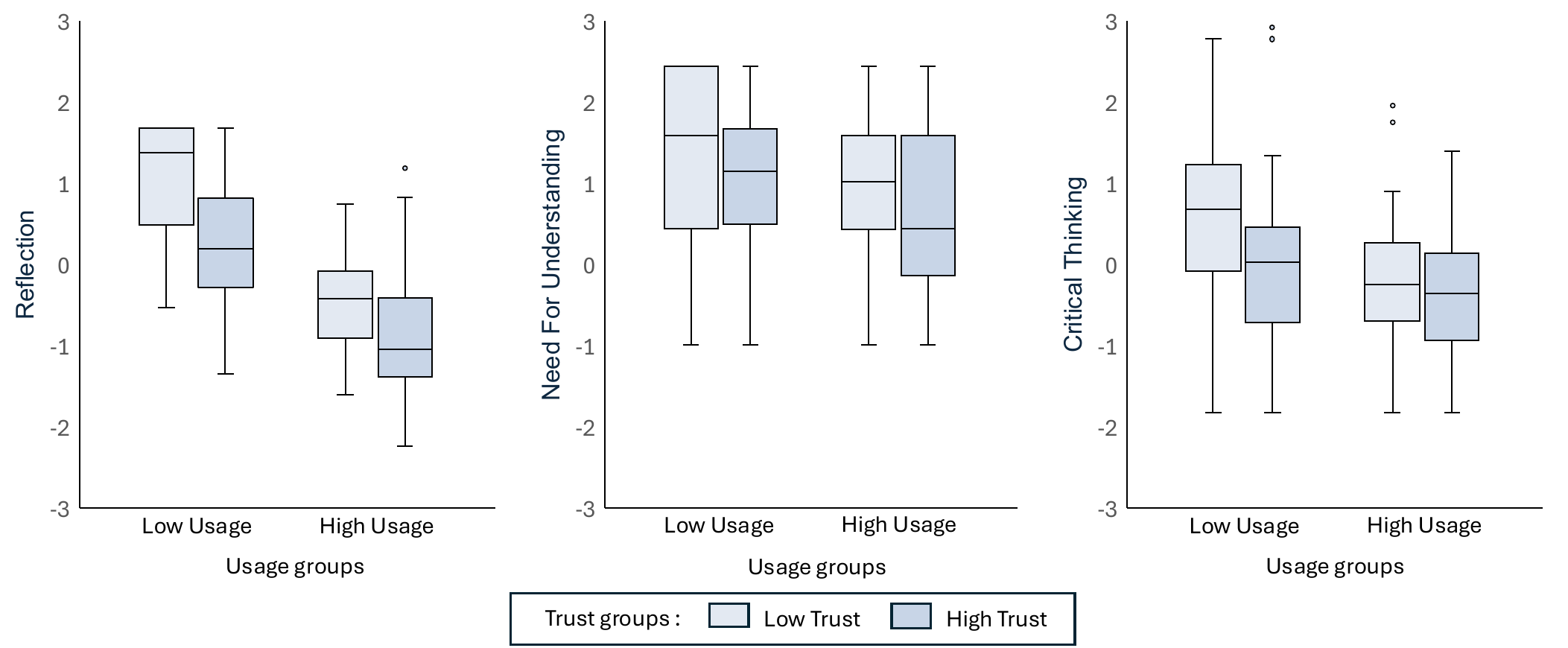}
% % \vspace{-5px}
\caption{Trust $\rightarrow$ usage $\rightarrow$ cognitive engagement distributions. 
Boxplots show standardized latent scores for Reflection, Need for Understanding, and Critical Thinking among participants grouped by genAI usage (median split) and trust in genAI (low vs. high; median split within each usage group).
% For \textit{every} usage group, the more participants trusted genAI, the less cognitive engagement they reported with their coursework.
}
% % \vspace{-5px}
\label{fig:usage-trust-boxplot}
% % \vspace{-7px}
\end{figure*}

Fig.~\ref{fig:usage-trust-boxplot} shows median-split box plot distributions by trust and usage. 
Across both low- and high-trust groups, participants with higher usage reported lower cognitive engagement (across all three dimensions). Moreover, within each usage group, participants in the ``high-trust'' group reported even lower cognitive engagement than their ``low-trust'' peers, indicating that trust amplified these outcomes. 

%---Each figure always goes right AFTER the first mention of it.
%----- trust -> usage -> cog engagement boxplots
%----- 
\begin{table}[!h] %MMB: use [!h] until just before ship
% \footnotesize
% \scriptsize
\centering
\caption{Effects of students' trust in genAI on routine usage and cognitive engagement habits (H1, H2b-H4b): Standardized path coefficients ($\beta$), standard deviations (SD), 95\% confidence intervals (CI), \textit{p}-values, and effect sizes ($f^2$). Transitive effects (H2b-H4b) capture how trust influences each cognitive engagement dimension mediated through routine usage. We consider $f^2< 0.02$ no effect, $f^2 \in [0.02, 0.15)$ small, $f^2 \in [0.15, 0.35)$ medium, and $f^2> 0.35$ large~\cite{cohen2013statistical}.}

% \vspace{-8px}
\robustify{\bfseries}
\sisetup{
    mode=text,
    group-digits = false ,
    input-signs ={-},
    input-symbols = ( ) [ ] - + *,
    detect-weight=true, 
    detect-family=true,
    table-format=0.2,
    add-decimal-zero=false, %% doesn't seem to work :-(
    add-integer-zero=false,
    round-mode=places, 
    round-precision=2, %% change this for precision.
    parse-numbers = true
}
% \begin{tabular}{P{4.3cm}
%                 S
%                 S[table-format=0.2]
%                 >{\centering\arraybackslash}p{1.3cm}
%                 S[table-format=0.3,round-precision=3]}
% \toprule
% & {\textit{B}} & {SD} & {95\% CI} & {\textit{p}}\\
% \midrule
\begin{tabular}{P{5.9cm}  % Adjusted width for the first column
                S
                S[table-format=0.2]
                >{\centering\arraybackslash}p{1.7cm}
                S[table-format=0.3,round-precision=3]
                S[table-format=0.3,round-precision=2]} % New column for f^2
\toprule
& {\textit{B}} & {SD} & {95\% CI} & {\textit{p}} & {$f^2$} \\
\midrule \midrule
%\textsc{Hypotheses}&&&&\\
%\midrule
\rowcolor{myBlue}\hangindent1em \textbf{H1} Trust$\rightarrow$Usage & .636 & .032 & (0.58, 0.69) & \bfseries .000 & .248\\
\midrule
\multirow{1}{*}{\textbf{Transitive effects}: Trust to cognitive engagement mediated via usage} &&&&&\\
\rowcolor{myBlue} \hangindent1em \textbf{H2b} Trust$\rightarrow$Usage$\rightarrow$Reflection & -.388 & .033 & (-0.54, -0.45) & \bfseries .000 & .16\\
%\addlinespace
\rowcolor{myBlue} \hangindent1em \textbf{H3b} Trust$\rightarrow$Usage$\rightarrow$Need for Understanding & -.133 & .039 & (0.07, 0.20) & \bfseries .000 & 0.11\\
\rowcolor{myBlue} \hangindent1em \textbf{H4b} Trust $\rightarrow$Usage$\rightarrow$Critical Thinking & -.261 & .047 & (-0.32,-0.21) & \bfseries .001 & 0.15\\

\bottomrule

\end{tabular}
\begin{tablenotes}
  % \footnotesize
\item \textit{Note: PLS-SEM estimates all structural paths jointly within a single model. Significance tests (and their Type-I error rates) are defined within this joint estimation; separate family-wise error corrections are not required}~\cite{hair2014primer, hair2019use}.
\end{tablenotes}
\label{tab:new_path_analysis}
\end{table}

% Mediation analysis confirmed these transitive effects (see Tab.~\ref{tab:new_path_analysis}). Following standard PLS mediation procedures~\cite{hair2014primer}, we tested the transitive effects of trust on each cognitive engagement dimension via usage. All three effects were significant, supporting \textbf{H2b–H4b}:

Mediation analysis confirmed these patterns (Tab.~\ref{tab:new_path_analysis}). Following standard PLS procedures~\cite{hair2014primer}, we tested the transitive effects of trust on each cognitive engagement dimension via routine usage. All three effects were significant, supporting \textbf{H2b–H4b}: participants with greater trust in genAI reported significantly less Reflection ($\beta = -0.39$, \textit{p}<.001, $f^2 = 0.16$, VAF $= 0.89$), less Need for Understanding ($\beta = -0.13$, \textit{p}<.001, $f^2 = 0.11$, VAF $= 0.85$), and less Critical thinking ($\beta = -0.26$, \textit{p} = .001, $f^2 = 0.15$, VAF $= 0.94$). All VAF (\ul{V}ariance \ul{A}ccounted \ul{F}or) values were $>$ 0.8, indicating \textit{full mediation}~\cite{hair2014primer}: trust in genAI affected each cognitive engagement dimension entirely through its effect on routine genAI usage.

\begin{takeawayBox}
\textbf{\textit{Takeaway}}: 
RQ1-How's findings point to a consistent pattern of cognitive disengagement among students using genAI. Participants \textit{who trusted and routinely used genAI} reported significantly lower levels of \textit{reflection, need for understanding, and critical thinking} in STEM coursework. 

\end{takeawayBox}

\vspace{2mm}
%-----------------------------------------------------
\subsection{RQ2-Who: Students' Cognitive Styles $\rightarrow$ Usage $\rightarrow$ Cognitive Engagement?}
\label{sec-RQ2}

Trust was a strong predictor of routine genAI usage---and, transitively, of cognitive engagement. 
However, it alone did not fully account for the observed variance in participants' routine genAI use ($R^2$ = .54 in Fig.~\ref{fig:RQ1model}).
Since individuals differ in how they approach and use technology~\cite{burnett2016gendermag, vorvoreanu2019-gmInMicrosoftApp-CHI, anderson2024-GMstylesInAI-TiiS}, here we consider whether some students were particularly prone to genAI-related reductions in cognitive engagement via increased routine usage.
If so, \textit{which} students?

RQ2-Who investigates this question using cognitive styles: whether and how students' cognitive styles affect their routine use of genAI, and, in turn, their cognitive engagement. 

% ----
\subsubsection{\textbf{Cognitive styles and routine genAI usage}} 

Prior work has reported certain cognitive styles to significantly influence genAI adoption among knowledge workers~\cite{choudhuri2025guides, choudhuri2025needs, choudhuri2025ai}: intrinsic enjoyment of engaging with technology (\textit{technophilic motivations}), comfort with uncertainty and technology-related risks (\textit{risk tolerance}), and confidence in one’s ability to use such tools effectively (\textit{computer self-efficacy})~\cite{burnett2016gendermag, martinez2017assessing, lyell2017automation, bandura1997self, anderson2024-GMstylesInAI-TiiS}. 
Thus, we investigated whether these same cognitive styles also predict \textit{students'} routine usage of genAI.

\begin{figure*}[h] %RQ2 model figure
\centering
% % \vspace{-10px}
\includegraphics[width=\textwidth]{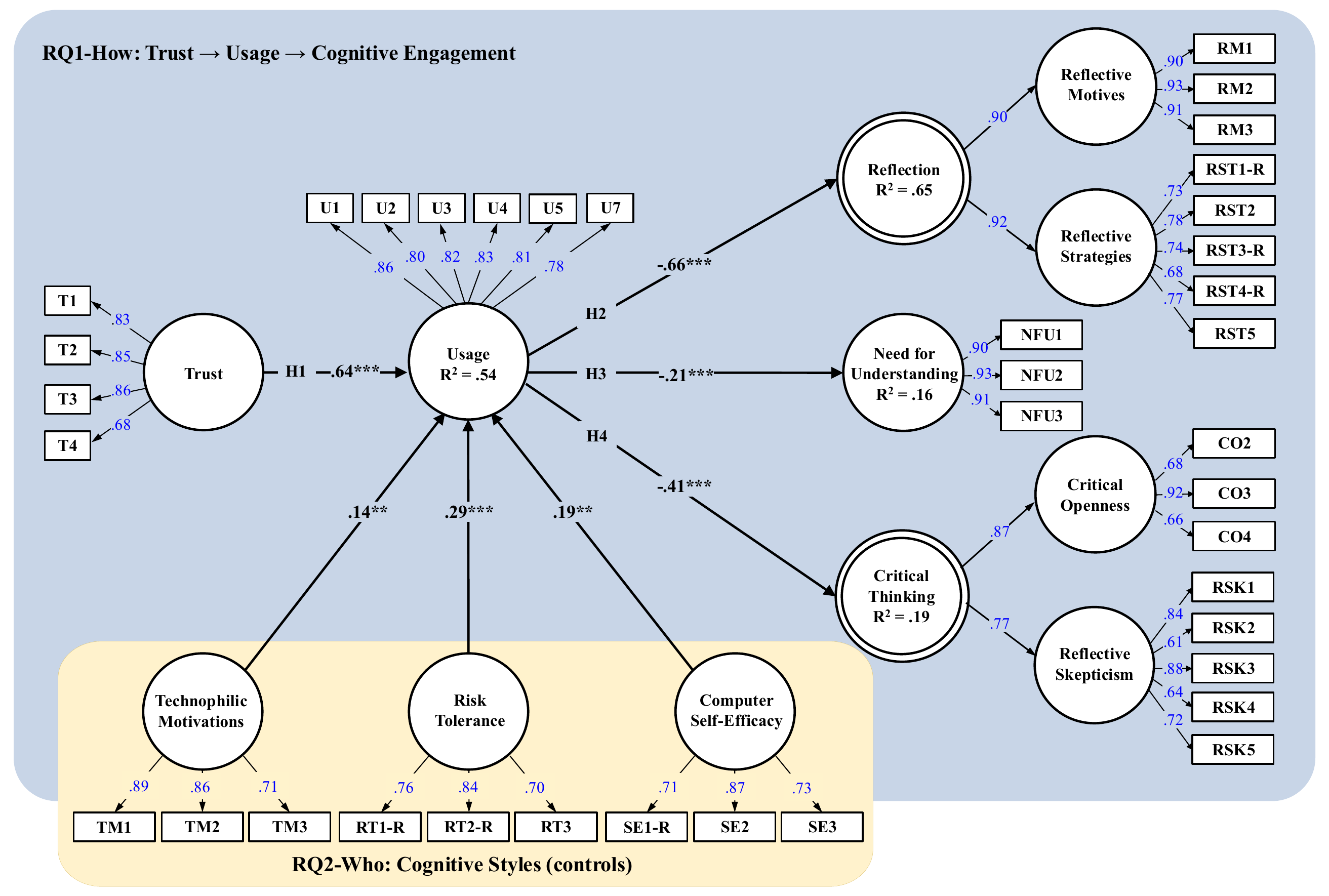}
% \vspace{-5px}
\caption{\small Structural model path associations (full model: RQs1\&2). Standardized path coefficients show direct effects of Trust and cognitive styles (Technophilic Motivations, Risk Tolerance, Computer Self-Efficacy; RQ2—yellow) on Usage, and of Usage on Reflection, Need for Understanding, and Critical Thinking, with significance ***p<.001, **p<.01, and *p<.05. Paths from LOCs (circles) to items (squares), and from HOCs (double circles) to LOCs, indicate factor loadings (in \textcolor{blue}{\textit{blue}}; all $\ge$ 0.6~\cite{hair2019use}). Transitive effects of cognitive styles on cognitive engagement are reported in Tab.~\ref{tab:RQ2-controls}.}
% \vspace{-5px}
\label{fig:RQ2model}
% \vspace{-7px}
\end{figure*}

The results were consistent with prior findings in knowledge work contexts.
%All three cognitive styles showed significant positive associations with students' routine genAI usage in coursework . 
Participants with stronger technophilic motivations, higher risk tolerance, and greater computer self-efficacy reported significantly more routine genAI use than their peers (see Tab.~\ref{tab:RQ2-controls-usage}, Fig.~\ref{fig:RQ2model} for path associations). 

%-------Tables go right AFTER the first ref to them
\begin{table}[!hbt]
% \footnotesize
% \scriptsize
\centering
\caption{Effects of students' cognitive styles on routine genAI usage: Standardized path coefficients ($\beta$), standard deviations (SD), 95\% confidence intervals (CI), \textit{p}-values, and effect sizes ($f^2$). We consider $f^2< 0.02$ no effect, $f^2 \in [0.02, 0.15)$ small, $f^2 \in [0.15, 0.35)$ medium, and $f^2> 0.35$ large~\cite{cohen2013statistical}.}

\label{tab:RQ2-controls-usage}
% \vspace{-8px}
\robustify{\bfseries}
\sisetup{
    mode=text,
    group-digits = false ,
    input-signs ={-},
    input-symbols = ( ) [ ] - + *,
    detect-weight=true, 
    detect-family=true,
    table-format=0.2,
    add-decimal-zero=false, %% doesn't seem to work :-(
    add-integer-zero=false,
    round-mode=places, 
    round-precision=2, %% change this for precision.
    parse-numbers = true
}
% \begin{tabular}{P{4.3cm}
%                 S
%                 S[table-format=0.2]
%                 >{\centering\arraybackslash}p{1.3cm}
%                 S[table-format=0.3,round-precision=3]}
% \toprule
% & {\textit{B}} & {SD} & {95\% CI} & {\textit{p}}\\
% \midrule
\begin{tabular}{P{7.9cm}  % Adjusted width for the first column
                S
                S[table-format=0.2]
                >{\centering\arraybackslash}p{1.7cm}
                S[table-format=0.3,round-precision=3]
                S[table-format=0.3,round-precision=2]} % New column for f^2
\toprule
& {\textit{B}} & {SD} & {95\% CI} & {\textit{p}} & {$f^2$} \\
\midrule \midrule
%\textsc{Hypotheses}&&&&\\
%\midrule
%\addlinespace
\rowcolor{myYellow}\hangindent1em Technophilic Motivations$\rightarrow$Usage & .140 & .058 & (0.07, 0.26) & \bfseries .011 & .171\\
\rowcolor{myYellow}\hangindent1em Risk Tolerance$\rightarrow$Usage & .291 & .049 & (0.19, 0.38) & \bfseries .000 & .214\\
\rowcolor{myYellow}\hangindent1em Computer Self-Efficacy$\rightarrow$Usage & .186 & .042 & (0.06, 0.14) & \bfseries .002 & .11\\
\bottomrule

\end{tabular}
\begin{tablenotes}
  % \footnotesize
\item \textit{Note: PLS-SEM estimates all structural paths jointly within a single model. Significance tests (and their Type-I error rates) are defined within this joint estimation; separate family-wise error corrections are not required}~\cite{hair2014primer, hair2019use}.
\end{tablenotes}
% \vspace{-4mm}
\end{table}

We find this result unsurprising. 
If a student is motivated by engaging with technology (cognitive style: technophilic motivations), it seems likely that they would likewise enjoy engaging with genAI, even when it is known to carry risks (cognitive style: risk tolerance), especially if the student is confident that they can handle those risks and use the tool effectively (cognitive style: computer self-efficacy). 
Indeed, these qualities are often touted in STEM education as indicators of career preparedness (e.g.,~\cite{US_Teens_Fear_Risk_Taking_2013, reagan2016, corbett2024psychological, spair2025techemployer, Getting_Students_Excited_about_STEM_2022, Finding_Success_in_Failure_2016, sakellariou2021self}).
Further, researchers have shown that these qualities \textit{are} often helpful for STEM~\cite{martinez2017assessing, giaccone2022unveiling}. 
However, it may be time to question whether this still holds with routine genAI use, which we investigate next.
%One interpretation is that students high in these traits naturally gravitate toward technological solutions. They find it intrinsically rewarding, feel comfortable acting despite uncertainty, and trust their capacity to manage the tool effectively~\cite{martinez2017assessing, margulieux2024self}. These are the qualities often associated with technological fluency and readiness~\cite{giaccone2022unveiling, corbett2024psychological}. But here is where the pattern becomes troubling.

%--------Figures go right AFTER the 1st ref
%---end RQ2 model figure

%-----
\subsubsection{\textbf{Cognitive styles and cognitive engagement}}

If higher technophilic motivations, risk tolerance, and computer self-efficacy remain helpful for STEM education even under conditions of routine genAI use, then students with these cognitive styles should also exhibit higher cognitive engagement in their coursework. 
% (e.g., spend more time reflecting upon it).
For example, among RQ1-How's findings, the participants in the High Usage/Trust group (recall outliers above the box in Fig.~\ref{fig:usage-trust-boxplot}) who exhibited fairly high reflection despite heavy genAI use would presumably be those with these cognitive styles.

\begin{figure}[!hbt] %fig motivations
\centering
% % % \vspace{-10px}
\includegraphics[width=\textwidth]{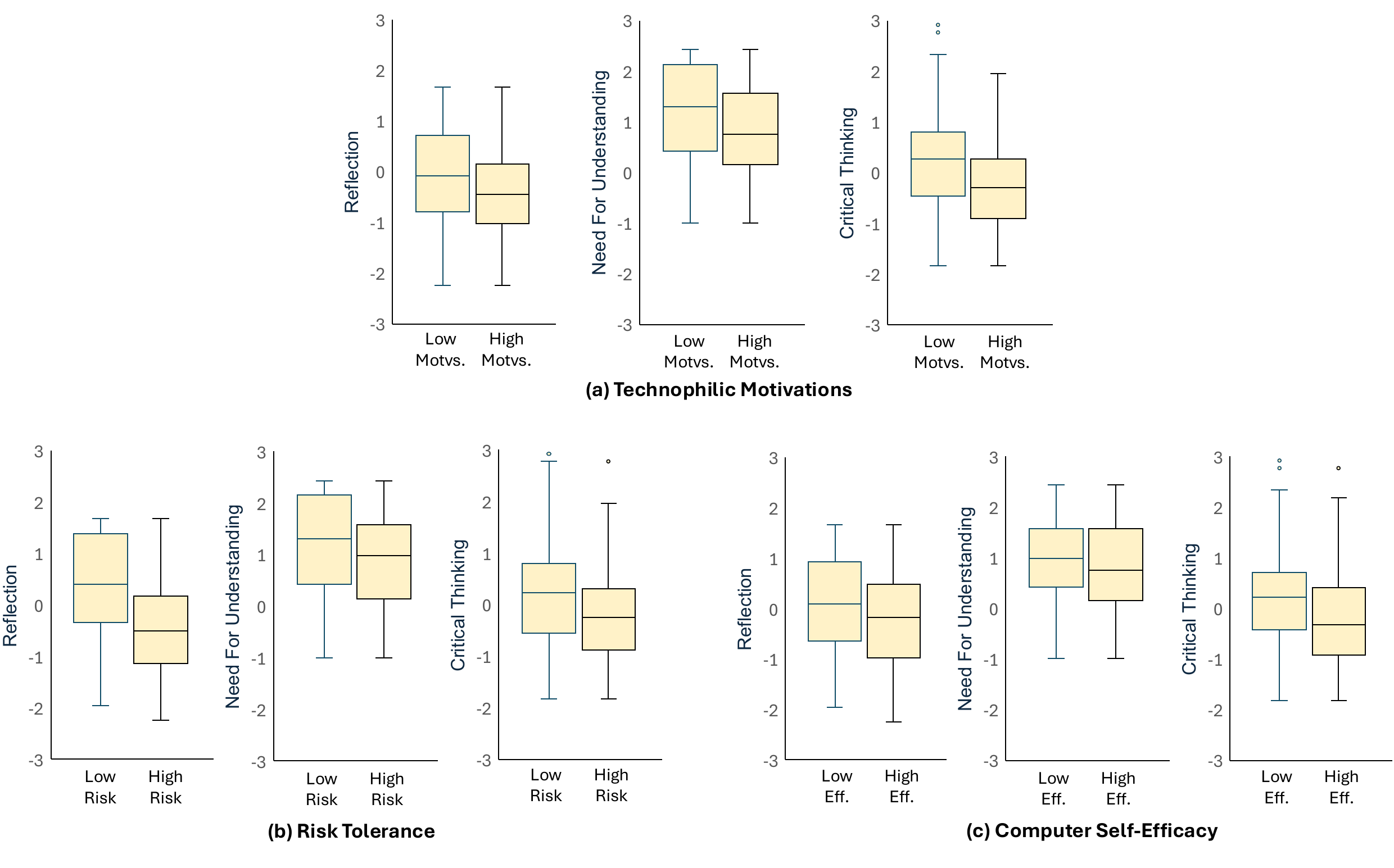}
% % \vspace{-5px}
\caption{Cognitive engagement scores by median-split groups on (a) Technophilic Motivations, (b) Risk Tolerance, and (c) Computer Self-Efficacy. Boxplots show standardized latent scores for Reflection, Need for Understanding, and Critical Thinking across all groups.}
% % \vspace{-5px}
\label{fig:boxplot-cognitive-styles}
% % \vspace{-7px}
\end{figure}

\textbf{Technophilic motivations:} 
However, the opposite was true. 
As Fig.~\ref{fig:boxplot-cognitive-styles} (a) illustrates, higher technophilic motivations corresponded with \textit{lower} median cognitive engagement across all three dimensions. 
Mediation analysis (Tab.~\ref{tab:RQ2-controls}) confirmed these patterns: participants with higher technophilic motivations reported significantly less Reflection ($\beta=-0.11$, \textit{p}=.010, $f^2=0.15$), significantly less Need for Understanding ($\beta=-0.12$, \textit{p} = .014, $f^2=0.07$), and significantly less Critical Thinking ($\beta=-0.05$, \textit{p} = .007, $f^2=0.09$).

%--- end of fig boxplot-motivations

%--------fig/table always goes RIGHT under the FIRST place it's mentioned.
\begin{table}[!hbt]
% \footnotesize
% \scriptsize
\centering
\caption{Transitive effects of students' cognitive styles on their cognitive engagement habits via routine genAI use: Standardized path coefficients ($\beta$), standard deviations (SD), confidence intervals (CI), \textit{p}-values, effect sizes ($f^2$), and Variance Accounted For (VAF $\ge 0.8$ indicates full mediation). We consider $f^2<$ 0.02 no effect, $f^2 \in$ [0.02, 0.15) small, $f^2 \in$ [0.15, 0.35) medium, and $f^2 >$ 0.35 large~\cite{cohen2013statistical}.}

\label{tab:RQ2-controls}
% \vspace{-8px}
\robustify{\bfseries}
\sisetup{
    mode=text,
    group-digits = false ,
    input-signs ={-},
    input-symbols = ( ) [ ] - + *,
    detect-weight=true, 
    detect-family=true,
    table-format=0.2,
    add-decimal-zero=false, %% doesn't seem to work :-(
    add-integer-zero=false,
    round-mode=places, 
    round-precision=2, %% change this for precision.
    parse-numbers = true
}
% \begin{tabular}{P{4.3cm}
%                 S
%                 S[table-format=0.2]
%                 >{\centering\arraybackslash}p{1.3cm}
%                 S[table-format=0.3,round-precision=3]}
% \toprule
% & {\textit{B}} & {SD} & {95\% CI} & {\textit{p}}\\
% \midrule
\begin{tabular}{P{7.9cm}  % Adjusted width for the first column
                S
                S[table-format=0.2]
                >{\centering\arraybackslash}p{1.7cm}
                S[table-format=0.3,round-precision=3]
                S[table-format=0.3,round-precision=2]
                S[table-format=0.3,round-precision=2]} % New column for VAF
\toprule
& {\textit{B}} & {SD} & {95\% CI} & {\textit{p}} & {$f^2$} & {VAF}\\
\midrule \midrule
%\textsc{Hypotheses}&&&&\\
%\midrule
%\addlinespace

\rowcolor{myYellow}\hangindent1em Technophilic Motivations$\rightarrow$Usage$\rightarrow$Reflection & -0.11 & .046 & (-0.21, -0.10) & \bfseries .010 & .15 & 0.83\\
\rowcolor{myYellow}\hangindent1em Technophilic Motivations$\rightarrow$Usage$\rightarrow$Need For Understanding & -0.12 & .014 & (-0.14, -0.11) & \bfseries .014 & .07 & 0.83\\
\rowcolor{myYellow}\hangindent1em Technophilic Motivations$\rightarrow$Usage$\rightarrow$Crticial Thinking & -0.05 & .026 & (-0.10, -0.02) & \bfseries .007 & .09 & 0.86\\
\grayline
\rowcolor{myYellow}\hangindent1em Risk Tolerance$\rightarrow$Usage$\rightarrow$Reflection & -0.148 & .039 & (-0.19, -0.12) & \bfseries .000 & .08 & 0.87\\
\rowcolor{myYellow}\hangindent1em Risk Tolerance$\rightarrow$Usage$\rightarrow$Need For Understanding & -0.113 & .021 & (-0.12, -0.06) & \bfseries .002 & .07 & 0.85\\
\rowcolor{myYellow}\hangindent1em Risk Tolerance$\rightarrow$Usage$\rightarrow$Crticial Thinking & -0.125 & .026 & (-0.14, -0.09) & \bfseries .000 & .08 & 0.82\\
\grayline
\rowcolor{myYellow}\hangindent1em Computer Self-Efficacy$\rightarrow$Usage$\rightarrow$Reflection & -0.167 & .033 & (-0.21, -0.12) & \bfseries .013 & .15 & 0.84\\
\rowcolor{myYellow}\hangindent1em Computer Self-Efficacy$\rightarrow$Usage$\rightarrow$Need For Understanding & -0.118 & .011 & (-0.17, -0.10) & \bfseries .005 & .07 & 0.86\\
\rowcolor{myYellow}\hangindent1em Computer Self-Efficacy$\rightarrow$Usage$\rightarrow$Crticial Thinking & -0.135 & .019 & (-0.16, -0.11) & \bfseries .020 & .13 & 0.87\\
\bottomrule

\end{tabular}
\begin{tablenotes}
  % \footnotesize
\item \textit{Note: PLS-SEM estimates all structural paths jointly within a single model. Significance tests (and their Type-I error rates) are defined within this joint estimation; separate family-wise error corrections are not required}~\cite{hair2014primer, hair2019use}.
\end{tablenotes}
% \vspace{-4mm}
\end{table}

\textbf{Risk tolerance and computer self-efficacy:} 
Similar patterns emerged for the other two cognitive styles (Fig.~\ref{fig:boxplot-cognitive-styles} (b) and (c)). Participants with higher risk tolerance reported significantly less Reflection ($\beta=-0.15$, \textit{p}<.001, $f^2=0.08$), significantly less Need for Understanding ($\beta=-0.11$, \textit{p}=.002, $f^2=0.07$), and significantly less Critical Thinking ($\beta=-0.13$, \textit{p}<.001, $f^2=0.08$). 
Research on attitude towards risk suggests one reason why~\cite{loewenstein2001risk}: risk tolerance captures willingness to proceed despite uncertainty. In HAI contexts, this can, alongside increasing usage, reduce users' inclination to question or scrutinize AI outputs~\cite{zhai2024effects}, even when such scrutiny would be warranted.

Participants with higher computer self-efficacy likewise reported significantly less Reflection ($\beta=-0.17$, \textit{p}=.013, $f^2=0.15$), significantly less Need for Understanding ($\beta=-0.12$, \textit{p}=.005, $f^2=0.07$), and significantly less Critical Thinking ($\beta=-0.14$, \textit{p}=.020, $f^2=0.13$). 
While prior work suggests that higher self-efficacy leads to greater success (e.g., as per self-efficacy as an effective surrogate for ability research~\cite{bandura1997self, compeau1995computer}), some education research indicates that greater self-confidence in genAI contexts may lead students to overestimate learning gains from such interactions~\cite{walker2025learningOutcomes, fan2025educational}.

\textbf{Group analysis}: Still, some research suggests that academic level and prior genAI experience may be important factors~\cite{strzelecki2025chatgpt, singer2025generative, choudhuri2025insights}. To investigate whether these factors affected the structural relationships in our model, we conducted standard PLS Multi-Group Analysis (MGA)~\cite{hair2014primer, russo2021pls}. We defined groups by (i) academic level (lower- vs. upper-division) and (ii) prior genAI experience (median split on self-reported experience). The analysis revealed no statistically significant differences across any paths, confirming that the model relationships held consistently regardless of participants' academic seniority or prior experience with genAI.

\begin{takeawayBox}
\textbf{\textit{Takeaway}}: 
Students that prevailing wisdom would classify as well-positioned for STEM work---those with \textit{higher technophilic motivations, risk tolerance, and computer self-efficacy}---were also the ones who were particularly susceptible to cognitively disengaging when they routinely used genAI.
\end{takeawayBox}

\subsection{Model adequacy assessment} 

Statistical and practical significance establish that relationships exist, but they do not indicate how well a model accounts for observed variance, whether it is well-specified and consistent with the data (fit), or whether it exhibits predictive relevance beyond the estimation sample. To address these questions, we assessed model adequacy using coefficients of determination ($R^2$, Adj.~$R^2$), model fit (SRMR), and out-of-sample predictive relevance ($Q^2$), following recommended PLS-SEM practices~\cite{russo2021pls, hair2019use}.

Coefficients of determination values ($R^2$, Adj.~$R^2$) capture how much variance in each outcome (e.g., Reflection) is explained by its predictors. As shown in Tab.~\ref{tab:r2-table}, the model explains a substantial proportion of variance in \textit{Usage} ($R^2=0.54$) and \textit{Reflection} ($R^2=0.65$). The model explains more moderate variance in \textit{Need for Understanding} ($R^2=0.16$) and \textit{Critical Thinking} ($R^2=0.19$). While these constructs are influenced by a broader set of factors beyond the model, the observed values exceed the common benchmark of 0.15 for exploratory model adequacy in behavioral sciences~\cite{chin1998partial, hair2014primer}.

\begin{table}[htb]
% \scriptsize
%\caption{Coefficient of determination ($R^2$, $\text{Adjusted R}^2$) and Predictive relevance ($Q^2_{predict}$)}
\caption{Explanatory power of the structural model. $R^2$ and Adj. $R^2$ reflect the variance in endogeneous constructs explained by its predictors; $Q^2_\text{predict}$ > 0 indicate the model's out-of-sample predictive relevance.}
\label{tab:r2-table}
% \vspace{-8px}
\centering
\begin{tabular}{P{3.5cm}  % Adjusted width for the first column
                S[table-format=1.5]  % Allocated space for R^2
                S[table-format=1.5]  % Allocated space for Adj. R^2
                S[table-format=1.5]} % Allocated space for Q^2
\toprule
\textbf{Construct} & \textit{$\boldsymbol{R^2}$} & \textit{\textbf{Adj.} $\boldsymbol{R^2}$} & \textit{$\boldsymbol{Q^2_\text{\textbf{predict}}}$} \\
\midrule
\midrule
Usage & 0.544 & 0.541 & 0.515 \\
Reflection & 0.652 & 0.649 & 0.478 \\
Need for Understanding & 0.164 & 0.163 & 0.131 \\
Critical Thinking & 0.192 & 0.188 & 0.267\\
\bottomrule
\end{tabular}
% \vspace{-3mm}
\end{table}

Model fit was assessed using the Standardized Root Mean Square Residual (SRMR), which is recommended for detecting structural misspecification in PLS-SEM~\cite{russo2021pls}. The observed SRMR of 0.067 falls below the conservative threshold of 0.08~\cite{henseler2016using}, indicating a well-specified model, i.e., the proposed relationships are consistent with the covariance structure of the data.

Finally, we assessed out-of-sample predictive relevance to evaluate whether the model generalizes beyond the estimation sample. Using Stone-Geisser’s $Q^2$~\cite{stone1974cross}, computed via 10-fold, 10-repeat PLS-predict, all dependent constructs yielded positive values (see Tab.~\ref{tab:r2-table}). This indicates that the model’s predictions on held-out data outperform a naive mean-based benchmark, supporting its predictive relevance beyond in-sample explanatory fit.

\begin{takeawayBox}
   \textbf{Takeaway:} The proposed model is adequate for its intended purpose: it captures sufficient variance in key outcomes, shows no evidence of structural misspecification, and demonstrates predictive relevance beyond in-sample fit.
\end{takeawayBox}

% % \vspace{-2mm}

\section{Discussion: The Cognitive Debt Cycle}
\label{sec:discussion}

%\subsection{The Cognitive Debt Cycle}

Our findings raise the possibility of a \textit{cognitive debt cycle} emerging from students’ trust-driven, routine use of genAI. Much like technical debt accumulates when short-term expedience substitutes for more principled design, cognitive debt may accrue when students repeatedly delegate relevant cognitive work onto genAI. The concern is that genAI may restructure the learning environment in ways that make delegation appear rational, gradually training the brain to reconsider what effort investment feels worthwhile. 
Further, these effects may be \textit{self-reinforcing}: as students repeatedly delegate cognitive work, their ability to perform these tasks independently may decline. GenAI may then become increasingly necessary, further reducing the already limited opportunities to hone these skills. The result may be a downward spiral of overreliance, in which understanding becomes increasingly shallow, and the capacity to apply knowledge in new or dynamic contexts gradually erodes.

Unfortunately, our findings suggest that responsible genAI use did not automatically emerge with academic seniority or prior genAI experience, challenging the assumption that students will ``grow into'' reflective users. This raises a fundamental question: \textit{How can students be expected to regulate their genAI use when the very capacities required to do so are progressively weakened by routine interaction with these tools?} 

\vspace{-2mm}
\subsection{Interrupting the Cycle: What can educators do?}
    
Recognizing the cognitive debt cycle is one thing; interrupting it is another. A common response has been to call for more AI literacy~\cite{gu2025ai}, with the expectation that better-informed learners will be more attuned to the risks of overreliance. Indeed, emerging frameworks, such as the Digital Education Council's AI literacy model~\cite{dec2025ailiteracy}, position critical thinking and judgment about AI as core competencies, emphasizing the ability to evaluate AI reasoning, identify bias, and recognize contextual gaps in AI outputs. 
We agree with the importance of these skills. Still, prior work suggests that awareness of AI's fallibility does not reliably deter reliance when answers are immediate and cognitively effortless~\cite{buccinca2021trust, Stone2024}. Our findings echo this pattern: neither prior AI experience nor academic seniority insulated students from cognitive disengagement, suggesting that \textit{awareness alone does not counteract the incentives to disengage}.

Why might this be? Anderson’s theory of rational analysis~\cite{anderson2013adaptive} suggests one reason why. According to this theory, humans exhibit a ``local maximum of adaptedness’’ to their environments, optimizing for maximum immediate reward and minimum immediate cost. In educational contexts, this implies that delegating cognitive work to genAI may be a rational response to environments that reward efficiency while imposing few short-term costs for shallow engagement. 

This theory then also suggests possibilities for adjusting students' perceived costs and rewards of AI delegation as a means to interrupt the cognitive debt cycle---affording educators a rethinking of educational experiences where delegating cognitive work becomes less attractive to students.

% \textit{Making engagement rewarding through meaningful work design}: 
One possibility centers on increasing the intrinsic rewards of cognitive engagement itself. Tasks that are personally meaningful or inherently enjoyable could reduce the appeal of delegation~\cite{lazarus1991emotion, roseman2001appraisal, lips2009discriminating}, as learners continue to derive satisfaction from the work. For example, designing coursework as opportunities for exploration through storytelling~\cite{clark2023learning}, gamified elements~\cite{hamari2014does, lieberoth2015shallow}, or curiosity-driven puzzles (e.g., ``Why did the AI get this wrong?'' or ``How might the answer differ depending on the point of view?'') could tap into intrinsic motivations that invite more deliberate engagement~\cite{costantini2025playful}. Notably, curiosity-based approaches may also make the consequences of delegation more visible by surfacing AI errors, omissions, or mismatches within students’ own contexts.

Another possibility is to consider increasing the costs of AI delegation through intentional task and activity design. 
One approach could involve alternating between AI-assisted and independent work. For instance, students might first generate code with AI assistance, then be required to independently critique, debug/repair, verify, and adapt that code to their project requirements without AI (e.g., during a supervised lab). Such designs can make delegation immediately costly, as knowledge gaps that were masked during the AI-assisted phase must be confronted during subsequent independent work. Cumulative assignments that build on previous assignments can amplify these costs further, as shallow cognitive engagement in early assignments could compound into visible struggles in later ones. 
Finally, combining these approaches with in-class assessments and/or collaborative/group activities~\cite{smith1996cooperative} can further reinforce engagement by leveraging social accountability~\cite{lerner1999accounting, wentzel1991social}, where shallow AI-generated contributions become quickly visible, potentially compromising individual and/or group reputation and outcomes.

\boldification{** But, this is not a one-stop solution. The challenge for educators remains adaptive as these tools and human mindsets coevolve. }

Importantly, these strategies are not ``silver bullets” nor miracle cures. As genAI capabilities, human practices, and mindsets co-evolve, responses to this challenge will need to adapt. A key ongoing question for \textit{educators} remains:
\textit{How to design educational experiences to foster cognitive engagement in ways that improve both (1) students’ learning of the subject matter and (2) their ability to use AI appropriately?}

\subsection{Interrupting the Cycle: What can AI designers do?}

Interrupting the debt cycle may require more than pedagogical adjustments; it raises questions about how genAI itself can be designed for educational use.
One possibility, drawing on Engelbart's early vision of intellectual augmentation~\cite{engelbart2023augmenting}, is to design genAI as ``bicycles for the mind''. Just as bicycles amplify physical efficiency while requiring the rider's active control; genAI, designed as mind-bicycles, could amplify cognitive reach (e.g., by critiquing, counter-arguing, nudging the user~\cite{reicherts2025ai}) while keeping humans responsible for exploration, steering, and judgment. 
The goal is for progress to emerge through \emph{complementarity}: the human and the AI each need to contribute distinct forms of cognitive work to achieve meaningful outcomes. 
% \textbf{** What this framing seeks to avoid is a single point of failure. Designing for complementarity sustains AI-augmented knowledge work. How do we do it effectively is the question.}
% Crucially, such systems preserve \emph{productive friction}. Effort is redistributed where machines excel, but retained where human judgment, interpretation, and evaluation remain essential.

What this framing seeks to avoid is a wholesale substitution of human agency. When genAI ``thinks'' on a student's behalf, they lose opportunities to build skill, and genAI, in turn, loses access to future high-quality human input shaping its evolution. 
Designing for complementarity is therefore not only pedagogically desirable; it is necessary for sustaining genAI-assisted knowledge work. 
% From this perspective, friction is not simply a usability flaw to be eliminated, but a potential design requirement. 
A key ongoing challenge for \textit{AI designers} then is: \textit{How to design genAI for complementarity in ways that meaningfully amplify human thinking instead of replacing it?}

% \textbf{** This may run counter to product metrics. After all, we are suggesting tools to have friction. But, friction has indeed proved to be helpful before in HAI contexts, because it makes people think.}

% Moreover, this orientation points to a broader design consideration: moving beyond optimization for maximal convenience toward designing for mutual competency and redundancy. 

% Such designs run counter to traditional product metrics optimizing for maximal convenience; after all, we are suggesting designing systems that incorporate friction for mutual competency. 

Incorporating ``friction''---intentional points of difficulty~\cite{cox2016design}---could offer one possibility toward designing such systems for educational use. 
Designing for friction runs counter to conventional product metrics that prioritize maximal convenience (e.g., fewer clicks). However, prior work shows that friction can be productive when it slows users down to provoke reflection on their emerging work~\cite{buccinca2021trust, drosos2025makes}. 
%beckwith2008-tinkering, but you don't need it...
% 
One way to introduce friction, for example, could be to design education-oriented genAI as a ``provocateur,'' repositioning assistance as questioning rather than confirming~\cite{drosos2025makes}.
Such a provocateur could prompt students to revisit their assumptions and interrogate genAI’s underlying rationales, enabling them to identify and correct flaws in that (and/or their own) reasoning.
Another way could be to design systems that require users to deliberately engage with multiple AI perspectives: for instance, by comparing outputs generated under differing viewpoints or competing claims and reconciling disagreements among them. Prior work shows that such designs can indeed strengthen users’ critical thinking and reasoning practices~\cite{liu2025perspectra}.

% \textbf{** Further, our findings point that this reimagining should be for a spectrum of users: not one-size-fits all. %Ironically, ``tech ready'' students ended up disengaging most. Accordingly, adaptivity in educational AI to include cognitive profiles can be beneficial. Here are some ways.}

Furthermore, our findings suggest that reimagining genAI for complementarity should not assume a one-size-fits-all. Recall, students who might have been regarded as most ``tech-ready''---those high in technophilic motivations, self-efficacy, and risk tolerance---were also the ones most prone to cognitive disengagement. 
% Thus, redesigning genAI to improve the latter group's quality of education should not come at the cost of undermining the education quality of the former group.

This raises possibilities for adaptivity in education-oriented genAI design to account for students’ diverse cognitive styles.
For instance, technophilic users may benefit from \textit{interface nudges} that demand comparison and justification (e.g., selecting among multiple AI-generated solutions and justifying their choice, or attempting to ``beat the AI''). Similarly, highly self-efficacious users may benefit from scaffolds that require an initial attempt (e.g., plans, assumptions, rationales) before AI assistance is provided, helping recalibrate their expectations and confidence in AI support.

That said, adaptivity that is too automatic risks undermining user agency and locking students into a fixed interaction style. Although human-AI interaction research is still relatively young, the principle of keeping users in control has been affirmed by many HAI researchers (e.g., ~\cite{anderson2024-GMstylesInAI-TiiS, amershi2019guidelines, choudhuri2024far, palanque2023multi, bhattacharya2025userControl, shneiderman2022humancenteredAI, schmidt2020interactiveHumanAI, hook2000beforeIUIsbecomeReal}). Accordingly, adaptivity should remain configurable, enabling students to steer how the system supports them as their goals, contexts, and skills evolve.

Ultimately, realizing this future will require coordinated effort among AI designers, researchers, and educators to support students in developing and sustaining essential intellectual habits when using genAI, rather than perpetuating the cognitive debt cycle. While the specific pathways forward remain open and evolving, the cost of inaction is clear: \textit{a future where apparent ``efficiency'' rises even as sustained thinking habits quietly erode}.

% % % \vspace{-2mm}
% \input{sec/7.Threat}
% % % \vspace{-2mm}
% % \vspace{0.5mm}
\section{Conclusion}
\label{sec:conclusion}

We began this investigation by asking: \textit{What happens to students' cognitive engagement when genAI becomes routine?} The answers we found were both clear and concerning. Our main findings were:

\begin{itemize}
    \item \textbf{RQ1-How}: Students who routinely used genAI reported statistically significantly less cognitive engagement, i.e., less reflection, less need for understanding, and less critical thinking in their coursework. Higher trust in genAI further amplified these disengagement effects through increased routine genAI use.
    \item \textbf{RQ2-Who}: Ironically, students commonly touted as ``tech-natives''---those with stronger technophilic motivations, greater computer self-efficacy, and higher risk tolerance---were the ones who were significantly more prone to genAI-related cognitive disengagement.
\end{itemize}

If routine reliance on genAI during formative years changes students’ willingness to engage in effortful thinking, many may enter professional life without having developed the intellectual habits that earlier generations developed through practice. 
Yet this need not be inevitable. The cognitive debt cycle can be interrupted, but doing so will take more than exhortation or awareness campaigns. It calls for deliberate redesign of both learning environments and genAI systems to preserve human agency and support genuine complementarity---a division of cognitive labor in which both human and AI contribute meaningfully, and neither is displaced. The choices educators and AI designers make now will shape not only what students learn, but how they learn to think.

%%
%% The acknowledgments section is defined using the "acks" environment
%% (and NOT an unnumbered section). This ensures the proper
%% identification of the section in the article metadata, and the
%% consistent spelling of the heading.

\begin{acks}

This work was supported in part by USDA-NIFA 2021-67021-35344 and by the National Science Foundation under Grant Nos. 2235601, 2236198, 2247929, 2303042, and 2303043. Any opinions, findings, conclusions, or recommendations expressed in this material are those of the authors and do not necessarily reflect the views of the sponsors.

\textbf{Declarations of interest:} No potential conflict of interest was reported by the author(s).

\end{acks}

% \bibliographystyle{ACM-Reference-Format}
% \bibliography{acmart}

% \appendix
% \pagebreak
% \bibliographystyle{IEEEtranN}
\bibliographystyle{ACM-Reference-Format}
% \bibliography{REFERENCES}
\bibliography{acmart}

%%% -*-BibTeX-*-
%%% Do NOT edit. File created by BibTeX with style
%%% ACM-Reference-Format-Journals [18-Jan-2012].

\begin{thebibliography}{144}

%%% ====================================================================
%%% NOTE TO THE USER: you can override these defaults by providing
%%% customized versions of any of these macros before the \bibliography
%%% command.  Each of them MUST provide its own final punctuation,
%%% except for \shownote{} and \showURL{}.  The latter two
%%% do not use final punctuation, in order to avoid confusing it with
%%% the Web address.
%%%
%%% To suppress output of a particular field, define its macro to expand
%%% to an empty string, or better, \unskip, like this:
%%%
%%% \newcommand{\showURL}[1]{\unskip}   % LaTeX syntax
%%%
%%% \def \showURL #1{\unskip}           % plain TeX syntax
%%%
%%% ====================================================================

\ifx \showCODEN    \undefined \def \showCODEN     #1{\unskip}     \fi
\ifx \showISBNx    \undefined \def \showISBNx     #1{\unskip}     \fi
\ifx \showISBNxiii \undefined \def \showISBNxiii  #1{\unskip}     \fi
\ifx \showISSN     \undefined \def \showISSN      #1{\unskip}     \fi
\ifx \showLCCN     \undefined \def \showLCCN      #1{\unskip}     \fi
\ifx \shownote     \undefined \def \shownote      #1{#1}          \fi
\ifx \showarticletitle \undefined \def \showarticletitle #1{#1}   \fi
\ifx \showURL      \undefined \def \showURL       {\relax}        \fi
% The following commands are used for tagged output and should be
% invisible to TeX
\providecommand\bibfield[2]{#2}
\providecommand\bibinfo[2]{#2}
\providecommand\natexlab[1]{#1}
\providecommand\showeprint[2][]{arXiv:#2}

\bibitem[sup({[n.\,d.]})]%
        {supplemental}
 \bibinfo{year}{[n.\,d.]}\natexlab{}.
\newblock \bibinfo{title}{Supplemental Package}.
\newblock \bibinfo{howpublished}{\url{https://zenodo.org/record/18420685}}.
\newblock


\bibitem[Abbas et~al\mbox{.}(2024)]%
        {abbas2024harmful}
\bibfield{author}{\bibinfo{person}{Muhammad Abbas}, \bibinfo{person}{Farooq~Ahmed Jam}, {and} \bibinfo{person}{Tariq~Iqbal Khan}.} \bibinfo{year}{2024}\natexlab{}.
\newblock \showarticletitle{Is it harmful or helpful? Examining the causes and consequences of generative AI usage among university students}.
\newblock \bibinfo{journal}{\emph{International Journal of Educational Technology in Higher Education}} \bibinfo{volume}{21}, \bibinfo{number}{1} (\bibinfo{year}{2024}), \bibinfo{pages}{10}.
\newblock


\bibitem[Amershi et~al\mbox{.}(2019)]%
        {amershi2019guidelines}
\bibfield{author}{\bibinfo{person}{Saleema Amershi}, \bibinfo{person}{Dan Weld}, \bibinfo{person}{Mihaela Vorvoreanu}, \bibinfo{person}{Adam Fourney}, \bibinfo{person}{Besmira Nushi}, \bibinfo{person}{Penny Collisson}, \bibinfo{person}{Jina Suh}, \bibinfo{person}{Shamsi Iqbal}, \bibinfo{person}{Paul~N Bennett}, \bibinfo{person}{Kori Inkpen}, {et~al\mbox{.}}} \bibinfo{year}{2019}\natexlab{}.
\newblock \showarticletitle{Guidelines for human-{AI} interaction}. In \bibinfo{booktitle}{\emph{2019 CHI Conference on Human Factors in Computing Systems}}. \bibinfo{pages}{1--13}.
\newblock


\bibitem[Amoozadeh et~al\mbox{.}(2024)]%
        {amoozadeh2024trust}
\bibfield{author}{\bibinfo{person}{Matin Amoozadeh}, \bibinfo{person}{David Daniels}, \bibinfo{person}{Daye Nam}, \bibinfo{person}{Aayush Kumar}, \bibinfo{person}{Stella Chen}, \bibinfo{person}{Michael Hilton}, \bibinfo{person}{Sruti Srinivasa~Ragavan}, {and} \bibinfo{person}{Mohammad~Amin Alipour}.} \bibinfo{year}{2024}\natexlab{}.
\newblock \showarticletitle{Trust in Generative AI among Students: An exploratory study}. In \bibinfo{booktitle}{\emph{Proceedings of the 55th ACM Technical Symposium on Computer Science Education V. 1}}. \bibinfo{pages}{67--73}.
\newblock


\bibitem[Anderson et~al\mbox{.}(2024)]%
        {anderson2024-GMstylesInAI-TiiS}
\bibfield{author}{\bibinfo{person}{Andrew Anderson}, \bibinfo{person}{Jimena~Noa Guevara}, \bibinfo{person}{Fatima Moussaoui}, \bibinfo{person}{Tianyi Li}, \bibinfo{person}{Mihaela Vorvoreanu}, {and} \bibinfo{person}{Margaret Burnett}.} \bibinfo{year}{2024}\natexlab{}.
\newblock \showarticletitle{Measuring user experience inclusivity in human-AI interaction via five user problem-solving styles}.
\newblock \bibinfo{journal}{\emph{ACM Transactions on Interactive Intelligent Systems}} \bibinfo{volume}{14}, \bibinfo{number}{3} (\bibinfo{year}{2024}), \bibinfo{pages}{1--90}.
\newblock


\bibitem[Anderson et~al\mbox{.}(2025)]%
        {anderson2025llm}
\bibfield{author}{\bibinfo{person}{Andrew Anderson}, \bibinfo{person}{David Piorkowski}, \bibinfo{person}{Margaret Burnett}, {and} \bibinfo{person}{Justin Weisz}.} \bibinfo{year}{2025}\natexlab{}.
\newblock \showarticletitle{An LLM's Attempts to Adapt to Diverse Software Engineers' Problem-Solving Styles: More Inclusive \& Equitable?}
\newblock \bibinfo{journal}{\emph{arXiv preprint arXiv:2503.11018}} (\bibinfo{year}{2025}).
\newblock


\bibitem[Anderson(2013)]%
        {anderson2013adaptive}
\bibfield{author}{\bibinfo{person}{John~R Anderson}.} \bibinfo{year}{2013}\natexlab{}.
\newblock \bibinfo{booktitle}{\emph{The adaptive character of thought}}.
\newblock \bibinfo{publisher}{Psychology Press}.
\newblock


\bibitem[Bandura(1997)]%
        {bandura1997self}
\bibfield{author}{\bibinfo{person}{Albert Bandura}.} \bibinfo{year}{1997}\natexlab{}.
\newblock \showarticletitle{Self-efficacy the exercise of control. New York: H}.
\newblock \bibinfo{journal}{\emph{Freeman \& Co. Student Success}}  \bibinfo{volume}{333} (\bibinfo{year}{1997}), \bibinfo{pages}{48461}.
\newblock


\bibitem[Barcaui(2025)]%
        {barcaui2025chatgpt}
\bibfield{author}{\bibinfo{person}{Andr{\'e} Barcaui}.} \bibinfo{year}{2025}\natexlab{}.
\newblock \showarticletitle{ChatGPT as a cognitive crutch: Evidence from a randomized controlled trial on knowledge retention}.
\newblock \bibinfo{journal}{\emph{Social Sciences \& Humanities Open}}  \bibinfo{volume}{12} (\bibinfo{year}{2025}), \bibinfo{pages}{102287}.
\newblock


\bibitem[Becker et~al\mbox{.}(2012)]%
        {becker2012hierarchical}
\bibfield{author}{\bibinfo{person}{Jan-Michael Becker}, \bibinfo{person}{Kristina Klein}, {and} \bibinfo{person}{Martin Wetzels}.} \bibinfo{year}{2012}\natexlab{}.
\newblock \showarticletitle{Hierarchical latent variable models in PLS-SEM: guidelines for using reflective-formative type models}.
\newblock \bibinfo{journal}{\emph{Long range planning}} \bibinfo{volume}{45}, \bibinfo{number}{5-6} (\bibinfo{year}{2012}), \bibinfo{pages}{359--394}.
\newblock


\bibitem[Bhattacharya et~al\mbox{.}(2025)]%
        {bhattacharya2025userControl}
\bibfield{author}{\bibinfo{person}{Aditya Bhattacharya}, \bibinfo{person}{Simone Stumpf}, {and} \bibinfo{person}{Katrien Verbert}.} \bibinfo{year}{2025}\natexlab{}.
\newblock \showarticletitle{Importance of User Control in Data-Centric Steering for Healthcare Experts}.
\newblock \bibinfo{journal}{\emph{arXiv preprint arXiv:2506.18770}} (\bibinfo{year}{2025}).
\newblock


\bibitem[Biggs et~al\mbox{.}(2001)]%
        {biggs2001revised}
\bibfield{author}{\bibinfo{person}{John Biggs}, \bibinfo{person}{David Kember}, {and} \bibinfo{person}{Doris~YP Leung}.} \bibinfo{year}{2001}\natexlab{}.
\newblock \showarticletitle{The revised two-factor study process questionnaire: R-SPQ-2F}.
\newblock \bibinfo{journal}{\emph{British Journal of Educational Psychology}} \bibinfo{volume}{71}, \bibinfo{number}{1} (\bibinfo{year}{2001}), \bibinfo{pages}{133--149}.
\newblock


\bibitem[Bloom et~al\mbox{.}(1956)]%
        {bloom1956taxonomy}
\bibfield{author}{\bibinfo{person}{Benjamin~S Bloom}, \bibinfo{person}{Max~D Engelhart}, \bibinfo{person}{Edward~J Furst}, \bibinfo{person}{Walker~H Hill}, \bibinfo{person}{David~R Krathwohl}, {et~al\mbox{.}}} \bibinfo{year}{1956}\natexlab{}.
\newblock \bibinfo{booktitle}{\emph{Taxonomy of educational objectives: The classification of educational goals. Handbook 1: Cognitive domain}}.
\newblock \bibinfo{publisher}{Longman New York}.
\newblock


\bibitem[Boekaerts et~al\mbox{.}(1999)]%
        {boekaerts1999handbook}
\bibfield{author}{\bibinfo{person}{Monique Boekaerts}, \bibinfo{person}{Moshe Zeidner}, {and} \bibinfo{person}{Paul~R Pintrich}.} \bibinfo{year}{1999}\natexlab{}.
\newblock \bibinfo{booktitle}{\emph{Handbook of self-regulation}}.
\newblock \bibinfo{publisher}{Elsevier}.
\newblock


\bibitem[Bryce and Withers(2003)]%
        {bryce2003engaging}
\bibfield{author}{\bibinfo{person}{Jennifer Bryce} {and} \bibinfo{person}{Graeme Withers}.} \bibinfo{year}{2003}\natexlab{}.
\newblock \showarticletitle{Engaging secondary school students in lifelong learning}.
\newblock  (\bibinfo{year}{2003}).
\newblock


\bibitem[Bu{\c{c}}inca et~al\mbox{.}(2021)]%
        {buccinca2021trust}
\bibfield{author}{\bibinfo{person}{Zana Bu{\c{c}}inca}, \bibinfo{person}{Maja~Barbara Malaya}, {and} \bibinfo{person}{Krzysztof~Z Gajos}.} \bibinfo{year}{2021}\natexlab{}.
\newblock \showarticletitle{To trust or to think: cognitive forcing functions can reduce overreliance on AI in AI-assisted decision-making}.
\newblock \bibinfo{journal}{\emph{Proceedings of the ACM on Human-computer Interaction}} \bibinfo{volume}{5}, \bibinfo{number}{CSCW1} (\bibinfo{year}{2021}), \bibinfo{pages}{1--21}.
\newblock


\bibitem[Burnett et~al\mbox{.}(2016a)]%
        {burnett2016-GMfieldStudy-CHI}
\bibfield{author}{\bibinfo{person}{Margaret Burnett}, \bibinfo{person}{Anicia Peters}, \bibinfo{person}{Charles Hill}, {and} \bibinfo{person}{Noha Elarief}.} \bibinfo{year}{2016}\natexlab{a}.
\newblock \showarticletitle{Finding gender-inclusiveness software issues with GenderMag: A field investigation}. In \bibinfo{booktitle}{\emph{Proceedings of the 2016 CHI Conference on Human Factors in Computing Systems}}. \bibinfo{pages}{2586--2598}.
\newblock


\bibitem[Burnett et~al\mbox{.}(2016b)]%
        {burnett2016gendermag}
\bibfield{author}{\bibinfo{person}{Margaret Burnett}, \bibinfo{person}{Simone Stumpf}, \bibinfo{person}{Jamie Macbeth}, \bibinfo{person}{Stephann Makri}, \bibinfo{person}{Laura Beckwith}, \bibinfo{person}{Irwin Kwan}, \bibinfo{person}{Anicia Peters}, {and} \bibinfo{person}{William Jernigan}.} \bibinfo{year}{2016}\natexlab{b}.
\newblock \showarticletitle{GenderMag: A method for evaluating software's gender inclusiveness}.
\newblock \bibinfo{journal}{\emph{Interacting with Computers}} \bibinfo{volume}{28}, \bibinfo{number}{6} (\bibinfo{year}{2016}), \bibinfo{pages}{760--787}.
\newblock


\bibitem[Butler et~al\mbox{.}(2025)]%
        {nfw2025}
\bibfield{author}{\bibinfo{person}{Jenna Butler}, \bibinfo{person}{Steven Jaffe}, \bibinfo{person}{Ralf Jan{\ss}en}, \bibinfo{person}{Nancy Baym}, \bibinfo{person}{Brent Hecht}, \bibinfo{person}{Jake Hofman}, \bibinfo{person}{Sean Rintel}, \bibinfo{person}{Bahareh Sarrafzadeh}, \bibinfo{person}{Abigail~M. Sellen}, \bibinfo{person}{Teodor Vorvoreanu}, {and} \bibinfo{person}{Jaime Teevan}.} \bibinfo{year}{2025}\natexlab{}.
\newblock \bibinfo{booktitle}{\emph{Microsoft New Future of Work Report 2025}}.
\newblock \bibinfo{type}{{T}echnical {R}eport} MSR-TR-2025-58. \bibinfo{institution}{Microsoft Research}.
\newblock
\urldef\tempurl%
\url{https://aka.ms/nfw2025}
\showURL{%
\tempurl}


\bibitem[Cacioppo and Petty(1982)]%
        {cacioppo1982need}
\bibfield{author}{\bibinfo{person}{John~T Cacioppo} {and} \bibinfo{person}{Richard~E Petty}.} \bibinfo{year}{1982}\natexlab{}.
\newblock \showarticletitle{The need for cognition.}
\newblock \bibinfo{journal}{\emph{Journal of personality and social psychology}} \bibinfo{volume}{42}, \bibinfo{number}{1} (\bibinfo{year}{1982}), \bibinfo{pages}{116}.
\newblock


\bibitem[Chen et~al\mbox{.}(2025)]%
        {chen2025more}
\bibfield{author}{\bibinfo{person}{Xinyue Chen}, \bibinfo{person}{Kunlin Ruan}, \bibinfo{person}{Kexin~Phyllis Ju}, \bibinfo{person}{Nathan Yap}, {and} \bibinfo{person}{Xu Wang}.} \bibinfo{year}{2025}\natexlab{}.
\newblock \showarticletitle{{More AI assistance reduces cognitive engagement: Examining the AI assistance dilemma in AI-supported note-taking}}.
\newblock \bibinfo{journal}{\emph{Proceedings of the ACM on Human-Computer Interaction}} \bibinfo{volume}{9}, \bibinfo{number}{7} (\bibinfo{year}{2025}), \bibinfo{pages}{1--29}.
\newblock


\bibitem[Chi(2009)]%
        {chi2009active}
\bibfield{author}{\bibinfo{person}{Michelene~TH Chi}.} \bibinfo{year}{2009}\natexlab{}.
\newblock \showarticletitle{Active-constructive-interactive: A conceptual framework for differentiating learning activities}.
\newblock \bibinfo{journal}{\emph{Topics in cognitive science}} \bibinfo{volume}{1}, \bibinfo{number}{1} (\bibinfo{year}{2009}), \bibinfo{pages}{73--105}.
\newblock


\bibitem[Chi and Wylie(2014)]%
        {chi2014icap}
\bibfield{author}{\bibinfo{person}{Michelene~TH Chi} {and} \bibinfo{person}{Ruth Wylie}.} \bibinfo{year}{2014}\natexlab{}.
\newblock \showarticletitle{The ICAP framework: Linking cognitive engagement to active learning outcomes}.
\newblock \bibinfo{journal}{\emph{Educational psychologist}} \bibinfo{volume}{49}, \bibinfo{number}{4} (\bibinfo{year}{2014}), \bibinfo{pages}{219--243}.
\newblock


\bibitem[Chin et~al\mbox{.}(1998)]%
        {chin1998partial}
\bibfield{author}{\bibinfo{person}{Wynne~W Chin} {et~al\mbox{.}}} \bibinfo{year}{1998}\natexlab{}.
\newblock \showarticletitle{The partial least squares approach to structural equation modeling}.
\newblock \bibinfo{journal}{\emph{Modern Methods for Business Research}} \bibinfo{volume}{295}, \bibinfo{number}{2} (\bibinfo{year}{1998}), \bibinfo{pages}{295--336}.
\newblock


\bibitem[Choudhuri et~al\mbox{.}(2025a)]%
        {choudhuri2025ai}
\bibfield{author}{\bibinfo{person}{Rudrajit Choudhuri}, \bibinfo{person}{Carmen Badea}, \bibinfo{person}{Christian Bird}, \bibinfo{person}{Jenna Butler}, \bibinfo{person}{Rob DeLine}, {and} \bibinfo{person}{Brian Houck}.} \bibinfo{year}{2025}\natexlab{a}.
\newblock \showarticletitle{AI Where It Matters: Where, Why, and How Developers Want AI Support in Daily Work}.
\newblock \bibinfo{journal}{\emph{arXiv preprint arXiv:2510.00762}} (\bibinfo{year}{2025}).
\newblock


\bibitem[Choudhuri et~al\mbox{.}(2024)]%
        {choudhuri2024far}
\bibfield{author}{\bibinfo{person}{Rudrajit Choudhuri}, \bibinfo{person}{Dylan Liu}, \bibinfo{person}{Igor Steinmacher}, \bibinfo{person}{Marco Gerosa}, {and} \bibinfo{person}{Anita Sarma}.} \bibinfo{year}{2024}\natexlab{}.
\newblock \showarticletitle{{How Far Are We? The Triumphs and Trials of Generative AI in Learning Software Engineering}}. In \bibinfo{booktitle}{\emph{Proceedings of the IEEE/ACM 46th international conference on software engineering}}. \bibinfo{pages}{1--13}.
\newblock


\bibitem[Choudhuri et~al\mbox{.}(2025b)]%
        {choudhuri2025insights}
\bibfield{author}{\bibinfo{person}{Rudrajit Choudhuri}, \bibinfo{person}{Ambareesh Ramakrishnan}, \bibinfo{person}{Amreeta Chatterjee}, \bibinfo{person}{Bianca Trinkenreich}, \bibinfo{person}{Igor Steinmacher}, \bibinfo{person}{Marco Gerosa}, {and} \bibinfo{person}{Anita Sarma}.} \bibinfo{year}{2025}\natexlab{b}.
\newblock \showarticletitle{{Insights from the Frontline: GenAI Utilization Among Software Engineering Students}}. In \bibinfo{booktitle}{\emph{2025 IEEE/ACM 37th International Conference on Software Engineering Education and Training (CSEE\&T)}}. IEEE, \bibinfo{pages}{1--12}.
\newblock


\bibitem[Choudhuri et~al\mbox{.}(2025c)]%
        {choudhuri2025guides}
\bibfield{author}{\bibinfo{person}{Rudrajit Choudhuri}, \bibinfo{person}{Bianca Trinkenreich}, \bibinfo{person}{Rahul Pandita}, \bibinfo{person}{Eirini Kalliamvakou}, \bibinfo{person}{Igor Steinmacher}, \bibinfo{person}{Marco Gerosa}, \bibinfo{person}{Christopher Sanchez}, {and} \bibinfo{person}{Anita Sarma}.} \bibinfo{year}{2025}\natexlab{c}.
\newblock \showarticletitle{{What Guides Our Choices? Modeling Developers' Trust and Behavioral Intentions Towards GenAI}}. In \bibinfo{booktitle}{\emph{2025 IEEE/ACM 47th International Conference on Software Engineering (ICSE)}}.
\newblock
\href{https://doi.org/10.1109/ICSE55347.2025.00087}{doi:\nolinkurl{10.1109/ICSE55347.2025.00087}}


\bibitem[Choudhuri et~al\mbox{.}(2025d)]%
        {choudhuri2025needs}
\bibfield{author}{\bibinfo{person}{Rudrajit Choudhuri}, \bibinfo{person}{Bianca Trinkenreich}, \bibinfo{person}{Rahul Pandita}, \bibinfo{person}{Eirini Kalliamvakou}, \bibinfo{person}{Igor Steinmacher}, \bibinfo{person}{Marco Gerosa}, \bibinfo{person}{Christopher Sanchez}, {and} \bibinfo{person}{Anita Sarma}.} \bibinfo{year}{2025}\natexlab{d}.
\newblock \showarticletitle{What Needs Attention? Prioritizing Drivers of Developers' Trust and Adoption of Generative AI}.
\newblock \bibinfo{journal}{\emph{arXiv preprint arXiv:2505.17418}} (\bibinfo{year}{2025}).
\newblock


\bibitem[Clark and Mayer(2023)]%
        {clark2023learning}
\bibfield{author}{\bibinfo{person}{Ruth~C Clark} {and} \bibinfo{person}{Richard~E Mayer}.} \bibinfo{year}{2023}\natexlab{}.
\newblock \bibinfo{booktitle}{\emph{E-learning and the science of instruction: Proven guidelines for consumers and designers of multimedia learning}}.
\newblock \bibinfo{publisher}{john Wiley \& sons}.
\newblock


\bibitem[Cohen(2013)]%
        {cohen2013statistical}
\bibfield{author}{\bibinfo{person}{Jacob Cohen}.} \bibinfo{year}{2013}\natexlab{}.
\newblock \bibinfo{booktitle}{\emph{Statistical Power Analysis for the Behavioral Sciences}}.
\newblock \bibinfo{publisher}{Routledge}.
\newblock


\bibitem[Compeau and Higgins(1995)]%
        {compeau1995computer}
\bibfield{author}{\bibinfo{person}{Deborah~R Compeau} {and} \bibinfo{person}{Christopher~A Higgins}.} \bibinfo{year}{1995}\natexlab{}.
\newblock \showarticletitle{Computer self-efficacy: Development of a measure and initial test}.
\newblock \bibinfo{journal}{\emph{MIS Quarterly}} (\bibinfo{year}{1995}), \bibinfo{pages}{189--211}.
\newblock


\bibitem[Corbett(2024)]%
        {corbett2024psychological}
\bibfield{author}{\bibinfo{person}{Holly~C. Corbett}.} \bibinfo{year}{2024}\natexlab{}.
\newblock \bibinfo{booktitle}{\emph{Psychological Safety: Try This Tip to Know When to Take a Risk at Work}}.
\newblock Forbes.
\newblock
\urldef\tempurl%
\url{https://www.forbes.com/sites/hollycorbett/2024/09/27/psychological-safety-try-this-tip-to-know-when-to-take-a-risk-at-work/}
\showURL{%
\tempurl}
\newblock
\shownote{Accessed: 2026-01-07}.


\bibitem[Costantini et~al\mbox{.}(2025)]%
        {costantini2025playful}
\bibfield{author}{\bibinfo{person}{Arianna Costantini}, \bibinfo{person}{Arnold~B Bakker}, {and} \bibinfo{person}{Yuri~S Scharp}.} \bibinfo{year}{2025}\natexlab{}.
\newblock \showarticletitle{Playful Study Design: A Novel Approach to Enhancing Student Well-Being and Academic Performance}.
\newblock \bibinfo{journal}{\emph{Educational Psychology Review}} \bibinfo{volume}{37}, \bibinfo{number}{2} (\bibinfo{year}{2025}), \bibinfo{pages}{1--42}.
\newblock


\bibitem[Cox et~al\mbox{.}(2016)]%
        {cox2016design}
\bibfield{author}{\bibinfo{person}{Anna~L Cox}, \bibinfo{person}{Sandy~JJ Gould}, \bibinfo{person}{Marta~E Cecchinato}, \bibinfo{person}{Ioanna Iacovides}, {and} \bibinfo{person}{Ian Renfree}.} \bibinfo{year}{2016}\natexlab{}.
\newblock \showarticletitle{Design frictions for mindful interactions: The case for microboundaries}. In \bibinfo{booktitle}{\emph{Proceedings of the 2016 CHI conference extended abstracts on human factors in computing systems}}. \bibinfo{pages}{1389--1397}.
\newblock


\bibitem[Dewey(1933)]%
        {dewey1933we}
\bibfield{author}{\bibinfo{person}{John Dewey}.} \bibinfo{year}{1933}\natexlab{}.
\newblock \showarticletitle{How we think: A restatement of the relation of reflective thinking to the educative process}.
\newblock  (\bibinfo{year}{1933}).
\newblock


\bibitem[{Digital Education Council}(2025)]%
        {dec2025ailiteracy}
\bibfield{author}{\bibinfo{person}{{Digital Education Council}}.} \bibinfo{year}{2025}\natexlab{}.
\newblock \bibinfo{title}{Digital Education Council AI Literacy Framework}.
\newblock \bibinfo{howpublished}{\url{https://www.digitaleducationcouncil.com/post/digital-education-council-ai-literacy-framework}}.
\newblock
\newblock
\shownote{Accessed December, 2025}.


\bibitem[Drosos et~al\mbox{.}(2025)]%
        {drosos2025makes}
\bibfield{author}{\bibinfo{person}{Ian Drosos}, \bibinfo{person}{Advait Sarkar}, \bibinfo{person}{Neil Toronto}, {et~al\mbox{.}}} \bibinfo{year}{2025}\natexlab{}.
\newblock \showarticletitle{" It makes you think": Provocations Help Restore Critical Thinking to AI-Assisted Knowledge Work}.
\newblock \bibinfo{journal}{\emph{arXiv preprint arXiv:2501.17247}} (\bibinfo{year}{2025}).
\newblock


\bibitem[Dunlosky et~al\mbox{.}(2005)]%
        {dunlosky2005self}
\bibfield{author}{\bibinfo{person}{John Dunlosky}, \bibinfo{person}{Christopher Hertzog}, \bibinfo{person}{M Kennedy}, {and} \bibinfo{person}{Keith~W Thiede}.} \bibinfo{year}{2005}\natexlab{}.
\newblock \showarticletitle{The self-monitoring approach for effective learning}.
\newblock \bibinfo{journal}{\emph{Cognitive Technology}} \bibinfo{volume}{10}, \bibinfo{number}{1} (\bibinfo{year}{2005}), \bibinfo{pages}{4--11}.
\newblock


\bibitem[Engelbart(2023)]%
        {engelbart2023augmenting}
\bibfield{author}{\bibinfo{person}{Douglas~C Engelbart}.} \bibinfo{year}{2023}\natexlab{}.
\newblock \showarticletitle{Augmenting human intellect: A conceptual framework}.
\newblock In \bibinfo{booktitle}{\emph{Augmented education in the global age}}. \bibinfo{publisher}{Routledge}, \bibinfo{pages}{13--29}.
\newblock


\bibitem[Ennis(1993)]%
        {ennis1993critical}
\bibfield{author}{\bibinfo{person}{Robert~H Ennis}.} \bibinfo{year}{1993}\natexlab{}.
\newblock \showarticletitle{Critical thinking assessment}.
\newblock \bibinfo{journal}{\emph{Theory into practice}} \bibinfo{volume}{32}, \bibinfo{number}{3} (\bibinfo{year}{1993}), \bibinfo{pages}{179--186}.
\newblock


\bibitem[Ennis(2018)]%
        {ennis2018critical}
\bibfield{author}{\bibinfo{person}{Robert~H Ennis}.} \bibinfo{year}{2018}\natexlab{}.
\newblock \showarticletitle{Critical thinking across the curriculum: A vision}.
\newblock \bibinfo{journal}{\emph{Topoi}} \bibinfo{volume}{37}, \bibinfo{number}{1} (\bibinfo{year}{2018}), \bibinfo{pages}{165--184}.
\newblock


\bibitem[Evans et~al\mbox{.}(2003)]%
        {evans2003approaches}
\bibfield{author}{\bibinfo{person}{Christina~J Evans}, \bibinfo{person}{John~R Kirby}, {and} \bibinfo{person}{Leandre~R Fabrigar}.} \bibinfo{year}{2003}\natexlab{}.
\newblock \showarticletitle{Approaches to learning, need for cognition, and strategic flexibility among university students}.
\newblock \bibinfo{journal}{\emph{British Journal of Educational Psychology}} \bibinfo{volume}{73}, \bibinfo{number}{4} (\bibinfo{year}{2003}), \bibinfo{pages}{507--528}.
\newblock


\bibitem[Facione(1990)]%
        {facione1990critical}
\bibfield{author}{\bibinfo{person}{Peter Facione}.} \bibinfo{year}{1990}\natexlab{}.
\newblock \showarticletitle{Critical thinking: A statement of expert consensus for purposes of educational assessment and instruction (The Delphi Report)}.
\newblock  (\bibinfo{year}{1990}).
\newblock


\bibitem[Facione et~al\mbox{.}(2011)]%
        {facione2011critical}
\bibfield{author}{\bibinfo{person}{Peter~A Facione} {et~al\mbox{.}}} \bibinfo{year}{2011}\natexlab{}.
\newblock \showarticletitle{Critical thinking: What it is and why it counts}.
\newblock \bibinfo{journal}{\emph{Insight assessment}} \bibinfo{volume}{1}, \bibinfo{number}{1} (\bibinfo{year}{2011}), \bibinfo{pages}{1--23}.
\newblock


\bibitem[Facione et~al\mbox{.}(1995)]%
        {facione1995disposition}
\bibfield{author}{\bibinfo{person}{Peter~A Facione}, \bibinfo{person}{Carol~A Sanchez}, \bibinfo{person}{Noreen~C Facione}, {and} \bibinfo{person}{Joanne Gainen}.} \bibinfo{year}{1995}\natexlab{}.
\newblock \showarticletitle{The disposition toward critical thinking}.
\newblock \bibinfo{journal}{\emph{The Journal of general education}} \bibinfo{volume}{44}, \bibinfo{number}{1} (\bibinfo{year}{1995}), \bibinfo{pages}{1--25}.
\newblock


\bibitem[Fan et~al\mbox{.}(2025a)]%
        {fan2025educational}
\bibfield{author}{\bibinfo{person}{Lei Fan}, \bibinfo{person}{Kunyang Deng}, {and} \bibinfo{person}{Fangxue Liu}.} \bibinfo{year}{2025}\natexlab{a}.
\newblock \showarticletitle{Educational impacts of generative artificial intelligence on learning and performance of engineering students in China}.
\newblock \bibinfo{journal}{\emph{Scientific reports}} \bibinfo{volume}{15}, \bibinfo{number}{1} (\bibinfo{year}{2025}), \bibinfo{pages}{26521}.
\newblock


\bibitem[Fan et~al\mbox{.}(2025b)]%
        {fan2025beware}
\bibfield{author}{\bibinfo{person}{Yizhou Fan}, \bibinfo{person}{Luzhen Tang}, \bibinfo{person}{Huixiao Le}, \bibinfo{person}{Kejie Shen}, \bibinfo{person}{Shufang Tan}, \bibinfo{person}{Yueying Zhao}, \bibinfo{person}{Yuan Shen}, \bibinfo{person}{Xinyu Li}, {and} \bibinfo{person}{Dragan Ga{\v{s}}evi{\'c}}.} \bibinfo{year}{2025}\natexlab{b}.
\newblock \showarticletitle{Beware of metacognitive laziness: Effects of generative artificial intelligence on learning motivation, processes, and performance}.
\newblock \bibinfo{journal}{\emph{British Journal of Educational Technology}} \bibinfo{volume}{56}, \bibinfo{number}{2} (\bibinfo{year}{2025}), \bibinfo{pages}{489--530}.
\newblock


\bibitem[Faul et~al\mbox{.}(2009)]%
        {faul2009statistical}
\bibfield{author}{\bibinfo{person}{Franz Faul}, \bibinfo{person}{Edgar Erdfelder}, \bibinfo{person}{Axel Buchner}, {and} \bibinfo{person}{Albert-Georg Lang}.} \bibinfo{year}{2009}\natexlab{}.
\newblock \showarticletitle{Statistical power analyses using G* Power 3.1: Tests for correlation and regression analyses}.
\newblock \bibinfo{journal}{\emph{Behav Res Methods}} \bibinfo{volume}{41}, \bibinfo{number}{4} (\bibinfo{year}{2009}), \bibinfo{pages}{1149--1160}.
\newblock


\bibitem[Fredricks et~al\mbox{.}(2004)]%
        {fredricks2004school}
\bibfield{author}{\bibinfo{person}{Jennifer~A Fredricks}, \bibinfo{person}{Phyllis~C Blumenfeld}, {and} \bibinfo{person}{Alison~H Paris}.} \bibinfo{year}{2004}\natexlab{}.
\newblock \showarticletitle{School engagement: Potential of the concept, state of the evidence}.
\newblock \bibinfo{journal}{\emph{Review of educational research}} \bibinfo{volume}{74}, \bibinfo{number}{1} (\bibinfo{year}{2004}), \bibinfo{pages}{59--109}.
\newblock


\bibitem[Gerlich(2025)]%
        {gerlich2025ai}
\bibfield{author}{\bibinfo{person}{Michael Gerlich}.} \bibinfo{year}{2025}\natexlab{}.
\newblock \showarticletitle{AI tools in society: Impacts on cognitive offloading and the future of critical thinking}.
\newblock \bibinfo{journal}{\emph{Societies}} \bibinfo{volume}{15}, \bibinfo{number}{1} (\bibinfo{year}{2025}), \bibinfo{pages}{6}.
\newblock


\bibitem[Giaccone and Magnusson(2022)]%
        {giaccone2022unveiling}
\bibfield{author}{\bibinfo{person}{Sonia~C Giaccone} {and} \bibinfo{person}{Mats Magnusson}.} \bibinfo{year}{2022}\natexlab{}.
\newblock \showarticletitle{Unveiling the role of risk-taking in innovation: antecedents and effects}.
\newblock \bibinfo{journal}{\emph{R\&D Management}} \bibinfo{volume}{52}, \bibinfo{number}{1} (\bibinfo{year}{2022}), \bibinfo{pages}{93--107}.
\newblock


\bibitem[Goddard et~al\mbox{.}(2012)]%
        {goddard2012automation}
\bibfield{author}{\bibinfo{person}{Kate Goddard}, \bibinfo{person}{Abdul Roudsari}, {and} \bibinfo{person}{Jeremy~C Wyatt}.} \bibinfo{year}{2012}\natexlab{}.
\newblock \showarticletitle{Automation bias: a systematic review of frequency, effect mediators, and mitigators}.
\newblock \bibinfo{journal}{\emph{Journal of the American Medical Informatics Association}} \bibinfo{volume}{19}, \bibinfo{number}{1} (\bibinfo{year}{2012}), \bibinfo{pages}{121--127}.
\newblock


\bibitem[Gong et~al\mbox{.}(2025)]%
        {gong2025impact}
\bibfield{author}{\bibinfo{person}{Yuhong Gong}, \bibinfo{person}{Shang Zhang}, \bibinfo{person}{Ting Zhang}, {and} \bibinfo{person}{Xinfa Yi}.} \bibinfo{year}{2025}\natexlab{}.
\newblock \showarticletitle{The impact of feedback literacy on reflective learning types in Chinese high school students: based on latent profile analysis}.
\newblock \bibinfo{journal}{\emph{Frontiers in Psychology}}  \bibinfo{volume}{16} (\bibinfo{year}{2025}), \bibinfo{pages}{1516253}.
\newblock


\bibitem[Grichting(1994)]%
        {grichting1994meaning}
\bibfield{author}{\bibinfo{person}{Wolfgang~L Grichting}.} \bibinfo{year}{1994}\natexlab{}.
\newblock \showarticletitle{The meaning of “I Don't Know” in opinion surveys: Indifference versus ignorance}.
\newblock \bibinfo{journal}{\emph{Aust Psychol}} \bibinfo{volume}{29}, \bibinfo{number}{1} (\bibinfo{year}{1994}).
\newblock


\bibitem[Gu and Ericson(2025)]%
        {gu2025ai}
\bibfield{author}{\bibinfo{person}{Xingjian Gu} {and} \bibinfo{person}{Barbara~J Ericson}.} \bibinfo{year}{2025}\natexlab{}.
\newblock \showarticletitle{AI literacy in K-12 and higher education in the wake of generative AI: An integrative review}. In \bibinfo{booktitle}{\emph{Proceedings of the 2025 ACM Conference on International Computing Education Research V. 1}}. \bibinfo{pages}{125--140}.
\newblock


\bibitem[Hair(2014)]%
        {hair2014primer}
\bibfield{author}{\bibinfo{person}{Joseph~F Hair}.} \bibinfo{year}{2014}\natexlab{}.
\newblock \bibinfo{booktitle}{\emph{A primer on partial least squares structural equation modeling (PLS-SEM)}}.
\newblock \bibinfo{publisher}{sage}.
\newblock


\bibitem[Hair et~al\mbox{.}(2019)]%
        {hair2019use}
\bibfield{author}{\bibinfo{person}{Joseph~F Hair}, \bibinfo{person}{Jeffrey~J Risher}, \bibinfo{person}{Marko Sarstedt}, {and} \bibinfo{person}{Christian~M Ringle}.} \bibinfo{year}{2019}\natexlab{}.
\newblock \showarticletitle{When to use and how to report the results of {PLS-SEM}}.
\newblock \bibinfo{journal}{\emph{Eur. Bus. Rev.}} (\bibinfo{year}{2019}).
\newblock


\bibitem[Hamari et~al\mbox{.}(2014)]%
        {hamari2014does}
\bibfield{author}{\bibinfo{person}{Juho Hamari}, \bibinfo{person}{Jonna Koivisto}, {and} \bibinfo{person}{Harri Sarsa}.} \bibinfo{year}{2014}\natexlab{}.
\newblock \showarticletitle{Does gamification work? A literature review of empirical studies on gamification}. In \bibinfo{booktitle}{\emph{2014 47th Hawaii international conference on system sciences}}. Ieee, \bibinfo{pages}{3025--3034}.
\newblock


\bibitem[Hamid et~al\mbox{.}(2024)]%
        {hamid2024-GMsurveyValidated}
\bibfield{author}{\bibinfo{person}{Md~Montaser Hamid}, \bibinfo{person}{Amreeta Chatterjee}, \bibinfo{person}{Mariam Guizani}, \bibinfo{person}{Andrew Anderson}, \bibinfo{person}{Fatima Moussaoui}, \bibinfo{person}{Sarah Yang}, \bibinfo{person}{Isaac Escobar}, \bibinfo{person}{Anita Sarma}, {and} \bibinfo{person}{Margaret Burnett}.} \bibinfo{year}{2024}\natexlab{}.
\newblock \showarticletitle{How to measure diversity actionably in technology}.
\newblock In \bibinfo{booktitle}{\emph{Equity, Diversity, and Inclusion in Software Engineering: Best Practices and Insights}}. \bibinfo{publisher}{Apress Berkeley, CA}, \bibinfo{pages}{469--485}.
\newblock


\bibitem[Hamid et~al\mbox{.}(2025)]%
        {hamid2025inclusive}
\bibfield{author}{\bibinfo{person}{Md~Montaser Hamid}, \bibinfo{person}{FatimA~A Moussaoui}, \bibinfo{person}{Jimena~Noa Guevara}, \bibinfo{person}{Andrew Anderson}, \bibinfo{person}{Puja Agarwal}, \bibinfo{person}{Jonathan Dodge}, {and} \bibinfo{person}{Margaret Burnett}.} \bibinfo{year}{2025}\natexlab{}.
\newblock \showarticletitle{Inclusive design of AI’s Explanations: Just for Those Previously Left Out?}
\newblock \bibinfo{journal}{\emph{ACM Transactions on Interactive Intelligent Systems}} (\bibinfo{year}{2025}).
\newblock


\bibitem[Henseler et~al\mbox{.}(2016)]%
        {henseler2016using}
\bibfield{author}{\bibinfo{person}{J{\"o}rg Henseler}, \bibinfo{person}{Geoffrey Hubona}, {and} \bibinfo{person}{Pauline~Ash Ray}.} \bibinfo{year}{2016}\natexlab{}.
\newblock \showarticletitle{Using PLS path modeling in new technology research: updated guidelines}.
\newblock \bibinfo{journal}{\emph{Industrial Management \& Data Systems}} \bibinfo{volume}{116}, \bibinfo{number}{1} (\bibinfo{year}{2016}), \bibinfo{pages}{2--20}.
\newblock


\bibitem[Henseler et~al\mbox{.}(2015)]%
        {henseler2015new}
\bibfield{author}{\bibinfo{person}{J{\"o}rg Henseler}, \bibinfo{person}{Christian~M Ringle}, {and} \bibinfo{person}{Marko Sarstedt}.} \bibinfo{year}{2015}\natexlab{}.
\newblock \showarticletitle{A new criterion for assessing discriminant validity in variance-based structural equation modeling}.
\newblock \bibinfo{journal}{\emph{J. Acad. Mark. Sci.}} \bibinfo{volume}{43}, \bibinfo{number}{1} (\bibinfo{year}{2015}), \bibinfo{pages}{115--135}.
\newblock


\bibitem[H{\"o}{\"o}k(2000)]%
        {hook2000beforeIUIsbecomeReal}
\bibfield{author}{\bibinfo{person}{Kristina H{\"o}{\"o}k}.} \bibinfo{year}{2000}\natexlab{}.
\newblock \showarticletitle{Steps to take before intelligent user interfaces become real}.
\newblock \bibinfo{journal}{\emph{Interacting with computers}} \bibinfo{volume}{12}, \bibinfo{number}{4} (\bibinfo{year}{2000}), \bibinfo{pages}{409--426}.
\newblock


\bibitem[Howard(2016)]%
        {howard2016review}
\bibfield{author}{\bibinfo{person}{Matt~C Howard}.} \bibinfo{year}{2016}\natexlab{}.
\newblock \showarticletitle{A review of exploratory factor analysis decisions and overview of current practices: What we are doing and how can we improve?}
\newblock \bibinfo{journal}{\emph{International Journal of Human-Computer Interaction}} \bibinfo{volume}{32}, \bibinfo{number}{1} (\bibinfo{year}{2016}), \bibinfo{pages}{51--62}.
\newblock


\bibitem[Kahneman(2011)]%
        {kahneman2011thinking}
\bibfield{author}{\bibinfo{person}{Daniel Kahneman}.} \bibinfo{year}{2011}\natexlab{}.
\newblock \showarticletitle{Thinking, fast and slow}.
\newblock \bibinfo{journal}{\emph{Farrar, Straus and Giroux}} (\bibinfo{year}{2011}).
\newblock


\bibitem[Kember et~al\mbox{.}(2000)]%
        {kember2000development}
\bibfield{author}{\bibinfo{person}{David Kember}, \bibinfo{person}{Doris~YP Leung}, \bibinfo{person}{Alice Jones}, \bibinfo{person}{Alice~Yuen Loke}, \bibinfo{person}{Jan McKay}, \bibinfo{person}{Kit Sinclair}, \bibinfo{person}{Harrison Tse}, \bibinfo{person}{Celia Webb}, \bibinfo{person}{Frances~Kam Yuet~Wong}, \bibinfo{person}{Marian Wong}, {et~al\mbox{.}}} \bibinfo{year}{2000}\natexlab{}.
\newblock \showarticletitle{Development of a questionnaire to measure the level of reflective thinking}.
\newblock \bibinfo{journal}{\emph{Assessment \& Evaluation in Higher Education}} \bibinfo{volume}{25}, \bibinfo{number}{4} (\bibinfo{year}{2000}), \bibinfo{pages}{381--395}.
\newblock


\bibitem[Kitchenham and Pfleeger(2008)]%
        {kitchenham2008personal}
\bibfield{author}{\bibinfo{person}{Barbara~A Kitchenham} {and} \bibinfo{person}{Shari~L Pfleeger}.} \bibinfo{year}{2008}\natexlab{}.
\newblock \showarticletitle{Personal opinion surveys}.
\newblock In \bibinfo{booktitle}{\emph{Guide to advanced empirical software engineering}}. \bibinfo{publisher}{Springer}, \bibinfo{pages}{63--92}.
\newblock


\bibitem[Ko et~al\mbox{.}(2015)]%
        {ko2015practical}
\bibfield{author}{\bibinfo{person}{Amy~J Ko}, \bibinfo{person}{Thomas~D LaToza}, {and} \bibinfo{person}{Margaret~M Burnett}.} \bibinfo{year}{2015}\natexlab{}.
\newblock \showarticletitle{A practical guide to controlled experiments of software engineering tools with human participants}.
\newblock \bibinfo{journal}{\emph{Empirical Software Engineering}} \bibinfo{volume}{20}, \bibinfo{number}{1} (\bibinfo{year}{2015}), \bibinfo{pages}{110--141}.
\newblock


\bibitem[Kobylarek et~al\mbox{.}(2022)]%
        {kobylarek2022critical}
\bibfield{author}{\bibinfo{person}{Aleksander Kobylarek}, \bibinfo{person}{Kamil B{\l}aszczy{\'n}ski}, \bibinfo{person}{Luba {\'S}l{\'o}sarz}, {and} \bibinfo{person}{Martyna Madej}.} \bibinfo{year}{2022}\natexlab{}.
\newblock \showarticletitle{Critical Thinking Questionnaire (CThQ)--construction and application of critical thinking test tool}.
\newblock \bibinfo{journal}{\emph{Andragogy Adult Education and Social Marketing}} \bibinfo{volume}{2}, \bibinfo{number}{2} (\bibinfo{year}{2022}), \bibinfo{pages}{1--1}.
\newblock


\bibitem[Kock(2014)]%
        {kock2014advanced}
\bibfield{author}{\bibinfo{person}{Ned Kock}.} \bibinfo{year}{2014}\natexlab{}.
\newblock \showarticletitle{Advanced mediating effects tests, multi-group analyses, and measurement model assessments in {PLS}-based {SEM}}.
\newblock \bibinfo{journal}{\emph{International Journal of e-Collaboration}} \bibinfo{volume}{10}, \bibinfo{number}{1} (\bibinfo{year}{2014}).
\newblock


\bibitem[Kock(2015)]%
        {kock2015common}
\bibfield{author}{\bibinfo{person}{Ned Kock}.} \bibinfo{year}{2015}\natexlab{}.
\newblock \showarticletitle{Common method bias in PLS-SEM: A full collinearity assessment approach}.
\newblock \bibinfo{journal}{\emph{International Journal of e-Collaboration (ijec)}} \bibinfo{volume}{11}, \bibinfo{number}{4} (\bibinfo{year}{2015}), \bibinfo{pages}{1--10}.
\newblock


\bibitem[Kosmyna et~al\mbox{.}(2025)]%
        {kosmyna2025your}
\bibfield{author}{\bibinfo{person}{Nataliya Kosmyna}, \bibinfo{person}{Eugene Hauptmann}, \bibinfo{person}{Ye~Tong Yuan}, \bibinfo{person}{Jessica Situ}, \bibinfo{person}{Xian-Hao Liao}, \bibinfo{person}{Ashly~Vivian Beresnitzky}, \bibinfo{person}{Iris Braunstein}, {and} \bibinfo{person}{Pattie Maes}.} \bibinfo{year}{2025}\natexlab{}.
\newblock \showarticletitle{Your brain on ChatGPT: Accumulation of cognitive debt when using an AI assistant for essay writing task}.
\newblock \bibinfo{journal}{\emph{arXiv preprint arXiv:2506.08872}} (\bibinfo{year}{2025}).
\newblock


\bibitem[Kreijkes et~al\mbox{.}(2025)]%
        {kreijkes2025effects}
\bibfield{author}{\bibinfo{person}{Pia Kreijkes}, \bibinfo{person}{Viktor Kewenig}, \bibinfo{person}{Martina Kuvalja}, \bibinfo{person}{Mina Lee}, \bibinfo{person}{Sylvia Vitello}, \bibinfo{person}{Jake~M Hofman}, \bibinfo{person}{Abigail Sellen}, \bibinfo{person}{Sean Rintel}, \bibinfo{person}{Daniel~G Goldstein}, \bibinfo{person}{David~M Rothschild}, {et~al\mbox{.}}} \bibinfo{year}{2025}\natexlab{}.
\newblock \showarticletitle{Effects of LLM use and note-taking on reading comprehension and memory: A randomised experiment in secondary schools}.
\newblock \bibinfo{journal}{\emph{Available at SSRN}} (\bibinfo{year}{2025}).
\newblock


\bibitem[Lamborn et~al\mbox{.}(1992)]%
        {lamborn1992significance}
\bibfield{author}{\bibinfo{person}{Susie Lamborn}, \bibinfo{person}{Fred Newmann}, {and} \bibinfo{person}{Gary Wehlage}.} \bibinfo{year}{1992}\natexlab{}.
\newblock \showarticletitle{The significance and sources of student engagement}.
\newblock \bibinfo{journal}{\emph{Student engagement and achievement in American secondary schools}} (\bibinfo{year}{1992}), \bibinfo{pages}{11--39}.
\newblock


\bibitem[Lazarus(1991)]%
        {lazarus1991emotion}
\bibfield{author}{\bibinfo{person}{Richard~S Lazarus}.} \bibinfo{year}{1991}\natexlab{}.
\newblock \bibinfo{booktitle}{\emph{Emotion and adaptation}}.
\newblock \bibinfo{publisher}{Oxford University Press}.
\newblock


\bibitem[Lee et~al\mbox{.}(2025)]%
        {lee2025impact}
\bibfield{author}{\bibinfo{person}{Hao-Ping Lee}, \bibinfo{person}{Advait Sarkar}, \bibinfo{person}{Lev Tankelevitch}, \bibinfo{person}{Ian Drosos}, \bibinfo{person}{Sean Rintel}, \bibinfo{person}{Richard Banks}, {and} \bibinfo{person}{Nicholas Wilson}.} \bibinfo{year}{2025}\natexlab{}.
\newblock \showarticletitle{The impact of generative AI on critical thinking: Self-reported reductions in cognitive effort and confidence effects from a survey of knowledge workers}. In \bibinfo{booktitle}{\emph{Proceedings of the 2025 CHI Conference on Human Factors in Computing Systems}}. \bibinfo{pages}{1--22}.
\newblock


\bibitem[Lee and See(2004)]%
        {lee2004trust}
\bibfield{author}{\bibinfo{person}{John~D Lee} {and} \bibinfo{person}{Katrina~A See}.} \bibinfo{year}{2004}\natexlab{}.
\newblock \showarticletitle{Trust in automation: Designing for appropriate reliance}.
\newblock \bibinfo{journal}{\emph{Human factors}} \bibinfo{volume}{46}, \bibinfo{number}{1} (\bibinfo{year}{2004}), \bibinfo{pages}{50--80}.
\newblock


\bibitem[Lepp and Kaimre(2025)]%
        {lepp2025does}
\bibfield{author}{\bibinfo{person}{Marina Lepp} {and} \bibinfo{person}{Joosep Kaimre}.} \bibinfo{year}{2025}\natexlab{}.
\newblock \showarticletitle{Does generative AI help in learning programming: Students’ perceptions, reported use and relation to performance}.
\newblock \bibinfo{journal}{\emph{Computers in Human Behavior Reports}}  \bibinfo{volume}{18} (\bibinfo{year}{2025}), \bibinfo{pages}{100642}.
\newblock


\bibitem[Lerner and Tetlock(1999)]%
        {lerner1999accounting}
\bibfield{author}{\bibinfo{person}{Jennifer~S Lerner} {and} \bibinfo{person}{Philip~E Tetlock}.} \bibinfo{year}{1999}\natexlab{}.
\newblock \showarticletitle{Accounting for the effects of accountability.}
\newblock \bibinfo{journal}{\emph{Psychological bulletin}} \bibinfo{volume}{125}, \bibinfo{number}{2} (\bibinfo{year}{1999}), \bibinfo{pages}{255}.
\newblock


\bibitem[Li et~al\mbox{.}(2023)]%
        {li2023study}
\bibfield{author}{\bibinfo{person}{Wei Li}, \bibinfo{person}{Ji-Yi Huang}, \bibinfo{person}{Cheng-Ye Liu}, \bibinfo{person}{Judy~CR Tseng}, {and} \bibinfo{person}{Shu-Pan Wang}.} \bibinfo{year}{2023}\natexlab{}.
\newblock \showarticletitle{A study on the relationship between student'learning engagements and higher-order thinking skills in programming learning}.
\newblock \bibinfo{journal}{\emph{Thinking Skills and Creativity}}  \bibinfo{volume}{49} (\bibinfo{year}{2023}), \bibinfo{pages}{101369}.
\newblock


\bibitem[Liao and Sundar(2022)]%
        {liao2022designing}
\bibfield{author}{\bibinfo{person}{Q~Vera Liao} {and} \bibinfo{person}{S~Shyam Sundar}.} \bibinfo{year}{2022}\natexlab{}.
\newblock \showarticletitle{Designing for responsible trust in {AI} systems: A communication perspective}.
\newblock \bibinfo{journal}{\emph{2022 ACM Conference on Fairness, Accountability, and Transparency}} (\bibinfo{year}{2022}), \bibinfo{pages}{1257--1268}.
\newblock


\bibitem[Lieberoth(2015)]%
        {lieberoth2015shallow}
\bibfield{author}{\bibinfo{person}{Andreas Lieberoth}.} \bibinfo{year}{2015}\natexlab{}.
\newblock \showarticletitle{Shallow gamification: Testing psychological effects of framing an activity as a game}.
\newblock \bibinfo{journal}{\emph{Games and Culture}} \bibinfo{volume}{10}, \bibinfo{number}{3} (\bibinfo{year}{2015}), \bibinfo{pages}{229--248}.
\newblock


\bibitem[Lips-Wiersma and Morris(2009)]%
        {lips2009discriminating}
\bibfield{author}{\bibinfo{person}{Marjolein Lips-Wiersma} {and} \bibinfo{person}{Lani Morris}.} \bibinfo{year}{2009}\natexlab{}.
\newblock \showarticletitle{Discriminating between ‘meaningful work’and the ‘management of meaning’}.
\newblock \bibinfo{journal}{\emph{Journal of business ethics}} \bibinfo{volume}{88}, \bibinfo{number}{3} (\bibinfo{year}{2009}), \bibinfo{pages}{491--511}.
\newblock


\bibitem[Liu et~al\mbox{.}(2025)]%
        {liu2025perspectra}
\bibfield{author}{\bibinfo{person}{Yiren Liu}, \bibinfo{person}{Viraj Shah}, \bibinfo{person}{Sangho Suh}, \bibinfo{person}{Pao Siangliulue}, \bibinfo{person}{Tal August}, {and} \bibinfo{person}{Yun Huang}.} \bibinfo{year}{2025}\natexlab{}.
\newblock \showarticletitle{Perspectra: Choosing Your Experts Enhances Critical Thinking in Multi-Agent Research Ideation}.
\newblock \bibinfo{journal}{\emph{arXiv preprint arXiv:2509.20553}} (\bibinfo{year}{2025}).
\newblock


\bibitem[Loewenstein et~al\mbox{.}(2001)]%
        {loewenstein2001risk}
\bibfield{author}{\bibinfo{person}{George~F Loewenstein}, \bibinfo{person}{Elke~U Weber}, \bibinfo{person}{Christopher~K Hsee}, {and} \bibinfo{person}{Ned Welch}.} \bibinfo{year}{2001}\natexlab{}.
\newblock \showarticletitle{Risk as feelings.}
\newblock \bibinfo{journal}{\emph{Psychological bulletin}} \bibinfo{volume}{127}, \bibinfo{number}{2} (\bibinfo{year}{2001}), \bibinfo{pages}{267}.
\newblock


\bibitem[Lyell and Coiera(2017)]%
        {lyell2017automation}
\bibfield{author}{\bibinfo{person}{David Lyell} {and} \bibinfo{person}{Enrico Coiera}.} \bibinfo{year}{2017}\natexlab{}.
\newblock \showarticletitle{Automation bias and verification complexity: a systematic review}.
\newblock \bibinfo{journal}{\emph{Journal of the American Medical Informatics Association}} \bibinfo{volume}{24}, \bibinfo{number}{2} (\bibinfo{year}{2017}), \bibinfo{pages}{423--431}.
\newblock


\bibitem[{Manufacturing.net (ASQ survey report)}(2013)]%
        {US_Teens_Fear_Risk_Taking_2013}
\bibfield{author}{\bibinfo{person}{{Manufacturing.net (ASQ survey report)}}.} \bibinfo{year}{2013}\natexlab{}.
\newblock \bibinfo{booktitle}{\emph{U.S. Teens Fear Risk-Taking; STEM Careers Demand It}}.
\newblock Manufacturing.net.
\newblock
\urldef\tempurl%
\url{https://www.manufacturing.net/industry40/article/13215403/us-teens-fear-risktaking-stem-careers-demand-it}
\showURL{%
\tempurl}
\newblock
\shownote{Accessed: 2026-01-08}.


\bibitem[Margulieux et~al\mbox{.}(2024)]%
        {margulieux2024self}
\bibfield{author}{\bibinfo{person}{Lauren~E Margulieux}, \bibinfo{person}{James Prather}, \bibinfo{person}{Brent~N Reeves}, \bibinfo{person}{Brett~A Becker}, \bibinfo{person}{Gozde Cetin~Uzun}, \bibinfo{person}{Dastyni Loksa}, \bibinfo{person}{Juho Leinonen}, {and} \bibinfo{person}{Paul Denny}.} \bibinfo{year}{2024}\natexlab{}.
\newblock \showarticletitle{Self-regulation, self-efficacy, and fear of failure interactions with how novices use llms to solve programming problems}.
\newblock In \bibinfo{booktitle}{\emph{Proceedings of the 2024 on Innovation and Technology in Computer Science Education V. 1}}. \bibinfo{pages}{276--282}.
\newblock


\bibitem[Mart{\'\i}nez-C{\'o}rcoles et~al\mbox{.}(2017)]%
        {martinez2017assessing}
\bibfield{author}{\bibinfo{person}{Mario Mart{\'\i}nez-C{\'o}rcoles}, \bibinfo{person}{Mare Teichmann}, {and} \bibinfo{person}{Mart Murdvee}.} \bibinfo{year}{2017}\natexlab{}.
\newblock \showarticletitle{Assessing technophobia and technophilia: Development and validation of a questionnaire}.
\newblock \bibinfo{journal}{\emph{Technology in Society}}  \bibinfo{volume}{51} (\bibinfo{year}{2017}), \bibinfo{pages}{183--188}.
\newblock


\bibitem[Miklian and Hoelscher(2025)]%
        {miklian2025new}
\bibfield{author}{\bibinfo{person}{Jason Miklian} {and} \bibinfo{person}{Kristian Hoelscher}.} \bibinfo{year}{2025}\natexlab{}.
\newblock \showarticletitle{A New Digital Divide? Coder Worldviews, the Slop Economy, and Democracy in the Age of AI}.
\newblock \bibinfo{journal}{\emph{arXiv preprint arXiv:2510.04755}} (\bibinfo{year}{2025}).
\newblock


\bibitem[Palanque et~al\mbox{.}(2023)]%
        {palanque2023multi}
\bibfield{author}{\bibinfo{person}{Philippe Palanque}, \bibinfo{person}{Fabio Patern{\`o}}, \bibinfo{person}{Virpi Roto}, \bibinfo{person}{Albrecht Schmidt}, \bibinfo{person}{Simone Stumpf}, {and} \bibinfo{person}{J{\"u}rgen Ziegler}.} \bibinfo{year}{2023}\natexlab{}.
\newblock \showarticletitle{A multi-perspective panel on user-centred transparency, explainability, and controllability in automations}. In \bibinfo{booktitle}{\emph{IFIP Conference on Human-Computer Interaction}}. Springer, \bibinfo{pages}{349--353}.
\newblock


\bibitem[Parasuraman and Riley(1997)]%
        {parasuraman1997humans}
\bibfield{author}{\bibinfo{person}{Raja Parasuraman} {and} \bibinfo{person}{Victor Riley}.} \bibinfo{year}{1997}\natexlab{}.
\newblock \showarticletitle{Humans and automation: Use, misuse, disuse, abuse}.
\newblock \bibinfo{journal}{\emph{Human Factors}} \bibinfo{volume}{39}, \bibinfo{number}{2} (\bibinfo{year}{1997}), \bibinfo{pages}{230--253}.
\newblock


\bibitem[{ParentsCanada Sponsored Content}(2022)]%
        {Getting_Students_Excited_about_STEM_2022}
\bibfield{author}{\bibinfo{person}{{ParentsCanada Sponsored Content}}.} \bibinfo{year}{2022}\natexlab{}.
\newblock \bibinfo{booktitle}{\emph{Getting Students Excited About STEM Requires Taking Risks}}.
\newblock ParentsCanada.
\newblock
\urldef\tempurl%
\url{https://parentscanada.com/sponsored/getting-students-excited-about-stem/}
\showURL{%
\tempurl}
\newblock
\shownote{Accessed: 2026-01-08}.


\bibitem[Paul et~al\mbox{.}(1997)]%
        {paul1997california}
\bibfield{author}{\bibinfo{person}{Richard~W Paul}, \bibinfo{person}{Linda Elder}, {and} \bibinfo{person}{Ted Bartell}.} \bibinfo{year}{1997}\natexlab{}.
\newblock \showarticletitle{California teacher preparation for instruction in critical thinking: Research findings and policy recommendations.}
\newblock  (\bibinfo{year}{1997}).
\newblock


\bibitem[Paul(2014)]%
        {paul2014assessment}
\bibfield{author}{\bibinfo{person}{Sheila~A Paul}.} \bibinfo{year}{2014}\natexlab{}.
\newblock \showarticletitle{Assessment of critical thinking: a Delphi study}.
\newblock \bibinfo{journal}{\emph{Nurse Education Today}} \bibinfo{volume}{34}, \bibinfo{number}{11} (\bibinfo{year}{2014}), \bibinfo{pages}{1357--1360}.
\newblock


\bibitem[Perrig et~al\mbox{.}(2023)]%
        {perrig2023trust}
\bibfield{author}{\bibinfo{person}{Sebastian~AC Perrig}, \bibinfo{person}{Nicolas Scharowski}, {and} \bibinfo{person}{Florian Br{\"u}hlmann}.} \bibinfo{year}{2023}\natexlab{}.
\newblock \showarticletitle{Trust issues with trust scales: examining the psychometric quality of trust measures in the context of {AI}}. In \bibinfo{booktitle}{\emph{Extended Abstracts of the 2023 CHI Conference on Human Factors in Computing Systems}}. \bibinfo{pages}{1--7}.
\newblock


\bibitem[Pintrich and De~Groot(1990)]%
        {pintrich1990motivational}
\bibfield{author}{\bibinfo{person}{Paul~R Pintrich} {and} \bibinfo{person}{Elisabeth~V De~Groot}.} \bibinfo{year}{1990}\natexlab{}.
\newblock \showarticletitle{Motivational and self-regulated learning components of classroom academic performance.}
\newblock \bibinfo{journal}{\emph{Journal of educational psychology}} \bibinfo{volume}{82}, \bibinfo{number}{1} (\bibinfo{year}{1990}), \bibinfo{pages}{33}.
\newblock


\bibitem[Podsakoff et~al\mbox{.}(2003)]%
        {podsakoff2003common}
\bibfield{author}{\bibinfo{person}{Philip~M Podsakoff}, \bibinfo{person}{Scott~B MacKenzie}, \bibinfo{person}{Jeong-Yeon Lee}, {and} \bibinfo{person}{Nathan~P Podsakoff}.} \bibinfo{year}{2003}\natexlab{}.
\newblock \showarticletitle{Common method biases in behavioral research: a critical review of the literature and recommended remedies.}
\newblock \bibinfo{journal}{\emph{Journal of Applied Psychology}} \bibinfo{volume}{88}, \bibinfo{number}{5} (\bibinfo{year}{2003}), \bibinfo{pages}{879}.
\newblock


\bibitem[Prather et~al\mbox{.}(2025)]%
        {prather2025beyond}
\bibfield{author}{\bibinfo{person}{James Prather}, \bibinfo{person}{Juho Leinonen}, \bibinfo{person}{Natalie Kiesler}, \bibinfo{person}{Jamie Gorson~Benario}, \bibinfo{person}{Sam Lau}, \bibinfo{person}{Stephen MacNeil}, \bibinfo{person}{Narges Norouzi}, \bibinfo{person}{Simone Opel}, \bibinfo{person}{Vee Pettit}, \bibinfo{person}{Leo Porter}, {et~al\mbox{.}}} \bibinfo{year}{2025}\natexlab{}.
\newblock \showarticletitle{Beyond the hype: A comprehensive review of current trends in generative AI research, teaching practices, and tools}.
\newblock \bibinfo{journal}{\emph{2024 Working Group Reports on Innovation and Technology in Computer Science Education}} (\bibinfo{year}{2025}), \bibinfo{pages}{300--338}.
\newblock


\bibitem[Prather et~al\mbox{.}(2023)]%
        {prather2023s}
\bibfield{author}{\bibinfo{person}{James Prather}, \bibinfo{person}{Brent~N Reeves}, \bibinfo{person}{Paul Denny}, \bibinfo{person}{Brett~A Becker}, \bibinfo{person}{Juho Leinonen}, \bibinfo{person}{Andrew Luxton-Reilly}, \bibinfo{person}{Garrett Powell}, \bibinfo{person}{James Finnie-Ansley}, {and} \bibinfo{person}{Eddie~Antonio Santos}.} \bibinfo{year}{2023}\natexlab{}.
\newblock \showarticletitle{“It’s weird that it knows what I want”: Usability and interactions with copilot for novice programmers}.
\newblock \bibinfo{journal}{\emph{ACM transactions on computer-human interaction}} \bibinfo{volume}{31}, \bibinfo{number}{1} (\bibinfo{year}{2023}), \bibinfo{pages}{1--31}.
\newblock


\bibitem[Prather et~al\mbox{.}(2024)]%
        {prather2024widening}
\bibfield{author}{\bibinfo{person}{James Prather}, \bibinfo{person}{Brent~N Reeves}, \bibinfo{person}{Juho Leinonen}, \bibinfo{person}{Stephen MacNeil}, \bibinfo{person}{Arisoa~S Randrianasolo}, \bibinfo{person}{Brett~A Becker}, \bibinfo{person}{Bailey Kimmel}, \bibinfo{person}{Jared Wright}, {and} \bibinfo{person}{Ben Briggs}.} \bibinfo{year}{2024}\natexlab{}.
\newblock \showarticletitle{The widening gap: The benefits and harms of generative AI for novice programmers}. In \bibinfo{booktitle}{\emph{Proceedings of the 2024 ACM Conference on International Computing Education Research-Volume 1}}. \bibinfo{pages}{469--486}.
\newblock


\bibitem[Qin et~al\mbox{.}(2025)]%
        {qin2025role}
\bibfield{author}{\bibinfo{person}{Qiaolin Qin}, \bibinfo{person}{Ronnie de~Souza Santos}, {and} \bibinfo{person}{Rodrigo Spinola}.} \bibinfo{year}{2025}\natexlab{}.
\newblock \showarticletitle{On the Role and Impact of GenAI Tools in Software Engineering Education}.
\newblock \bibinfo{journal}{\emph{arXiv preprint arXiv:2512.04256}} (\bibinfo{year}{2025}).
\newblock


\bibitem[Qu et~al\mbox{.}(2025)]%
        {qu2025generative}
\bibfield{author}{\bibinfo{person}{Xiaodong Qu}, \bibinfo{person}{Joshua Sherwood}, \bibinfo{person}{Peiyan Liu}, {and} \bibinfo{person}{Nawwaf Aleisa}.} \bibinfo{year}{2025}\natexlab{}.
\newblock \showarticletitle{Generative AI Tools in Higher Education: A Meta-Analysis of Cognitive Impact}. In \bibinfo{booktitle}{\emph{Proceedings of the Extended Abstracts of the CHI Conference on Human Factors in Computing Systems}}. \bibinfo{pages}{1--9}.
\newblock


\bibitem[{Qualtrics}(2025)]%
        {qualtrics2025}
\bibfield{author}{\bibinfo{person}{{Qualtrics}}.} \bibinfo{year}{2025}\natexlab{}.
\newblock \bibinfo{title}{Qualtrics Survey Platform}.
\newblock \bibinfo{howpublished}{\url{https://www.qualtrics.com}}.
\newblock
\newblock
\shownote{Accessed August 11, 2025}.


\bibitem[Raykov and Marcoulides(2011)]%
        {raykov2011introduction}
\bibfield{author}{\bibinfo{person}{Tenko Raykov} {and} \bibinfo{person}{George~A Marcoulides}.} \bibinfo{year}{2011}\natexlab{}.
\newblock \bibinfo{booktitle}{\emph{Introduction to psychometric theory}}.
\newblock \bibinfo{publisher}{Routledge}.
\newblock


\bibitem[Reagan(2016)]%
        {reagan2016}
\bibfield{author}{\bibinfo{person}{Miranda~Talley Reagan}.} \bibinfo{year}{2016}\natexlab{}.
\newblock \showarticletitle{The Benefits of the STEM Mindset: Risk Taking and Resilience}.
\newblock  (\bibinfo{year}{2016}).
\newblock


\bibitem[Reicherts et~al\mbox{.}(2025)]%
        {reicherts2025ai}
\bibfield{author}{\bibinfo{person}{Leon Reicherts}, \bibinfo{person}{Zelun~Tony Zhang}, \bibinfo{person}{Elisabeth von Oswald}, \bibinfo{person}{Yuanting Liu}, \bibinfo{person}{Yvonne Rogers}, {and} \bibinfo{person}{Mariam Hassib}.} \bibinfo{year}{2025}\natexlab{}.
\newblock \showarticletitle{AI, help me think—but for myself: Assisting people in complex decision-making by providing different kinds of cognitive support}. In \bibinfo{booktitle}{\emph{Proceedings of the 2025 CHI Conference on Human Factors in Computing Systems}}. \bibinfo{pages}{1--19}.
\newblock


\bibitem[Risko and Gilbert(2016)]%
        {risko2016cognitive}
\bibfield{author}{\bibinfo{person}{Evan~F Risko} {and} \bibinfo{person}{Sam~J Gilbert}.} \bibinfo{year}{2016}\natexlab{}.
\newblock \showarticletitle{Cognitive offloading}.
\newblock \bibinfo{journal}{\emph{Trends in Cognitive Sciences}} \bibinfo{volume}{20}, \bibinfo{number}{9} (\bibinfo{year}{2016}), \bibinfo{pages}{676--688}.
\newblock


\bibitem[Roseman and Smith(2001)]%
        {roseman2001appraisal}
\bibfield{author}{\bibinfo{person}{Ira~J Roseman} {and} \bibinfo{person}{Craig~A Smith}.} \bibinfo{year}{2001}\natexlab{}.
\newblock \showarticletitle{Appraisal theory}.
\newblock \bibinfo{journal}{\emph{Appraisal processes in emotion: Theory, methods, research}} (\bibinfo{year}{2001}), \bibinfo{pages}{3--19}.
\newblock


\bibitem[Russo(2024)]%
        {russo2024navigating}
\bibfield{author}{\bibinfo{person}{Daniel Russo}.} \bibinfo{year}{2024}\natexlab{}.
\newblock \showarticletitle{Navigating the complexity of generative {AI} adoption in software engineering}.
\newblock \bibinfo{journal}{\emph{ACM Transactions on Software Engineering and Methodology}} (\bibinfo{year}{2024}).
\newblock


\bibitem[Russo and Stol(2021)]%
        {russo2021pls}
\bibfield{author}{\bibinfo{person}{Daniel Russo} {and} \bibinfo{person}{Klaas-Jan Stol}.} \bibinfo{year}{2021}\natexlab{}.
\newblock \showarticletitle{{PLS-SEM} for Software Engineering Research: An Introduction and Survey}.
\newblock \bibinfo{journal}{\emph{ACM Comput Surv}} \bibinfo{volume}{54}, \bibinfo{number}{4} (\bibinfo{year}{2021}).
\newblock


\bibitem[Sakellariou and Fang(2021)]%
        {sakellariou2021self}
\bibfield{author}{\bibinfo{person}{Chris Sakellariou} {and} \bibinfo{person}{Zheng Fang}.} \bibinfo{year}{2021}\natexlab{}.
\newblock \showarticletitle{Self-efficacy and interest in STEM subjects as predictors of the STEM gender gap in the US: The role of unobserved heterogeneity}.
\newblock \bibinfo{journal}{\emph{International Journal of Educational Research}}  \bibinfo{volume}{109} (\bibinfo{year}{2021}), \bibinfo{pages}{101821}.
\newblock


\bibitem[Sarstedt and Cheah(2019)]%
        {sarstedt2019partial}
\bibfield{author}{\bibinfo{person}{Marko Sarstedt} {and} \bibinfo{person}{Jun-Hwa Cheah}.} \bibinfo{year}{2019}\natexlab{}.
\newblock \bibinfo{title}{Partial least squares structural equation modeling using Smart{PLS}: a software review}.
\newblock


\bibitem[Sarstedt et~al\mbox{.}(2019)]%
        {sarstedt2019specify}
\bibfield{author}{\bibinfo{person}{Marko Sarstedt}, \bibinfo{person}{Joseph~F. Hair}, \bibinfo{person}{Jun-Hwa Cheah}, {et~al\mbox{.}}} \bibinfo{year}{2019}\natexlab{}.
\newblock \showarticletitle{How to specify, estimate, and validate higher-order constructs in PLS-SEM}.
\newblock \bibinfo{journal}{\emph{Australas. Mark. J.}} \bibinfo{volume}{27}, \bibinfo{number}{3} (\bibinfo{year}{2019}), \bibinfo{pages}{197--211}.
\newblock


\bibitem[Sarstedt et~al\mbox{.}(2017)]%
        {sarstedt2017treating}
\bibfield{author}{\bibinfo{person}{Marko Sarstedt}, \bibinfo{person}{Christian~M Ringle}, {and} \bibinfo{person}{Joseph~F Hair}.} \bibinfo{year}{2017}\natexlab{}.
\newblock \showarticletitle{Treating unobserved heterogeneity in {PLS-SEM}: A multi-method approach}.
\newblock In \bibinfo{booktitle}{\emph{Partial Least Squares Path Modeling}}. \bibinfo{publisher}{Springer}, \bibinfo{pages}{197--217}.
\newblock


\bibitem[Schmidt(2020)]%
        {schmidt2020interactiveHumanAI}
\bibfield{author}{\bibinfo{person}{Albrecht Schmidt}.} \bibinfo{year}{2020}\natexlab{}.
\newblock \showarticletitle{Interactive human centered artificial intelligence: a definition and research challenges}. In \bibinfo{booktitle}{\emph{Proceedings of the 2020 International Conference on Advanced Visual Interfaces}}. \bibinfo{pages}{1--4}.
\newblock


\bibitem[Shneiderman(2022)]%
        {shneiderman2022humancenteredAI}
\bibfield{author}{\bibinfo{person}{Ben Shneiderman}.} \bibinfo{year}{2022}\natexlab{}.
\newblock \bibinfo{booktitle}{\emph{Human-centered AI}}.
\newblock \bibinfo{publisher}{Oxford University Press}.
\newblock


\bibitem[Shull et~al\mbox{.}(2007)]%
        {shull2007guide}
\bibfield{author}{\bibinfo{person}{Forrest Shull}, \bibinfo{person}{Janice Singer}, {and} \bibinfo{person}{Dag~IK Sj{\o}berg}.} \bibinfo{year}{2007}\natexlab{}.
\newblock \bibinfo{title}{Guide to Advanced Empirical Software Engineering}.
\newblock


\bibitem[Silver et~al\mbox{.}(2023)]%
        {silver2023using}
\bibfield{author}{\bibinfo{person}{Naomi Silver}, \bibinfo{person}{Matthew Kaplan}, \bibinfo{person}{Danielle LaVaque-Manty}, {and} \bibinfo{person}{Deborah Meizlish}.} \bibinfo{year}{2023}\natexlab{}.
\newblock \bibinfo{booktitle}{\emph{Using reflection and metacognition to improve student learning: Across the disciplines, across the academy}}.
\newblock \bibinfo{publisher}{Taylor \& Francis}.
\newblock


\bibitem[Singer-Freeman et~al\mbox{.}(2025)]%
        {singer2025generative}
\bibfield{author}{\bibinfo{person}{Karen~E Singer-Freeman}, \bibinfo{person}{Kristi Verbeke}, {and} \bibinfo{person}{Betsy Barre}.} \bibinfo{year}{2025}\natexlab{}.
\newblock \showarticletitle{Generative AI Use Among University Students Depends on Academic Level and Task}.
\newblock \bibinfo{journal}{\emph{Higher Learning Research Communications}} \bibinfo{volume}{15}, \bibinfo{number}{2} (\bibinfo{year}{2025}), \bibinfo{pages}{8}.
\newblock


\bibitem[{SmartPLS Team}(2024)]%
        {smartpls_website}
\bibfield{author}{\bibinfo{person}{{SmartPLS Team}}.} \bibinfo{year}{2024}\natexlab{}.
\newblock \bibinfo{title}{SmartPLS (Version 4.1.0) [Computer software]}.
\newblock
\urldef\tempurl%
\url{https://smartpls.com/}
\showURL{%
\tempurl}
\newblock
\shownote{Accessed: 2024-07-07}.


\bibitem[Smith(1996)]%
        {smith1996cooperative}
\bibfield{author}{\bibinfo{person}{Karl~A Smith}.} \bibinfo{year}{1996}\natexlab{}.
\newblock \showarticletitle{Cooperative learning: Making" groupwork" work}.
\newblock \bibinfo{journal}{\emph{New directions for teaching and learning}} \bibinfo{number}{67} (\bibinfo{year}{1996}).
\newblock


\bibitem[Sosu(2013)]%
        {sosu2013development}
\bibfield{author}{\bibinfo{person}{Edward~M Sosu}.} \bibinfo{year}{2013}\natexlab{}.
\newblock \showarticletitle{The development and psychometric validation of a Critical Thinking Disposition Scale}.
\newblock \bibinfo{journal}{\emph{Thinking Skills and Creativity}}  \bibinfo{volume}{9} (\bibinfo{year}{2013}), \bibinfo{pages}{107--119}.
\newblock


\bibitem[Spair(2025)]%
        {spair2025techemployer}
\bibfield{author}{\bibinfo{person}{Rick Spair}.} \bibinfo{year}{2025}\natexlab{}.
\newblock \bibinfo{booktitle}{\emph{It Ain't Your Father's Tech Employer Anymore: The Shift in Job Security and Workplace Culture}}.
\newblock LinkedIn.
\newblock
\urldef\tempurl%
\url{https://www.linkedin.com/pulse/aint-your-fathers-tech-employer-anymore-shift-job-security-rick-spair-hw3ze/}
\showURL{%
\tempurl}
\newblock
\shownote{Accessed: 2026-01-07}.


\bibitem[Sparrow et~al\mbox{.}(2011)]%
        {sparrow2011google}
\bibfield{author}{\bibinfo{person}{Betsy Sparrow}, \bibinfo{person}{Jenny Liu}, {and} \bibinfo{person}{Daniel~M Wegner}.} \bibinfo{year}{2011}\natexlab{}.
\newblock \showarticletitle{Google effects on memory: Cognitive consequences of having information at our fingertips}.
\newblock \bibinfo{journal}{\emph{Science}} \bibinfo{volume}{333}, \bibinfo{number}{6043} (\bibinfo{year}{2011}), \bibinfo{pages}{776--778}.
\newblock


\bibitem[Stadler et~al\mbox{.}(2024)]%
        {stadler2024cognitive}
\bibfield{author}{\bibinfo{person}{Matthias Stadler}, \bibinfo{person}{Maria Bannert}, {and} \bibinfo{person}{Michael Sailer}.} \bibinfo{year}{2024}\natexlab{}.
\newblock \showarticletitle{Cognitive ease at a cost: LLMs reduce mental effort but compromise depth in student scientific inquiry}.
\newblock \bibinfo{journal}{\emph{Computers in Human Behavior}}  \bibinfo{volume}{160} (\bibinfo{year}{2024}), \bibinfo{pages}{108386}.
\newblock


\bibitem[Sternberg and Grigorenko(1997)]%
        {sternberg1997cognitive}
\bibfield{author}{\bibinfo{person}{Robert~J Sternberg} {and} \bibinfo{person}{Elena~L Grigorenko}.} \bibinfo{year}{1997}\natexlab{}.
\newblock \showarticletitle{Are cognitive styles still in style?}
\newblock \bibinfo{journal}{\emph{American Psychologist}} \bibinfo{volume}{52}, \bibinfo{number}{7} (\bibinfo{year}{1997}), \bibinfo{pages}{700}.
\newblock


\bibitem[Stol and Fitzgerald(2018)]%
        {stol2018abc}
\bibfield{author}{\bibinfo{person}{Klaas-Jan Stol} {and} \bibinfo{person}{Brian Fitzgerald}.} \bibinfo{year}{2018}\natexlab{}.
\newblock \showarticletitle{The {ABC} of software engineering research}.
\newblock \bibinfo{journal}{\emph{ACM TOSEM}} \bibinfo{volume}{27}, \bibinfo{number}{3} (\bibinfo{year}{2018}).
\newblock


\bibitem[Stone(2024)]%
        {Stone2024}
\bibfield{author}{\bibinfo{person}{Brian~W. Stone}.} \bibinfo{year}{2024}\natexlab{}.
\newblock \showarticletitle{Generative AI in Higher Education: Uncertain Students, Ambiguous Use Cases, and Mercenary Perspectives}.
\newblock \bibinfo{journal}{\emph{Teaching of Psychology}} (\bibinfo{year}{2024}).
\newblock
\urldef\tempurl%
\url{https://www.researchgate.net/publication/387284657_Generative_AI_in_Higher_Education_Uncertain_Students_Ambiguous_Use_Cases_and_Mercenary_Perspectives}
\showURL{%
\tempurl}
\newblock
\shownote{41\% of students reported using AI in ways explicitly banned by policies}.


\bibitem[Stone(1974)]%
        {stone1974cross}
\bibfield{author}{\bibinfo{person}{Mervyn Stone}.} \bibinfo{year}{1974}\natexlab{}.
\newblock \showarticletitle{Cross-validatory choice and assessment of statistical predictions}.
\newblock \bibinfo{journal}{\emph{J R Stat Soc Series B Stat Methodol}}  \bibinfo{volume}{36} (\bibinfo{year}{1974}).
\newblock


\bibitem[Stringer(2016)]%
        {Finding_Success_in_Failure_2016}
\bibfield{author}{\bibinfo{person}{Kate Stringer}.} \bibinfo{year}{2016}\natexlab{}.
\newblock \bibinfo{booktitle}{\emph{Finding Success in Failure: STEM Educators Say Student Risk-Taking Is Key to Real-World Learning}}.
\newblock The 74.
\newblock
\urldef\tempurl%
\url{https://www.the74million.org/article/finding-success-in-failure-stem-educators-say-student-risk-taking-is-key-to-real-world-learning/}
\showURL{%
\tempurl}
\newblock
\shownote{Accessed: 2026-01-08}.


\bibitem[Strzelecki(2025)]%
        {strzelecki2025chatgpt}
\bibfield{author}{\bibinfo{person}{Artur Strzelecki}.} \bibinfo{year}{2025}\natexlab{}.
\newblock \showarticletitle{ChatGPT in higher education: Investigating bachelor and master students’ expectations towards AI tool}.
\newblock \bibinfo{journal}{\emph{Education and Information Technologies}} \bibinfo{volume}{30}, \bibinfo{number}{8} (\bibinfo{year}{2025}), \bibinfo{pages}{10231--10255}.
\newblock


\bibitem[Trinkenreich et~al\mbox{.}(2023)]%
        {trinkenreich2023belong}
\bibfield{author}{\bibinfo{person}{Bianca Trinkenreich}, \bibinfo{person}{Klaas-Jan Stol}, \bibinfo{person}{Anita Sarma}, \bibinfo{person}{Daniel~M German}, \bibinfo{person}{Marco~A Gerosa}, {and} \bibinfo{person}{Igor Steinmacher}.} \bibinfo{year}{2023}\natexlab{}.
\newblock \showarticletitle{Do I belong? modeling sense of virtual community among linux kernel contributors}. In \bibinfo{booktitle}{\emph{2023 IEEE/ACM 45th International Conference on Software Engineering (ICSE)}}. IEEE, \bibinfo{pages}{319--331}.
\newblock


\bibitem[Vereschak et~al\mbox{.}(2021)]%
        {vereschak2021evaluate}
\bibfield{author}{\bibinfo{person}{Oleksandra Vereschak}, \bibinfo{person}{Gilles Bailly}, {and} \bibinfo{person}{Baptiste Caramiaux}.} \bibinfo{year}{2021}\natexlab{}.
\newblock \showarticletitle{How to evaluate trust in {AI}-assisted decision making? A survey of empirical methodologies}.
\newblock \bibinfo{journal}{\emph{Human-Computer Interaction}} \bibinfo{volume}{5}, \bibinfo{number}{CSCW2} (\bibinfo{year}{2021}), \bibinfo{pages}{1--39}.
\newblock


\bibitem[Vorvoreanu et~al\mbox{.}(2019)]%
        {vorvoreanu2019-gmInMicrosoftApp-CHI}
\bibfield{author}{\bibinfo{person}{Mihaela Vorvoreanu}, \bibinfo{person}{Lingyi Zhang}, \bibinfo{person}{Yun-Han Huang}, \bibinfo{person}{Claudia Hilderbrand}, \bibinfo{person}{Zoe Steine-Hanson}, {and} \bibinfo{person}{Margaret Burnett}.} \bibinfo{year}{2019}\natexlab{}.
\newblock \showarticletitle{From gender biases to gender-inclusive design: An empirical investigation}. In \bibinfo{booktitle}{\emph{Proceedings of the 2019 CHI Conference on Human Factors in Computing Systems}}. \bibinfo{pages}{1--14}.
\newblock


\bibitem[Walker and Vorvoreanu(2025)]%
        {walker2025learningOutcomes}
\bibfield{author}{\bibinfo{person}{Kathleen Walker} {and} \bibinfo{person}{Mihaela Vorvoreanu}.} \bibinfo{year}{2025}\natexlab{}.
\newblock \bibinfo{booktitle}{\emph{Learning Outcomes with GenAI in the Classroom: A Review of Empirical Evidence}}.
\newblock \bibinfo{type}{{T}echnical {R}eport}. \bibinfo{institution}{Microsoft Aether Psi Working Group}.
\newblock


\bibitem[Wang et~al\mbox{.}(2023)]%
        {wang2023investigating}
\bibfield{author}{\bibinfo{person}{Ruotong Wang}, \bibinfo{person}{Ruijia Cheng}, \bibinfo{person}{Denae Ford}, {and} \bibinfo{person}{Thomas Zimmermann}.} \bibinfo{year}{2023}\natexlab{}.
\newblock \showarticletitle{Investigating and designing for trust in {AI}-powered code generation tools}.
\newblock \bibinfo{journal}{\emph{arXiv preprint arXiv:2305.11248}} (\bibinfo{year}{2023}).
\newblock


\bibitem[Wentzel(1991)]%
        {wentzel1991social}
\bibfield{author}{\bibinfo{person}{Kathryn~R Wentzel}.} \bibinfo{year}{1991}\natexlab{}.
\newblock \showarticletitle{Social competence at school: Relation between social responsibility and academic achievement}.
\newblock \bibinfo{journal}{\emph{Review of educational research}} \bibinfo{volume}{61}, \bibinfo{number}{1} (\bibinfo{year}{1991}), \bibinfo{pages}{1--24}.
\newblock


\bibitem[Wischnewski et~al\mbox{.}(2023)]%
        {wischnewski2023measuring}
\bibfield{author}{\bibinfo{person}{Magdalena Wischnewski}, \bibinfo{person}{Nicole Kr{\"a}mer}, {and} \bibinfo{person}{Emmanuel M{\"u}ller}.} \bibinfo{year}{2023}\natexlab{}.
\newblock \showarticletitle{Measuring and understanding trust calibrations for automated systems: a survey of the state-of-the-art and future directions}. In \bibinfo{booktitle}{\emph{Proceedings of the 2023 CHI Conference on Human Factors in Computing Systems}}. \bibinfo{pages}{1--16}.
\newblock


\bibitem[Yan et~al\mbox{.}(2025)]%
        {yan2025beyond}
\bibfield{author}{\bibinfo{person}{Lixiang Yan}, \bibinfo{person}{Viktoria Pammer-Schindler}, \bibinfo{person}{Caitlin Mills}, \bibinfo{person}{Andy Nguyen}, {and} \bibinfo{person}{Dragan Ga{\v{s}}evi{\'c}}.} \bibinfo{year}{2025}\natexlab{}.
\newblock \bibinfo{title}{Beyond efficiency: Empirical insights on generative AI's impact on cognition, metacognition and epistemic agency in learning}.
\newblock \bibinfo{numpages}{1675--1685}~pages.
\newblock


\bibitem[Zhai et~al\mbox{.}(2024)]%
        {zhai2024effects}
\bibfield{author}{\bibinfo{person}{Chunpeng Zhai}, \bibinfo{person}{Santoso Wibowo}, {and} \bibinfo{person}{Lily~D Li}.} \bibinfo{year}{2024}\natexlab{}.
\newblock \showarticletitle{The effects of over-reliance on AI dialogue systems on students' cognitive abilities: a systematic review}.
\newblock \bibinfo{journal}{\emph{Smart Learning Environments}} \bibinfo{volume}{11}, \bibinfo{number}{1} (\bibinfo{year}{2024}), \bibinfo{pages}{28}.
\newblock


\bibitem[Zhang and Xu(2025)]%
        {zhang2025paradox}
\bibfield{author}{\bibinfo{person}{Ling Zhang} {and} \bibinfo{person}{Junzhou Xu}.} \bibinfo{year}{2025}\natexlab{}.
\newblock \showarticletitle{The paradox of self-efficacy and technological dependence: Unraveling generative AI's impact on university students' task completion}.
\newblock \bibinfo{journal}{\emph{The Internet and Higher Education}}  \bibinfo{volume}{65} (\bibinfo{year}{2025}), \bibinfo{pages}{100978}.
\newblock


\bibitem[Zuriguel-P{\'e}rez et~al\mbox{.}(2022)]%
        {zuriguel2022nursing}
\bibfield{author}{\bibinfo{person}{Esperanza Zuriguel-P{\'e}rez}, \bibinfo{person}{Mar{\'\i}a-Teresa Lluch-Canut}, \bibinfo{person}{Montserrat Puig-Llobet}, \bibinfo{person}{Luis Basco-Prado}, \bibinfo{person}{Adri{\`a} Almazor-Sirvent}, \bibinfo{person}{Ainoa Biurrun-Garrido}, \bibinfo{person}{Mariela~Patricia Aguayo-Gonz{\'a}lez}, \bibinfo{person}{Olga Mestres-Soler}, {and} \bibinfo{person}{Juan Rold{\'a}n-Merino}.} \bibinfo{year}{2022}\natexlab{}.
\newblock \showarticletitle{The nursing critical thinking in clinical practice questionnaire for nursing students: A psychometric evaluation study}.
\newblock \bibinfo{journal}{\emph{Nurse Education in Practice}}  \bibinfo{volume}{65} (\bibinfo{year}{2022}), \bibinfo{pages}{103498}.
\newblock


\end{thebibliography}

\end{document}